\documentclass[preprint,11pt]{elsarticle}

 \usepackage{epsfig}

\usepackage{amssymb}
\usepackage{graphics,amssymb,amsmath}
\usepackage{graphicx}
\usepackage{subfigure}
\newproof{pf}{Proof}
\newdefinition{rmk}{Remark}
\newtheorem{thm}{Theorem}
\newdefinition{prop}{Proposition}

\begin{document}

\begin{frontmatter}



\title{Exponentiated Weibull Power Series Distributions and its Applications}


\author{Eisa Mahmoudi\corref{cor1}}
\ead{emahmoudi@yazduni.ac.ir}
\author{Mitra Shiran}

\cortext[cor1]{Corresponding author}

\address{Department of Statistics, Yazd University,
P.O. Box 89175-741, Yazd, Iran}

\begin{abstract}
In this paper we introduce the exponentiated Weibull power series (EWPS)
class of distributions which is obtained by compounding exponentiated Weibull
and power series distributions, where the compounding procedure follows
same way that was previously carried out by Roman et al. (2010) and
Cancho et al. (2011) in introducing the complementary
exponential-geometric (CEG) and the two-parameter
Poisson-exponential (PE) lifetime distributions, respectively. This
distribution contains several lifetime models such as: exponentiated
weibull-geometric (EWG), exponentiated weibull-binomial (EWB),
exponentiated weibull-poisson (EWP), exponentiated weibull-logarithmic (EWL)
distributions as a special case.

The hazard rate function of the EWPS distribution can be increasing, decreasing,
bathtub-shaped and unimodal failure rate among others. We obtain several properties
of the EWPS distribution such as its probability density function, its reliability and
failure rate functions, quantiles and moments. The maximum
likelihood estimation procedure via a EM-algorithm is presented in
this paper. Sub-models of the EWPS distribution are studied in
details. In the end, Applications to two real data sets are given to
show the flexibility and potentiality of the EWPS distribution.

\end{abstract}


\end{frontmatter}


\section{Introduction}

The Weibull and exponentiated Weibull (EW) distributions in spite of
their simplicity in solving many problems in lifetime and
reliability studies, do not provide a reasonable parametric fit to
some practical applications.

Recently, attempts have been made to define new families of
probability distributions that extend well-known families of
distributions and at the same time provide great flexibility in
modeling data in practice. One such class of distributions generated
by compounding the well-known lifetime distributions such as
exponential, Weibull, generalized exponential, exponentiated Weibull
and etc with some discrete distributions such as binomial,
geometric, zero-truncated Poisson, logarithmic and the power series
distributions in general. The non-negative random variable $Y$
denoting the lifetime of such a system is defined by ${{Y=\min
}_{1\le i\le N} X_i\ }$ or ${{Y=\max}_{1\le i\le N} X_i\ }$, where
the distribution of $X_i$ belongs to one of the lifetime
distributions and the random variable $N$ can have some discrete
distributions, mentioned above.

This new class of distributions has been received considerable
attention over the last years. The exponential geometric (EG),
exponential Poisson (EP), exponential logarithmic (EL), exponential
power series (EPS), Weibull geometric (WG), Weibull power series
(WPS), exponentiated exponential-Poisson (EEP), complementary
exponential geometric (CEG), two-parameter Poisson-exponential,
generalized exponential power series (GEPS), exponentiated
Weibull-Poisson (EWP) and generalized inverse Weibull-Poisson (GIWP)
distributions were introduced and studied by Adamidis and Loukas
\cite{Adamidis }, Kus \cite{Kus }, Tahmasbi and Rezaei
\cite{Tahmasbi }, Chahkandi and Ganjali \cite{Chahkandi},
Barreto-Souza et al. \cite{Barreto-Souza2011 }, Morais and
Barreto-Souza et al. \cite{Morais }, Barreto-Souza and Cribari-Neto
\cite{Barreto-Souza2009 }, Louzada-Neto et al. \cite{Louzada},
Cancho et al. \cite{Cancho }, Mahmoudi and Jafari
\cite{Mahmoudi2011a }, Mahmoudi and Sepahdar \cite{Mahmoudi2011b }
and Mahmoudi and Torki \cite{Mahmoudi2011c }.

In this paper we introduce the exponentiated Weibull power series (EWPS)
class of distributions which is obtained by compounding exponentiated Weibull
and power series distributions, where the compounding procedure follows
same way that was previously carried out by Roman et al. (2010) and
Cancho et al. (2011) in introducing the complementary
exponential-geometric (CEG) and the two-parameter
Poisson-exponential (PE) lifetime distributions, respectively. This
distribution contains several lifetime models such as: exponentiated
weibull-geometric (EWG), exponentiated weibull-binomial (EWB),
exponentiated weibull-poisson (EWP), exponentiated weibull-logarithmic (EWL)
distributions as a special case.

\section{ Exponentiated Weibull distribution: A brief review }
Mudholkar and Srivastava \cite{Mudholkar1993 } introduced the EW
family as extension of the Weibull family, which contains
distributions with bathtub-shaped and unimodal failure rates besides
a broader class of monotone failure rates. One can see Mudholkar et
al. \cite{Mudholkar1995 }, Mudholkar and Huston \cite{Mudholkar1996
}, Gupta and Kundu \cite{Gupta2001 }, Nassar and Eissa \cite{Nassar}
and Choudhury \cite{Choudhury} for applications of the EW
distribution in reliability and survival studies.

The random variable $X$ has an EW distribution if its cumulative
distribution function (cdf) takes the form
\begin{equation}\label{cdf ExW}
G_{X}(x)=\left(1-e^{-(\beta x)^{\gamma}}\right)^{\alpha},~~x>0,
\end{equation}
where $\gamma>0$, $\alpha>0$ and $\beta>0$, which is denoted by
$EW(\alpha,\beta,\gamma)$. The corresponding probability density
function (pdf) is
\begin{equation}\label{pdf ExW}
g_X(x)=\alpha\gamma\beta^{\gamma}x^{\gamma-1}e^{-(\beta
x)^{\gamma}}\left(1-e^{-(\beta x)^{\gamma}}\right)^{\alpha-1}.
\end{equation}
The survival and hazard rate functions of the EW distribution are
\begin{equation*}
S(x)=1-\left(1-e^{-(\beta x)^{\gamma}}\right)^{\alpha},
\end{equation*}
and
\begin{equation*}
h(x)= \alpha\gamma\beta^{\gamma}x^{\gamma-1}e^{-(\beta
x)^{\gamma}}\left(1-e^{-(\beta
x)^{\gamma}}\right)^{\alpha-1}\Big\{\left[1-\left(1-e^{-(\beta
x)^{\gamma}}\right)^{\alpha}\right]\Big\}^{-1},
\end{equation*}
respectively. The $k$th moment about zero of the EW distribution is
given by
\begin{equation}\label{mean EW}
E(X^{k})=\alpha\beta^{-k}\Gamma\left(\frac{k}{\gamma}+1\right)\sum_{j=0}^{\infty}(-1)^j
{\alpha-1 \choose j}(j+1)^{-(\frac{k}{\gamma}+1)}.
\end{equation}
Note that for positive integer values of $\alpha$, the index $j$ in
previous sum stops at $\alpha-1$, and the above expression takes the
closed form
\begin{equation}
E(X^{k})=\alpha\beta^{-k}\Gamma\left(\frac{k}{\gamma}+1\right)A_{k}(\gamma),
\end{equation}
where
\begin{equation}\label{v.p}
A_{k}(\gamma)=1+\sum_{j=1}^{\alpha-1}(-1)^j {\alpha-1 \choose
j}(j+1)^{-(\frac{k}{\gamma}+1)},~~~k=1,2,3,\cdots,
\end{equation}
in which $\Gamma(\frac{k}{\gamma}+1)$ denotes the gamma function
(see, Nassar and Eissa (2003) for more detail).

\begin{table}[htp!]
\centering \caption{Useful quantities of some power series distributions.}
\begin{small}
\begin{tabular}{|l|lcccccc|}
\hline
Distribution  & ~~~ $a‎_{n}$‎ & $C(‎\theta‎)$  & $C^{\prime‎}(‎\theta‎)  $ &$C^{\prime \prime}(‎\theta‎)$& $C(‎\theta‎)^{-1}$ & $S $&  \\
\hline Poisson & $\begin{array}{l}
n!^{-1}
\end{array}$
&$e^{‎\theta‎}-1$ &$e^{‎\theta‎}$  &$e^{‎\theta‎}$ & $‎\log‎(‎\theta+1‎)$&$‎\infty‎$&\\

Logarithmic & $\begin{array}{l}
n^{-1}
\end{array}$
&$-‎‎\log‎(1-‎\theta‎)$ &$‎(1-‎\theta‎)^{-1}$ &$‎(1-‎\theta‎)^{-2}$ &$1-e^{-‎\theta‎}$&1& \\

Geometric & $\begin{array}{l}
 1
\end{array}$
&$‎\theta‎‎(1-‎\theta‎)^{-1}$ &$‎(1-‎\theta‎)^{-2}$  &$2‎(1-‎\theta‎)^{-3}$& $‎\theta‎‎(1+‎\theta‎)^{-1}$&1& \\

Binomial & ${m \choose n}$& $‎‎(‎\theta+1‎‎)^{m}-1$  &$m‎‎(‎\theta+1‎‎)^{m-1}$ & $‎\frac{m(m-1)}{(‎\theta+1‎‎)^{2-m}}‎$ & $(‎\theta-1‎)^{‎\frac{1}{m}‎}-1$ &$‎\infty‎$&  \\

\hline
\end{tabular}
\end{small}
\end{table}

\section{ The class of EWPS distribution }
Consider the random variable $X$ having the EW distribution where
its cdf and pdf are given in (\ref{cdf ExW}) and (\ref{pdf ExW}).\\
Given $N$, let $X_{1},\cdots,X_{N}$ be independent and identically
distributed (iid) random variables from EW distribution. Let the
random variable $N$ is distributed according to the power series
distribution with pdf
\begin{equation*}
P(N=n)=\frac{a_{n}\theta^{n}}{C(\theta)},~n=1,2
,\cdots,
\end{equation*}
where $a_{n}\geq 0$ depends only on n, $C(\theta)=\Sigma _{n=1}^{\infty}a_{n}\theta^{n}$, $\theta\in (0,s)$ is such that $C(\theta)$ is finite. For more details on the power series class of distributions, see Noack (1950).Table 1 shows useful quantities of some power series distributions (truncated at zero) such as poisson, logarithmic, geometric and binomial (with m being the number of replicas) distributions. \\
Let $Y=\max(X_{1},\cdots,X_{N})$, then the conditional cdf of
$Y|N=n$ is given  by
\begin{equation}\label{dist y given N}
G_{Y|N=n}(y)=(G(y))^{n}=\left(1-e^{-(\beta y)^{\gamma}}\right)^{n\alpha},
\end{equation}
which is the EW distribution with parameters $n\alpha$, $\beta$,
$\gamma$, and denoted by EW$(n\alpha,\beta,\gamma)$. The
exponentiated Weibull power series (EWPS) distribution, denoted by
EWPS $(\alpha,\beta,\gamma,\theta)$, is defined by the marginal cdf of
$Y$, i.e.,
\begin{equation}\label{cdf EWPS}
F_Y(y)=\sum _{n=1}^{\infty} \frac{a_{n}\theta^{n}}{C(\theta)}(G(y))^{n}=\frac{C\left(\theta\left(1-e^{-(\beta y)^{\gamma}}\right)^{\alpha}\right)}{C(\theta)},~~y>0.
\end{equation}

\begin{rmk}
Let $Y=\min (X_{1},\cdots,X_{N})$, then the cdf of $Y$ is given by
\begin{equation}\label{cdf2 EWPS}
F_{Y}(y)=1-\frac{C\left(\theta\left(1-e^{-(\beta y)^{\gamma}}\right)^{\alpha}\right)}{C(\theta)},~~y>0.
\end{equation}
If $‎\alpha‎=1$, then the cdf of $Y$ is  $F_{Y}(y)=1-\frac{C\left(\theta\left(1-e^{-(\beta y)^{\gamma}}\right)\right)}{C(\theta)}$, which is called Weibull Power Series distributions (Morais and Barreto-Souza, 2011) and this family includes the life time distribution presented by Barreto-Souza et al. (2010a), Barreto-Souza et al. (2010b).   which
$X_{i}$'s has the exponentiated Weibull distribution is obtained. The EG distribution (Adamidis and Loukas, 1998) is obtained by taking $C(‎\theta‎)=‎\theta‎‎(1-‎\theta‎)^{-1}$ with $‎\theta‎\in (0,1)‎‎$ and $‎\alpha=1,‎\gamma=1‎‎$ in  (\ref{cdf2 EWPS}). Moreover, for $‎\alpha=1,‎\gamma=1‎‎$, we obtain the EP distribution (Kus, 2007) and the EL distribution (Tahmasbi and Rezaei, 2008) by taking $C(‎\theta‎)=‎e^{‎\theta‎}-1, ‎\theta‎\>0‎‎$, and $C(‎\theta‎)=‎-‎‎\log‎(1-‎\theta‎), \theta‎\in (0,1)$, respectively. The WG distribution (Barreto-Souza et al. (2010a), Barreto-Souza et al. (2010b) ) is obtained by taking $C(‎\theta‎)=‎\theta‎‎(1-‎\theta‎)^{-1}$ with $‎\theta‎\in (0,1)‎‎$ and $‎\alpha=1‎$ in  (\ref{cdf2 EWPS}).     The EWG distribution is obtained by considering $C(‎\theta‎)=‎\theta‎‎(1-‎\theta‎)^{-1}$ with $‎\theta‎\in (0,1)‎‎$ in  (\ref{cdf2 EWPS}).
\end{rmk}

The pdf of the EWPS distribution is given by
\begin{equation}\label{pdf EWPS}
f_Y(y)=\theta\alpha\gamma\beta^{\gamma}y^{\gamma-1} e^{-(\beta y)^{\gamma}}\left(1-e^{-(\beta y)^{\gamma}}\right)^{\alpha-1} \frac{C^{\prime}\left(\theta \left(1-e^{-(\beta y)^{\gamma}}\right)^{\alpha}\right)}{C(\theta)},
\end{equation}
where $\alpha,\beta,\gamma>0$ and $\theta\in (0,s)$.\\
The survival function and hazard rate function of the EWPS
distribution are given, respectively, by
\begin{equation}\label{servival EWPS}
S(y)=1-\frac{C \left(\theta \left(1-e^{-(\beta y)^{\gamma}}\right)^{\alpha}\right)}{C(\theta)},~~y>0,
\end{equation}
and
\begin{equation}
h(y)=\theta\alpha\gamma\beta^{\gamma}y^{\gamma-1} e^{-(\beta y)^{\gamma}}\left(1-e^{-(\beta y)^{\gamma}}\right)^{\alpha-1} \frac{C^{\prime}\left(\theta \left(1-e^{-(\beta y)^{\gamma}}\right)^{\alpha}\right)}{C(\theta)-C \left(\theta \left(1-e^{-(\beta y)^{\gamma}}\right)^{\alpha}\right)}.
\end{equation}
Consider $C(\theta)=‎\theta+‎\theta‎ ^{20}‎ ‎‎$. If $‎‎\beta=1‎‎$ and $‎\theta=1‎$, the plots of this density and its hazard rate function, for $‎\alpha=0.5, ‎\gamma=1, ‎\alpha=1, ‎\gamma=0.5,‎\alpha=2,‎\gamma=3,$ and $‎\alpha=2, ‎\gamma=1$ are given in Fig 1.

\begin{prop}
The limiting distribution of $EWPS(‎\alpha,‎\beta,‎\gamma,‎\theta‎‎‎‎)$ when $‎\theta‎\rightarrow 0^{+}‎‎$ is
\begin{equation*}
\begin{array}[b]{ll}
\lim‎_{\theta‎\rightarrow 0^{+}}‎‎‎ F(y)&=\lim‎_{\theta‎\rightarrow 0^{+}}‎\frac{C(\theta G(y))}{C(\theta)}=\lim‎_{\theta‎\rightarrow 0^{+}}‎\frac{\sum_{n=1}^{\infty} a_{n}‎\theta^{n}(G(y))^{n}‎}{\sum_{n=1}^{\infty} a_{n}‎\theta^{n}} \medskip\\
&=‎\lim‎_{\theta‎\rightarrow 0^{+}}‎‎\frac{ a_{c}(G(y))^{c}+\sum_{n=c+1}^{\infty} a_{n}‎\theta^{n-c}(G(y))^{n}}{a_{c}+\sum_{n=c+1}^{\infty} a_{n}‎\theta^{n-c}}‎\medskip\\ &=(G(y))^{c}=\left(1-e^{-(\beta y)^{\gamma}}\right)^{c\alpha}
\end{array}
\end{equation*}
which is a EW distribution with parameters $c‎\alpha‎$, $‎\gamma‎$ and $‎\beta‎$, where $c=min‎\lbrace‎n‎\in N‎: a_{n}‎\>0‎ ‎\rbrace‎$.
\end{prop}

\begin{prop}
The densities of EWPS class can be expressed as infinite linear combination of density of order distribution. We know that $$C^{\prime}(\theta)=\Sigma _{n=1}^{\infty}na_{n}\theta^{n-1}$$.\\Therefore,
\begin{equation*}
f_Y(y)=\theta\alpha\gamma\beta^{\gamma}y^{\gamma-1}e^{-(\beta
y)^{\gamma}}\left(1-e^{-(\beta y)^{\gamma}}\right)^{\alpha-1}\sum_{n=1}^{\infty}\frac{na_{n}}{C(\theta)}\left(\theta \left(1-e^{-(\beta y)^{\gamma}}\right)^{\alpha}\right)^{n-1}.
\end{equation*}
Using the EW density given before, we obtain
\begin{equation}\label{new EWPS}
f_{EWPS}(y;\alpha,\beta,\gamma,\theta)=\theta\sum_{n=1}^{\infty}\frac{\theta^{n-1} a_{n}}{C(\theta)}f_{EW}(y;n \alpha ,\beta,\gamma).
\end{equation}
Various mathematical properties (cdf, moments, percentiles, moment‎\textbf{•}‎
generating function, factorial moments, among others) of the EWPS
distribution for $|\theta|<1$ can be obtained from Eq. (\ref{new
EWPS}) and the corresponding properties of the EW distribution.
\end{prop}
\begin{prop}
The density of EWPS distribution can be expressed as infinite linear
combination of density of the biggest order statistic of
$X_{1},\cdots,X_{n}$, where $X_{i}\sim EW(\alpha,\beta,\gamma)$ for
$i=1,2,\cdots,n$. we have
\begin{equation*}
f_{EWPS}(y)=\sum^{\infty}_{n=1}(G(y))^{n}P(N=n)=\sum^{\infty}_{n=1}g_{X_{(n)}}(y)P(N=n),
\end{equation*}
in which $g_{X_{(n)}}(y)$ is the pdf of
$X_{(n)}=\max(X_{1},\cdots,X_{n})$.
\end{prop}

\begin{figure}[t]
\centering
\includegraphics[scale=0.45]{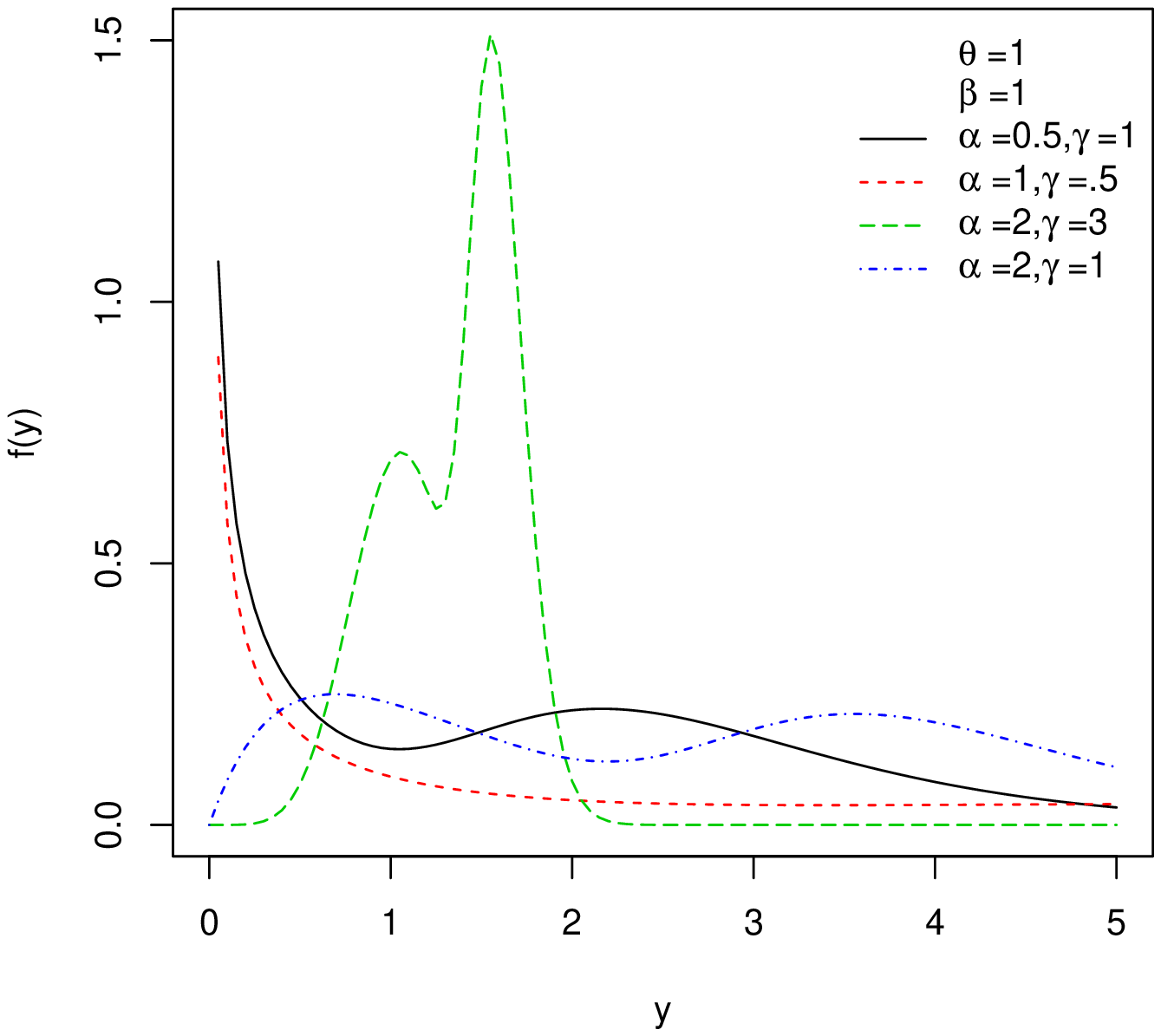}
\includegraphics[scale=0.45]{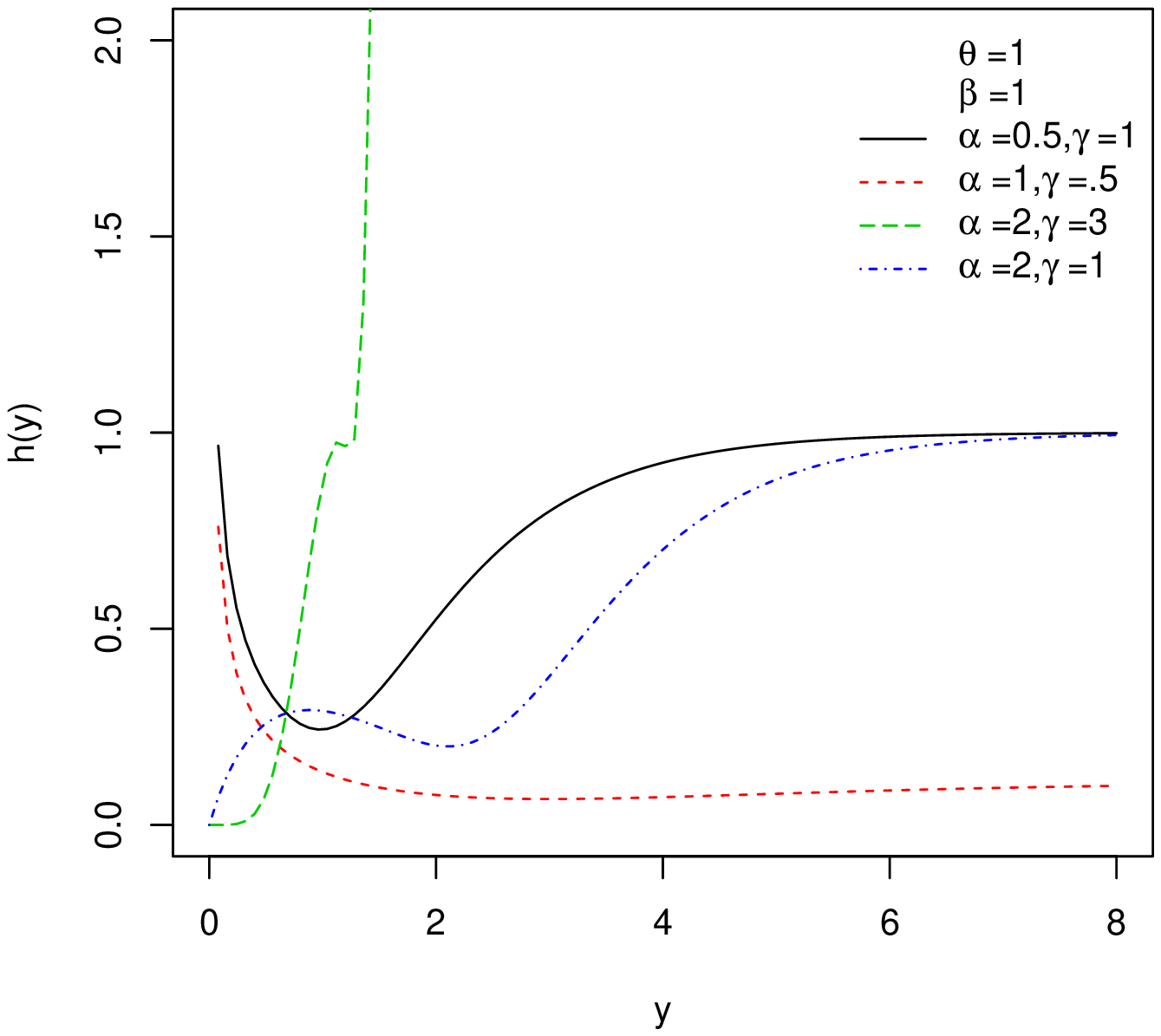}
\caption[]{Plots of pdf and hazard rate function of EWPS where $C(\theta)=‎\theta+‎\theta^{20}‎‎$.}
\end{figure}
 ‎‎‎‎‎
\section{ Quantiles and moments of the EWPS distribution }
The $q$th quantile of the EWPS  distribution is given by
\begin{equation*}
y_{q}=G^{-1}\left(\frac{C^{-1}(q C(\theta)}{\theta}\right),
\end{equation*}
Where $G^{-1}(y)=\frac{1}{\beta}\left(-\ln \left(1-y^{\frac{1}{\alpha}}\right)\right)^{\frac{1}{\gamma}}$, and $C^{-1}(.)$ is the inverse function of $C(.)$.
The $q$th quantile of the EWPS  distribution is used for data generation from the EWPS distribution. In
particular, the median of the EWPS distribution is given by
\begin{equation*}
y_{0.5}=G^{-1}\left(\frac{C^{-1}(0.5 C(\theta))}{\theta}\right).
\end{equation*}
\
Suppose that $Y\sim EWPS(\alpha,\beta,\gamma,\theta)$, and
$X_{(n)}=\max(X_{1},\cdots,X_{n})$, where $X_{i}\sim
EW(\alpha,\beta,\gamma)$ for $i=1,2,\cdots,n$,  then the $k$th
moment of $Y$ is given by
\begin{equation}\label{meank EWPS}
\begin{array}[b]{ll}
E(Y^{k})=E(E(Y^{k}|N))&=\sum^{\infty}_{n=1}P(N=n)E(Y^{k}|n)=\sum^{\infty}_{n=1}P(N=n)E(X^{k}_{(n)})\medskip\\
&=\alpha  \beta^{-k}\Gamma\left(\frac{k}{\gamma}+1\right)\sum^{\infty}_{n=1}\sum^{\infty}_{j=0}(-1)^j
\frac{a_{n}\theta^{n}}{C(\theta)}{n\alpha-1 \choose j} n (j+1)^{-(\frac{k}{\gamma}+1)}.
\end{array}
\end{equation}
For positive integer values of $\alpha$, the index $j$ in above
expression stops at $n\alpha-1$

Now we obtain the moment generating function of the EWPS  distribution
using the Eq. (\ref{meank EWPS}), as follow
\begin{equation}
\begin{array}[b]{l}\label{mgf EWPS}
M_{Y}(t)=\sum^{\infty}_{i=0}\frac{t^{i}}{i!}E(Y^{i})\medskip\\
=\sum^{\infty}_{i=0}\frac{t^{i}}{i!}\left[(1-\theta)\alpha
\beta^{-i}\Gamma\left(\frac{i}{\gamma}+1\right)\sum^{\infty}_{n=1}\sum^{\infty}_{j=0}(-1)^j
{n\alpha-1 \choose j}n
\theta^{n-1}(j+1)^{-(\frac{i}{\gamma}+1)}\right]\medskip\\
= \sum^{\infty}_{i=0}\sum^{\infty}_{n=1}\sum^{\infty}_{j=0}‎\frac{(-1)^j}{i!}‎(t/\beta)^{i}
 \frac{a_{n}\theta^{n}}{C(\theta)} {n\alpha-1 \choose
j}‎\frac{‎n ‎\alpha ‎\Gamma\left(\frac{i}{\gamma}+1\right)}{(j+1)^{-(\frac{i}{\gamma}+1)}}
.
\end{array}
\end{equation}
According to the Eq. (\ref{meank EWPS}), the mean and variance of the
EWPS  distribution are given, respectively, by
\begin{equation}\label{mean EWPS}
E(Y)=\alpha  \beta^{-1}\Gamma\left(\frac{1}{\gamma}+1\right)\sum^{\infty}_{n=1}\sum^{\infty}_{j=0}(-1)^j
\frac{a_{n}\theta^{n}}{C(\theta)}{n\alpha-1 \choose j} n (j+1)^{-(\frac{1}{\gamma}+1)},
\end{equation}
and
\begin{equation}\label{var EWPS}
Var(Y)=\alpha  \beta^{-2}\Gamma\left(\frac{2}{\gamma}+1\right)\sum^{\infty}_{n=1}\sum^{\infty}_{j=0}(-1)^j
\frac{a_{n}\theta^{n}}{C(\theta)}{n\alpha-1 \choose j} n (j+1)^{-(\frac{2}{\gamma}+1)}-E^{2}(Y),
\end{equation}
Where $E(Y)$ is given in Eq. (\ref{mean EWPS})

\section{ R\'{e}nyi and Shannon entropies}
Entropy has been used in various situations in science and
engineering. The entropy of a random $Y$ is a measure of variation
of the uncertainty. For a random variable with the pdf $f$, the
R\'{e}nyi entropy is defined by
$I_{R}(r)=\frac{1}{1-r}\log\{\int_{\mathbb{R}}f^{r}(y)dy\}$, for
$r>0$ and $r\neq 1$. For the EWPS  distribution, the power series
expansion gives

\begin{equation*}
\begin{array}[b]{ll}
\int^{\infty}_{0} f^{r}(y)dy&=\left[\frac{\alpha \theta \gamma\beta^{\gamma}}{C(\theta)}\right]^{r} \int^{\infty}_{0}e^{-r(\beta y)^\gamma}
(1-e^{-(\beta y)^\gamma})^{r \alpha -r}‎\frac{‎\left[C‎^{\prime}‎ (‎\theta(1-e^{-(\beta y)^\gamma})^{\alpha }‎‎)‎‎\right]‎^{r}‎ ‎}{‎\left[ ‎C(‎\theta‎)‎\right] ‎^{r}‎‎}‎dy
\end{array},
\end{equation*}
Applying the Equation
$‎\left(‎\sum^{\infty}_{i=0}w‎_{i}u‎^{i}‎‎\right)‎^{j}‎ ‎=‎\sum^{\infty}_{i=0}c‎_{i,j}u‎^{i}‎$,
where the coefficients $c‎_{i,j}$ for $i=1,2,...$ can be easily obtained from the recurrence relation
$c‎_{i,j}=(iw‎_{0}‎)‎^{-1}‎\sum^{i}_{m=1}(jm-i+m)w‎_{m}c‎_{i-m,j},$
white $c‎_{0,j}=w‎_{0}‎^{j}‎$ for $‎\left[C‎^{\prime}‎ (‎\theta(1-e^{-(\beta y)^\gamma})^{\alpha }‎‎)‎‎\right]‎^{r}‎ $and series expansion for $\left(1-(1-e^{-(\beta
x)^{\gamma}} )^{\alpha}\right)^{-(r+2)}$ gives
\begin{equation*}
\begin{array}[b]{ll}
\int^{\infty}_{0} f^{r}(y)dy&=\left[\frac{\alpha \theta \gamma\beta^{\gamma}}{C(\theta)}\right]^{r}\sum^{\infty}_{i=0} \sum^{\infty}_{j=0}(-1)‎^{j}‎c‎_{i,j}‎\theta ‎^{i}{\alpha(r+i)-r \choose j}‎‎\int^{\infty}_{0}e^{-(r+j)(\beta y)^\gamma}dy
\end{array},
\end{equation*}
But setting $u=(j+r)(\beta y)^\gamma$, gives
\begin{equation}\label{f^r}
\begin{array}[b]{ll}
\int^{\infty}_{0} f^{r}(y)dy&=\left[\frac{\alpha \theta}{C(\theta)}\right]^{r}‎\gamma‎ ^{r-1}‎‎\beta^{r\gamma-1}\Gamma(\frac{1}{\gamma})\sum_{i=0}^{\infty}\sum_{j=0}^{\infty}c‎_{i,j}‎\frac{‎(-1)^{j}\theta^{i}}{(j+r)^{\frac{1}{\gamma}}}{\alpha(r+i)-r \choose j}
\end{array}.
\end{equation}
Substituting from (\ref{f^r}), we obtain
\begin{equation}
I_{R}(r)=\frac{1}{1-r} \log \left\{\left[\frac{\alpha \theta}{C(\theta)}\right]^{r}‎\gamma‎ ^{r-1}‎‎\beta^{r\gamma-1}\Gamma(\frac{1}{\gamma})\sum_{i=0}^{\infty}\sum_{j=0}^{\infty}c‎_{i,j}‎\frac{‎(-1)^{j}\theta^{i}}{(j+r)^{\frac{1}{\gamma}}}{\alpha(r+i)-r \choose j} \right\}.
\end{equation}
The Shannon entropy which is defined by $E[-\log (f(Y))]$, is
derived from  $\lim_{r\rightarrow 1}I_{R}(r)$.

\section{ Moments of order statistics}
Order statistics make their appearance in many areas of statistical
theory and practice. Let the random variable $Y_{r:n}$ be the $r$th
order statistic $(Y_{1:n}\leq Y_{2:n}\leq \cdots\leq Y_{n:n})$ in a
sample of size $n$ from the EWPS   distribution. The pdf of $Y_{r:n}$
for $r=1,\cdots,n$, is given by
\begin{equation}\label{ordr EWPS}
f_{r:n}(y)=\frac{1}{B(r,n-r+1)}f(y)F(y)^{r-1}[1-F(y)]^{n-r},~~~y>0.
\end{equation}
where $F(y)$ and f(y) are given in (\ref{cdf EWPS}) and (\ref{pdf
EWPS}). Substituting from (\ref{cdf EWPS}) and (\ref{pdf EWPS}) into
(\ref{ordr EWPS}) gives

\begin{equation}
\begin{array}[b]{ll}
f_{r:n}(y)&=\frac{\alpha\theta \gamma
\beta^{\gamma}}{B(r,n-r+1)C(\theta)^{n}}y^{\gamma-1}e^{-(\beta
y)^{\gamma}} (1-e^{-(\beta y)^{\gamma}})^{\alpha-1}C^{\prime}(\theta (1-e^{-(\beta
y)^{\gamma}})^{\alpha})\\&(C(\theta (1-e^{-(\beta
y)^{\gamma}})^{\alpha}))^{r-1}(C(\theta)-C(\theta (1-e^{-(\beta
y)^{\gamma}})^{\alpha}))^{n-r},
\end{array}
\end{equation}
 Also the cdf of $Y_{r:n}$ is
 given by
\begin{equation}\label{cdf ordr EWPS}
\begin{array}[b]{ll}
F_{r:n}(y)&=\sum^{n}_{k=r}{{n}\choose{k}}[F(y)]^{k}[1-F(y)]^{n-k}\medskip\\&=\sum^{n}_{k=r}{{n}\choose{k}}\frac{(C(\theta (1-e^{-(\beta
y)^{\gamma}})^{\alpha}))^{k}(C(\theta)-C(\theta (1-e^{-(\beta
y)^{\gamma}})^{\alpha}))^{n-k}}{C(\theta)^{n}}.
\end{array}
\end{equation}

Expression for the rth moment of the order statistics $(Y_{1:n}\leq Y_{2:n}\leq \cdots\leq Y_{n:n})$, with a cdf in the form (\ref{cdf ordr EWPS}), are obtained by using a result due to Barakat and Abdelkader (2004) and becomes
\begin{equation}
\begin{array}[b]{ll}
E(Y_{r:n} ^{k})&=k\sum^{\infty}_{j=n-r+1}(-1)^{j-n+r-1}{{j-1}\choose{n-r}}{{n}\choose{j}}\int^{\infty}_{0} y^{k-1}S(y)^{j}dy\medskip\\&= k \sum^{\infty}_{j=n-r+1}\sum^{\infty}_{s=0}\frac{(-1)^{j-n+r+s-1}}{C(\theta)^{s}}{{j-1}\choose{n-r}}{{n}\choose{j}}{{j}\choose{s}}\int^{\infty}_{0} y^{k-1}C(\theta (1-e^{-(\beta
y)^{\gamma}})^{\alpha})^{s}dy.
\end{array}
\end{equation}

\section{ Residual life function of the EWPS distribution}
Given that a component survives up to time $t\geq0$, the residual
life is the period beyond $t$ until the time of failure and defined
by the conditional random variable $X|X > t$. In reliability, it is
well known that the mean residual life function and ratio of two
consecutive moments of residual life determine the distribution
uniquely (Gupta and Gupta, 1983). Therefore, we obtain the $r$th
order moment of the residual life via the general formula
\begin{equation*}
m_{r}(t)=E\left[(Y-t)^r|Y>t\right]=\frac{1}{S(t)}\int_{t}^{\infty}(y-t)^{r}f(y)dy,
\end{equation*}
where $S(t)=1-F(t)$, is the survival function.\newline Applying
series expansion (9), the binomial expansion to $(y-t)^r$ and
substituting $S(t)$ given by (\ref{servival EWPS}) into the above formula gives the
$r$th order moment of the residual life of the EWPS as
\begin{equation}\label{r res}
\begin{array}[b]{ll}
m_{r}(t)&=\frac{\alpha }{C(\theta)-C \left(\theta \left(1-e^{-(\beta t)^{\gamma}}\right)^{\alpha}\right)}\sum_{i=0}^{r}\sum_{n=1}^{\infty}\sum_{j=0}^{\infty}\frac{(-1)^{i+j}t^{i}n a_{n}\theta^{n}}{(j+1)^
{\frac{r+\gamma-i}{\gamma}}\beta^{r-i}}
{{n\alpha-1}\choose{j}}\\
&~~\times‎\Phi ‎\left(\frac{r+\gamma-i}{\gamma};(j+1)(\beta
t)^{\gamma}\right),~~~r\geq 1,
\end{array}
\end{equation}
where $‎\Phi‎(s; t)$ is the upper incomplete gamma function given by
$‎\Phi‎(s; t)= \int_{t}^{\infty}x^{s-1}e^{-x}dx$. \\ Another
important representation for the EWPS  is the mean Residual life (MRL)
function obtain by setting $r=1$ in Eq. (\ref{r res}). MRL function
as well as failure rate (FR) function is very important since each
of them can be used to determine a unique corresponding life time
distribution. Life times can exhibit IMRL (increasing MRL) or DMRL
(decreasing MRL). MRL functions that first decreases (increases) and
then increases (decreases) are usually called bathtub (upside-down
bathtub) shaped, BMRL (UMRL). The relationship between the behaviors
of the two functions of a distribution was studied by many authors
such as Ghitany (1998), Mi (1995), Park (1985), Shanbhag (1970), and
Tang et al. (1999). For the EWPS  distribution the MRL function is
given in the following theorem.

\begin{thm}
The MRL function of the EWPS  distribution with cdf (\ref{cdf EWPS}) is
\begin{equation}\label{MRL EWPS}
m(t)=(\mu_{1}+I(t)-t)/S(t),~~t\geq 0
\end{equation}
where $I(t)=\int^{t}_{0}F(y)dy$, $S(t)$ is the survival function
given in (\ref{servival EWPS}), and $\mu_{1}$ is the mean of the EWPS  in Eq. (\ref{mean
EWPS}).
\end{thm}
\begin{pf}
For more detail about the proof of this theorem see Nassar and Eissa
(2003).
\end{pf}
According to Theorem 1, for the EWPS distribution with $f(y)$ given
by (\ref{pdf EWPS}), we have
\begin{equation}\label{int MRL EWPS}
I(t)=\frac{1}{\beta \gamma C(\theta)}\sum_{n=1}^{\infty}\sum_{k=0}^{\infty}\frac{(-1)^{k}a_{n}\theta^{n}}{k^{\frac{1}{\gamma}}}{{\alpha n}\choose{k}}
\Psi(1/\gamma;l(\beta t)^{\gamma}),
\end{equation}
where $\Psi(s; t)$ is the lower incomplete gamma function given by
$\Psi(s; t)= \int_{0}^{t}x^{s-1}e^{-x}dx$. Substituting Eqs.
(\ref{mean EWPS}), (\ref{mean EWPS}) and (\ref{int MRL EWPS}) into (\ref{MRL EWPS})
gives the MRL function of the EWPS distribution.
\begin{equation*}\label{MLR}
\begin{array}[b]{ll}
m‎_{1}‎(t)&=\frac{1 }{C(\theta)‎\beta‎}\sum_{n=1}^{\infty}\sum_{k=0}^{\infty}‎\frac{(-1)^{k} a_{n}\theta^{n}}{s(t)}‎‎\left[ {{n\alpha-1}\choose{k}}‎\frac{n ‎\alpha‎ ‎\Gamma(1+\frac{1}{\gamma})}{(k+1)‎^{1+‎\frac{1}{‎\gamma‎}‎}‎}‎‎‎-{{n\alpha}\choose{k}}e^{-k(\beta t)^{\gamma}}‎‎t+{{n\alpha}\choose{k}}‎\frac{‎\Psi‎(\frac{1}{\gamma},k(‎\beta t‎)‎^{‎\gamma‎}‎)}{k‎^{‎\frac{1}{‎\gamma‎}‎}‎\gamma‎‎}\right] -t
\end{array}
\end{equation*}

\section{Reversed residual life function of the EWPS distribution}
Given that a component survives up to time $t\geq0$, the residual
life is the period beyond $t$ until the time of failure and defined
by the conditional random variable $X|X > t$.Therefore, we obtain the $r$th
order moment of the residual life via the general formula

\begin{equation*}
m_{r}(t)=E\left[(Y-t)^r|Y>t\right]=\frac{1}{F(t)}\int_{t}^{\infty}(y-t)^{r}f(y)dy,
\end{equation*}
where $F(t)$, is The
exponentiated Weibull power series (EWPS) distribution.\newline Applying
series expansion (9), the binomial expansion to $(t-y)^r$ and
substituting $F(t)$ given by (\ref{cdf EWPS}) into the above formula gives the
$r$th order moment of the reversed residual life of the EWPS as
\begin{equation}\label{rev res}
\begin{array}[b]{ll}
M{r}(t)&=\frac{\alpha }{F(t)}\sum_{i=0}^{r}\sum_{n=1}^{\infty}\sum_{j=0}^{\infty}\frac{(-1)^{i+j}t^{r-i}n a_{n}\theta^{n}}{(j+1)^{1+\frac{i}{\gamma}}C(‎\theta‎)\beta^{i}}
{{n\alpha-1}\choose{j}}\\
&~~\times‎\Psi‎ ‎\left(1+\frac{i}{\gamma};(j+1)(\beta
t)^{\gamma}\right),~~~r\geq 1,
\end{array}
\end{equation}
where $‎\Phi‎(s; t)$ is the upper incomplete gamma function given by
$‎\Phi‎(s; t)= \int_{t}^{\infty}x^{s-1}e^{-x}dx$.

\section{ Probability weighted moments }
Probability weighted moments (PWMs) are expectations of certain
functions of a random variable defined when the ordinary moments of
the random variable exist. The probability weighted moments method
can generally be used for estimating parameters of a distribution
whose inverse form cannot be expressed explicitly. We calculate the
PWMs of the EWPS  distribution since they can be used to obtain the
ordinary moments of the EWPS  distribution.

The PWMs of a random variable $X$ are formally defined by
\begin{equation}\label{PWMs1}
\tau _{s,r}=E[X^{s}F(X)^{r}]=\int_{0}^{\infty
}x^{s}F(x)^{r}f(x)dx,
\end{equation}
where $r$ and $s$ are positive integers and $F(x)$ and $f(x)$ are
the cdf and pdf of the random variable $X$. The following theorem
gives the PWMs of the EWPS  distribution.
\begin{thm}
The PWMs of the EWPS  distribution with cdf (\ref{cdf EWPS}) and pdf
(\ref{pdf EWPS}), are given by
\begin{equation}\label{PWMs2}
\tau_{s,r}=‎\frac{‎\alpha
‎\theta \Gamma(1+\frac{s}{\gamma})}{‎\beta^{s}(C(‎\theta‎))‎^{r+1}‎} \sum ^{\infty}_{n=1}\sum^{\infty}_{i=0}\sum^{\infty}_{j=0}(-1)^{j}‎\frac{n a‎_{n}‎\theta ‎^{i+n-1}‎‎‎}{(j+1)^{1+\frac{s}{\gamma}}}‎
{\alpha(n+i ) -1 \choose j }.
\end{equation}
\end{thm}

\begin{pf}
Substituting from (\ref{cdf EWPS}) and (\ref{pdf EWPS}) into
(\ref{PWMs1}) gives
\begin{equation*}
\tau_{s,r}=‎\frac{\alpha\gamma‎\theta‎\beta^{\gamma}}{(C(‎\theta‎))‎^{r+1}}‎\int\limits_{0}^{\infty
}x^{s+\gamma-1}e^{-(\beta x)^{\gamma}} (1-e^{-(\beta x)^{\gamma}}
)^{\alpha -1}C‎^{\prime}(\theta(1-e^{-(\beta x)^{\gamma}}
)^{\alpha})‎(C(\theta(1-e^{-(\beta x)^{\gamma}})^{\alpha}))‎^{r}‎dx.
\end{equation*}
Applying the Equation
$‎\left(‎\sum^{\infty}_{i=0}w‎_{i}u‎^{i}‎‎\right)‎^{j}‎ ‎=‎\sum^{\infty}_{i=0}c‎_{i,j}u‎^{i}‎$,
where the coefficients $c‎_{i,j}$ for $i=1,2,...$ can be easily obtained from the recurrence relation
$c‎_{i,j}=(iw‎_{0}‎)‎^{-1}‎\sum^{i}_{m=1}(jm-i+m)w‎_{m}c‎_{i-m,j},$
white $c‎_{0,j}=w‎_{0}‎^{j}‎$ for $‎\left[C‎^{\prime}‎ (‎\theta(1-e^{-(\beta y)^\gamma})^{\alpha }‎‎)‎‎\right]‎^{r}‎ $and series expansion for $\left(1-(1-e^{-(\beta
x)^{\gamma}} )^{\alpha}\right)^{-(r+2)}$ gives
\begin{equation*}
\tau_{s,r}=‎\frac{\alpha\gamma‎\theta‎\beta^{\gamma}}{(C(‎\theta‎))‎^{r+1}}\sum^{\infty}_{n=1}\sum^{\infty}_{i=0}\sum^{\infty}_{j=0}(-1)‎^{j}‎n a‎_{n}‎‎\theta‎^{i+n-1}
{\alpha( n +i ) -1 \choose j}‎‎\int\limits_{0}^{\infty
}x^{s+\gamma-1}e^{-(j+1)(\beta x)^{\gamma}}  dx.
\end{equation*}
Setting $u=(j+1)(\beta x)^{\gamma}$ the above integral reduces to
\begin{equation*}
\int\limits_{0}^{\infty }x^{s+\gamma-1}e^{-(k+1)(\beta x)^{\gamma}}
dx=\frac{1}{\gamma\beta^{\gamma+s}(k+1)^{1+\frac{s}{\gamma}}}\Gamma(1+\frac{s}{\gamma}),
\end{equation*}
and the proof is completed.
\end{pf}

\begin{rmk}
The $\textit{s}$th moment of EWPS  distribution can be obtained by
putting $r=0$ in Eq. (\ref{PWMs2}). Therefore
\begin{equation*}
E(X^{s}) =\alpha
(1-\theta)\beta^{-s}\Gamma(\frac{s}{\gamma}+1)\sum^{\infty}_{j=0}\sum^{\infty}_{k=0}(-1)^{k}
(j+1){\alpha( j +1) -1 \choose k }
(k+1)^{-(\frac{s}{\gamma}+1)}\theta^{j},
\end{equation*}
which is equal with Eq. (\ref{meank EWPS}) if $s$ is replaced by $k$.
Also, the mean and variance of the EWPS  distribution can be obtained.
\end{rmk}

\section{ Mean deviations}
The amount of scatter in a population can be measured by the
totality of deviations from the mean and median. For a random
variable $X$ with pdf $f(.)$, cdf $F(.)$, mean $\mu= E(X)$ and $ M =
Median(X)$ the mean deviation about the mean and the mean deviation
about the median, respectively, are defined by
$$\delta_{1}(X)=\int_{0}^{\infty}|x-\mu|f(x)dx=2\mu F(\mu)-2\mu + 2L(\mu),$$
and
$$\delta_{2}(X)=\int_{0}^{\infty}|x-M|f(x)dx=2M F(M)-M-\mu +
2L(M),$$ where $L(\mu)=\int^{\infty}_{\mu}xf(x)dx$ and
$L(M)=\int^{\infty}_{M}xf(x)dx.$

For the EWPS  distribution we have
\begin{equation}\label{L(b)}
L(b)=\alpha
\beta^{-1}\sum^{\infty}_{n=1}\sum^{\infty}_{j=0}(-1)^{j}\frac{n a_{n}\theta^{n}}{C(\theta)}
{{n \alpha-1} \choose
{j}}(j+1)^{-(\frac{1}{\gamma}+1)}‎\Phi ‎\left((\frac{1}{1+\gamma});(j+1)(\beta
b)^\gamma\right).
\end{equation}

\begin{thm}
The Mean deviations of the EWPS  distribution are
\begin{equation*}
\delta_{1}=2\mu
\left(\frac{C\left(\theta\left(1-e^{-(\beta \mu)^{\gamma}}\right)^{\alpha}\right)-C(\theta)}{C(\theta)}\right)+2L(\mu),
\end{equation*}
and
\begin{equation*}
\delta_{2}=2L(M)-\mu,
\end{equation*}
respectively, where $\mu$ is the mean of EWPS in Eq. (\ref{mean
EWPS}), $L(\mu)$ and $L(M)$ are obtained by substituting $\mu$ and
$M=G^{-1}\left(\frac{C^{-1}(0.5 C(\theta))}{\theta}\right)$
instead of $b$ in Eq. (\ref{L(b)}).
\end{thm}

\section{ Bonferoni and Lorenz curves}
The Bonferroni and Lorenz curves and Gini index have many
applications not only in economics to study income and poverty, but
also in other fields like reliability, medicine and insurance. The
Bonferroni curve $B_{F}[F(x)]$ is given by
\begin{equation*}
B_{F}[F(x)]=\frac{1}{\mu F(x)}\int^{x}_{0}u f(u) du.
\end{equation*}
The Bonferroni curve of the EWPS  distribution is given by
\begin{equation*}\label{B_{F} EWPS}
\begin{array}{ll}
B_{F}[F(x)]=\frac{\alpha \beta^{-1}}{\mu C\left(\theta\left(1-e^{-(\beta x)^{\gamma}}\right)^{\alpha}\right)}\sum^{\infty}_{n=1}\sum^{\infty}_{j=0}na_{n}\theta^{n}(-1)^{j}{{n\alpha-1}
\choose
{j}}&\\~~~~~~~~~~~~~~~\times(j+1)^{-(\frac{1}{\gamma}+1)}\Psi\left((\frac{1}{\gamma}+1);(j+1)(x
\beta)^{\gamma}\right).&
\end{array}
\end{equation*}

Also, the Lorenz curve of EWPS  distribution can be obtained via the
expression
\begin{equation*}
\begin{array}{ll}
L_{F} [F(x)] =\frac{\alpha \beta^{-1}}{\mu C(\theta)}\sum^{\infty}_{n=1}\sum^{\infty}_{j=0}na_{n}\theta^{n}(-1)^{j}{{n\alpha-1}
\choose
{j}}&\\~~~~~~~~~~~~~~~\times(j+1)^{-(\frac{1}{\gamma}+1)}\Psi\left((\frac{1}{\gamma}+1);(j+1)(x
\beta)^{\gamma}\right).&
\end{array}
\end{equation*}

The scaled total time on test transform of a distribution function
$F$ (Pundir et al., 2005) is defined by
\begin{equation*}
S_{F}[F(t)]=\frac{1}{\mu}\int^{t}_{0}\bar F(u)du.
\end{equation*}
If $F(t)$ denotes the cdf of EWPS   distribution then
\begin{equation*}
S_{F}[F(t)]=\frac{t}{\mu}-\frac{\beta^{-1}}{\mu \gamma  C(\theta)}\sum^{\infty}_{n=1}\sum^{\infty}_{j=0}(-1)^{j}\frac{  a_{n}\theta^{n}}{(j)^{(\frac{1}{\gamma})}}
{{ n \alpha} \choose {j}}\Psi
\left(\frac{1}{\gamma};j(t \beta)^{\gamma}\right).
\end{equation*}
The cumulative total time can be obtained by using formula
$C_F=\int_{0}^{1}S_{F}[F(t)]f(t)dt$ and the Gini index can be
derived from the relationship $G =1-C_{F}$.
\section{ Estimation and inference}
In what follows, we discuss the estimation of the parameters for the
EWPS  distribution. Let $Y_{1},Y_{2},\cdots,Y_{n}$ be a random sample
with observed values $y_{1},y_{2},\cdots,y_{n}$ from EWPS
distribution with parameters $\alpha,\beta,\gamma$ and $\theta$. Let
$\Theta=(\alpha,\beta,\gamma,\theta)^{T}$ be the parameter vector.
The total log-likelihood function is given by
\begin{equation*}
\begin{array}[b]{ll}
l_{n}\equiv l_{n}(y;\Theta)&= n[\log \alpha +\log \gamma + \gamma \log \beta + \log \theta]+ (\gamma-1)\sum^{n}_{i=1}
\log y_{i}-\sum^{n}_{i=1}(\beta y_{i})^{\gamma}\medskip \\
&~~~ +(\alpha-1)\sum^{n}_{i=1}\log (1-e^{-(\beta
y_{i})^{\gamma}})-n \log C(\theta)+\sum^{n}_{i=1}\log[C^{\prime}(\theta\left(1-e^{-(\beta
y_{i})^{\gamma}}\right)^{\alpha})].
\end{array}
\end{equation*}
The associated score function is given by $U_{n}(\Theta)=(\partial
l_{n}/\partial \alpha,\partial l_{n}/\partial \beta, \partial
l_{n}/\partial \gamma,\partial l_{n}/\partial \lambda)^{T}$, where

\begin{equation*}
\begin{array}{ll}
\frac{\partial l_{n}}{\partial
\alpha}&=\frac{n}{\alpha}+\sum^{n}_{i=1}\log (1-e^{-(\beta y_{i})^{\gamma}})+\theta \sum^{n}_{i=1}\log (1-e^{-(\beta y_{i})
^{\gamma}})(1-e^{-(\beta y_{i})^{\gamma}})^{\alpha}~\frac{C^{\prime\prime}(\theta(1-e^{-(\beta y_{i})^{\gamma}})^{\alpha})}{C^{\prime}(\theta(1-e^{-(\beta y_{i})^{\gamma}})^{\alpha})},\medskip \\

\frac{\partial l_{n}}{\partial \beta}&=\frac{n\gamma}{\beta}-\gamma\beta^{\gamma-1} \sum^{n}_{i=1}y_{i}^{\gamma}+
(\alpha-1)\gamma\beta^{\gamma-1} \sum^{n}_{i=1}\frac{y_{i}^{\gamma}e^{-(\beta y_{i})^{\gamma}}}{1-e^{-(\beta y_{i})^{\gamma}}}\medskip \\
&~~+\theta\alpha\gamma\beta^{\gamma-1}\sum^{n}_{i=1}\frac{y_{i}^{\gamma}e^{-(\beta
y_{i})^{\gamma}}(1-e^{-(\beta y_{i})^{\gamma}})
^{\alpha-1}C^{\prime\prime}(\theta(1-e^{-(\beta y_{i})^{\gamma}})^{\alpha})}{C^{\prime}(\theta(1-e^{-(\beta y_{i})^{\gamma}})^{\alpha})},\medskip \\

\frac{\partial l_{n}}{\partial
\gamma}&=\frac{n}{\gamma}+n\log\beta+\sum^{n}_{i=1}\log
y_{i}-\sum^{n}_{i=1}\log(\beta y_{i})(\beta y_{i})^{\gamma}
\medskip \\
 &~~+(\alpha-1)\sum^{n}_{i=1}\frac{\log(\beta y_{i})(\beta y_{i})^{\gamma}e^{-(\beta y_{i})^{\gamma}}}{1-e^{-(\beta y_{i})^{\gamma}}}
 + \theta \alpha \sum^{n}_{i=1}\frac{\log(\beta y_{i})(\beta y_{i})^{\gamma}e^{-(\beta y_{i})^{\gamma}}(1-e^{-(\beta y_{i})
 ^{\gamma}})^{\alpha-1}C^{\prime\prime}(\theta(1-e^{-(\beta y_{i})^{\gamma}})^{\alpha})}{C^{\prime}(\theta(1-e^{-(\beta y_{i})^{\gamma}})^{\alpha})} ,\medskip \\

\frac{\partial l_{n}}{\partial
\theta}&=\frac{n}{\theta}+\sum^{n}_{i=1}\frac{(1-e^{-(\beta
y_{i})^{\gamma}})^{\alpha}C^{\prime\prime}(\theta(1-e^{-(\beta y_{i})^{\gamma}})^{\alpha})}{C^{\prime}(\theta(1-e^{-(\beta y_{i})^{\gamma}})^{\alpha})}-n\frac{C^{\prime}(\theta)}{C(\theta)}.
\end{array}
\end{equation*}
The maximum likelihood estimation (MLE) of $\Theta $, say
$\widehat{\Theta }$, is obtained by solving the nonlinear system
$U_n\left(\Theta \right)=\textbf{0}$. The solution of this nonlinear
system of equation has not a closed form. For interval estimation
and hypothesis tests on the model parameters, we require the
information matrix. The $4\times 4$ observed information matrix is
\[I_n\left(\Theta \right)=-\left[ \begin{array}{cccc}
I_{\alpha \alpha } & I_{\alpha \beta } & I_{\alpha \gamma } & I_{\alpha \theta }\\
I_{\alpha \beta } & I_{\beta \beta } & I_{\beta \gamma }& I_{\beta \theta } \\
I_{\alpha \gamma } & I_{\beta \gamma } & I_{\gamma \gamma}& I_{\gamma \theta } \\
I_{\alpha \theta } & I_{\beta \theta } & I_{\gamma \theta}& I_{\theta \theta } \\
\end{array} \right],\] whose elements are given in Appendix.

Applying the usual large sample approximation, MLE of $\Theta $ i.e.
$\widehat{\Theta }$ can be treated as being approximately
$N_4(\Theta ,{J_n(\Theta )}^{-1}{\mathbf )}$, where $J_n\left(\Theta
\right)=E\left[I_n\left(\Theta \right)\right]$. Under conditions
that are fulfilled for parameters in the interior of the parameter
space but not on the boundary, the asymptotic distribution of
$\sqrt{n}(\widehat{\Theta }{\rm -}\Theta {\rm )}$ is $N_4({\mathbf
0},{J(\Theta )}^{{\mathbf -}{\mathbf 1}}{\mathbf )}$, where
$J\left(\Theta \right)={\mathop{\lim }_{n\to \infty }
{n^{-1}I}_n(\Theta )\ }$ is the unit information matrix. This
asymptotic behavior remains valid if $J(\Theta )$ is replaced by the
average sample information matrix evaluated at $\widehat{\Theta }$,
say ${n^{-1}I}_n(\widehat{\Theta })$. The estimated asymptotic
multivariate normal $N_4(\Theta ,{I_n(\widehat{\Theta })}^{{\mathbf
-}{\mathbf 1}}{\mathbf )}$ distribution of $\widehat{\Theta }$ can
be used to construct approximate confidence intervals for the
parameters and for the hazard rate and survival functions. An
$100(1-\gamma )$ asymptotic confidence interval for each parameter
${\Theta }_{{\rm r}}$ is given by
\[{ACI}_r=({\widehat{\Theta
}}_r-Z_{\frac{\gamma }{2}}\sqrt{{\hat{I}}^{rr}},{\widehat{\Theta
}}_r+Z_{\frac{\gamma }{2}}\sqrt{{\hat{I}}^{rr}}),\] where
${\hat{I}}^{rr}$ is the (\textit{r, r}) diagonal element of
${I_n\left(\widehat{\Theta }\right)}^{{\mathbf -}{\mathbf 1}}$ for
$r=1,~2,~3,~4,$ and $Z_{\frac{\gamma }{2}}$ is the quantile
$1-\gamma /2$ of the standard normal distribution.




\section{EM Algorithm    }
Let the complete-data be $Y_{1},\cdots,Y_{n}$ with observed values
$y_{1},\cdots,y_{n}$ and the hypothetical random variable
$Z_{1},\cdots,Z_{n}$. The joint probability density function is such
that the marginal density of $Y_{1},\cdots,Y_{n}$ is the likelihood
of interest. Then, we define a hypothetical complete-data
distribution for each $(Y_{i},Z_{i})~~i=1,\cdots,n$ with a joint
probability density function in the form
\begin{equation}
g(y,z;\Theta)=z\alpha \gamma \beta^{\gamma} y^{\gamma-1}e^{-(\beta y)^{\gamma}}(1-e^{-(\beta y)^{\gamma}})^{z\alpha-1}\frac{a_{z}\theta^{z}}{C(\theta)},
\end{equation}
where $\Theta=(\alpha,\beta,\gamma,\theta)$, $y>0$ and $z\in
\mathbb{N}$.\\
Under the formulation, the E-step of an EM cycle requires the
expectation of $(Z|Y;\Theta^{(r)})$ where
$\Theta^{(r)}=(\alpha^{(r)},\beta^{(r)},\gamma^{(r)},\theta^{(r)})$
is the current estimate (in the $r$th iteration) of $\Theta$.\\
The pdf of $Z$ given $Y$, say $g(z|y)$ is given by
\begin{equation*}
g(z|y)=\frac{z\left[(1-e^{-(\beta y)^{\gamma}})^{\alpha}\right]^{z-1} a_{z}\theta^{z-1}}{C^{\prime}\left[\theta(1-e^{-(\beta y)^{\gamma}})^{\alpha}\right]}.
\end{equation*}
Thus, its expected value is given by
\begin{equation*}
\begin{array}[b]{ll}
E[Z|Y=y]&=1+\frac{\theta(1-e^{-(\beta
y)^{\gamma}})^{\alpha}C^{\prime \prime}\left[\theta(1-e^{-(\beta y)^{\gamma}})^{\alpha}\right]}{C^{\prime}\left[\theta(1-e^{-(\beta y)^{\gamma}})^{\alpha}\right]}.
\end{array}
\end{equation*}

The EM cycle is completed  with the M-step by using the maximum
likelihood estimation over $\Theta$, with the missing Z's
replaced by  their conditional expectations given above.\\
The log-likelihood for the complete-data is
\begin{equation*}
\begin{array}{ll}
l^{*}_{n}(y_{1},\cdots,y_{n};z_{1},\cdots,z_{n};\Theta)&\propto
\sum^{n}_{i=1} \log z_{i} + n [\log \alpha + \log \gamma+\gamma \log
\beta ]-\sum^{n}_{i=1} (\beta y_{i})^{\gamma}\medskip \\ &~~ +
\sum^{n}_{i=1} (\gamma-1)  \log y_{i}+\sum^{n}_{i=1}\log
(1-e^{-(\beta y_{i})^{\gamma}})(\alpha z_{i}-1)\medskip \\ &~~-n\log
C(\theta)+\sum^{n}_{i=1}(z_{i})\log\theta.
\end{array}
\end{equation*}

The components of the score function $U^{*}_{n}(\Theta)=
(\frac{\partial l^{*}_{n}}{\partial \alpha},\frac{\partial l^{*}_{n}
}{\partial \beta},\frac{\partial l^{*}_{n}}{\partial \gamma
},\frac{\partial l^{*}_{n}}{\partial \theta})^{T}$ are given by
\begin{equation*}
\begin{array}{ll}
\frac{\partial
l^{*}_{n}}{\partial\alpha}&=\frac{n}{\alpha}+\sum^{n}_{i=1}\log (1-e^{-(\beta y_{i})^{\gamma}})z_{i},\medskip\\
\frac{\partial
l^{*}_{n}}{\partial\beta}&=\frac{n\gamma}{\beta}-\gamma\beta^{\gamma-1}\sum^{n}_{i=1}y_{i}^{\gamma}\frac{(1-\alpha z_{i}e^{-(\beta y_{i})^{\gamma}})}{1-e^{-(\beta y_{i})^{\gamma}}},\medskip\\
\frac{l^{*}_{n}}{\partial\gamma}&=\frac{n}{\gamma}+ n \log \beta +\sum^{n}_{i=1}\log  y_{i} -\sum^{n}_{i=1}\frac{\log(\beta y_{i})(\beta y_{i})^{\gamma}(1-\alpha z_{i}e^{-(\beta y_{i})^{\gamma}}) }{1-e^{-(\beta y_{i})^{\gamma}}} ,\medskip\\
\frac{\partial
l^{*}_{n}}{\partial\theta}&=\frac{-n C^{\prime}(\theta)}{C(\theta)}+\frac{\sum^{n}_{i=1}(z_{i})}{\theta}.
\end{array}
\end{equation*}

From a nonlinear system of equations $U^{*}_{n}(\Theta)=\textbf{0}$,
we obtain the iterative procedure of the EM algorithm as
\begin{equation*}
\begin{array}{l}
\hat\theta^{(t+1)}=\frac{C(\hat\theta^{(t+1))}}{n C^{\prime}(\hat\theta^{(t+1)) }}\sum^{n}_{i=1} z_{i}^{(t)},\medskip\\
\hat\alpha^{(t+1)}=\frac{-n}{\sum^{n}_{i=1} z_{i}^{(t)}\left[\log (1-e^{-(\hat\beta^{(t)} y_{i}})^{\hat\gamma^{(t)}})\right]},\medskip\\
\frac{n\hat\gamma^{(t)}}{\hat\beta^{(t+1)}}-\hat\gamma^{(t)}(\hat\beta^{(t+1)})^{(\hat\gamma^{(t)}-1)}\sum^{n}_{i=1}
y_{i}^{\hat\gamma^{(t)}}\frac{(1-\hat\alpha^{(t)}z_{i}^{(t)}e^{-(\hat\beta^{(t+1)} y_{i})^{\hat\gamma^{(t)}}})}{1-e^{-(\hat\beta^{(t+1)} y_{i})^{\hat\gamma^{(t)}}}}=0,\medskip\\
\frac{n}{\hat\gamma^{(t+1)}}+ n \log \hat\beta^{(t)}
+\sum^{n}_{i=1}\log  y_{i} -\sum^{n}_{i=1}\frac{\log(\hat\beta^{(t)}
y_{i})(\hat\beta^{(t)}
y_{i})^{\hat\gamma^{(t+1)}}(1-\hat\alpha^{(t)}
z_{i}^{(t)}e^{-(\hat\beta^{(t)} y_{i})^{\hat\gamma^{(t+1)}}})
}{1-e^{-(\hat\beta^{(t)} y_{i})^{\hat\gamma^{(t+1)}}}}=0,
\end{array}
\end{equation*}
where $\hat\gamma^{(t+1)}$, $\hat\beta^{(t+1)}$ and
are found numerically. Hence, for
$i=1,\cdots,n$, we have that
\begin{equation*}
z^{(t)}_{i}=1+\frac{\hat\theta^{(t)}(1-e^{-(\hat\beta^{(t)}
y_{i})^{\hat\gamma^{(t)}}})^{\hat\alpha^{(t)}}C^{\prime \prime}\left[\hat\theta^{(t)}(1-e^{-(\hat\beta^{(t)}
y_{i})^{\hat\gamma^{(t)}}})^{\hat\alpha^{(t)}}\right]}{C^{\prime}\left[\hat\theta^{(t)}(1-e^{-(\hat\beta^{(t)}
y_{i})^{\hat\gamma^{(t)}}})^{\hat\alpha^{(t)}}\right]}.
\end{equation*}

\section{Special cases of the EWPS distribution   }
In this section we study in detail cases of the EWPS class of distributions. To illustrate
the flexibility of the distributions, plots of the pdf and hazard function for some values of the parameters are
presented.

\subsection{Exponentiated weibull binomial distribution }
The exponentiated weibull binomial distribution is a special case of power series distributions with
$a_{n}={m \choose n}$ and $C(‎\theta‎)=‎‎(‎\theta+1‎‎)^{m}-1~(‎\theta‎>‎0‎‎)$, where m (n$‎\leq‎$m) is the number of replicas.
Using the cdf in (\ref{cdf EWPS}), the cdf of exponentiated weibull binomial (EWB) distribution is given by

\begin{equation*}\label{cdf EWB}
F(y)=‎\frac{(‎\theta (1-e^{-(\beta y)^{\gamma}})^{‎\alpha‎}+1 ‎)^{m}-1}{(‎\theta+1‎)^{m}-1}‎‎,~~~~y>0.
\end{equation*}

\begin{equation*}\label{pdf EWB}
f(y)=m‎\alpha \theta‎ ‎\gamma  ‎\beta^{‎\gamma‎‎‎‎‎} y^{‎\gamma-1‎}e^{-(\beta y)^{\gamma}}
(1-e^{-(\beta y)^{\gamma}})^{‎\alpha -1‎}‎‎\frac{\left(‎\theta‎ (1-e^{-(\beta y)^{\gamma}})^{‎\alpha ‎}‎+1\right)‎^{m-1}‎ ‎‎}{(‎\theta+1‎)^{m}-1},‎
\end{equation*}
and
\begin{equation*}\label{hazard WG}
h(y)=m‎\alpha ‎\theta ‎\gamma ‎\beta^{‎\gamma‎‎‎‎‎} y^{‎\gamma-1‎}e^{-(\beta y)^{\gamma}}
(1-e^{-(\beta y)^{\gamma}})^{‎\alpha -1‎}‎‎\frac{\left(‎\theta‎ (1-e^{-(\beta y)^{\gamma}})^{‎\alpha ‎}‎+1\right)‎^{m-1}‎ ‎‎}{(‎\theta+1‎)^{m}-\left(‎\theta‎ (1-e^{-(\beta y)^{\gamma}})^{‎\alpha ‎}‎+1\right)‎^{m}}.
\end{equation*}
The plots of density and hazard rate function of EWB distribution for some values of $‎\alpha, ‎\beta, ‎\gamma‎‎‎,‎\theta‎$ and $m=10$ are given in Fig. 2. \\
\begin{figure}[t]
\centering
\includegraphics[scale=0.25]{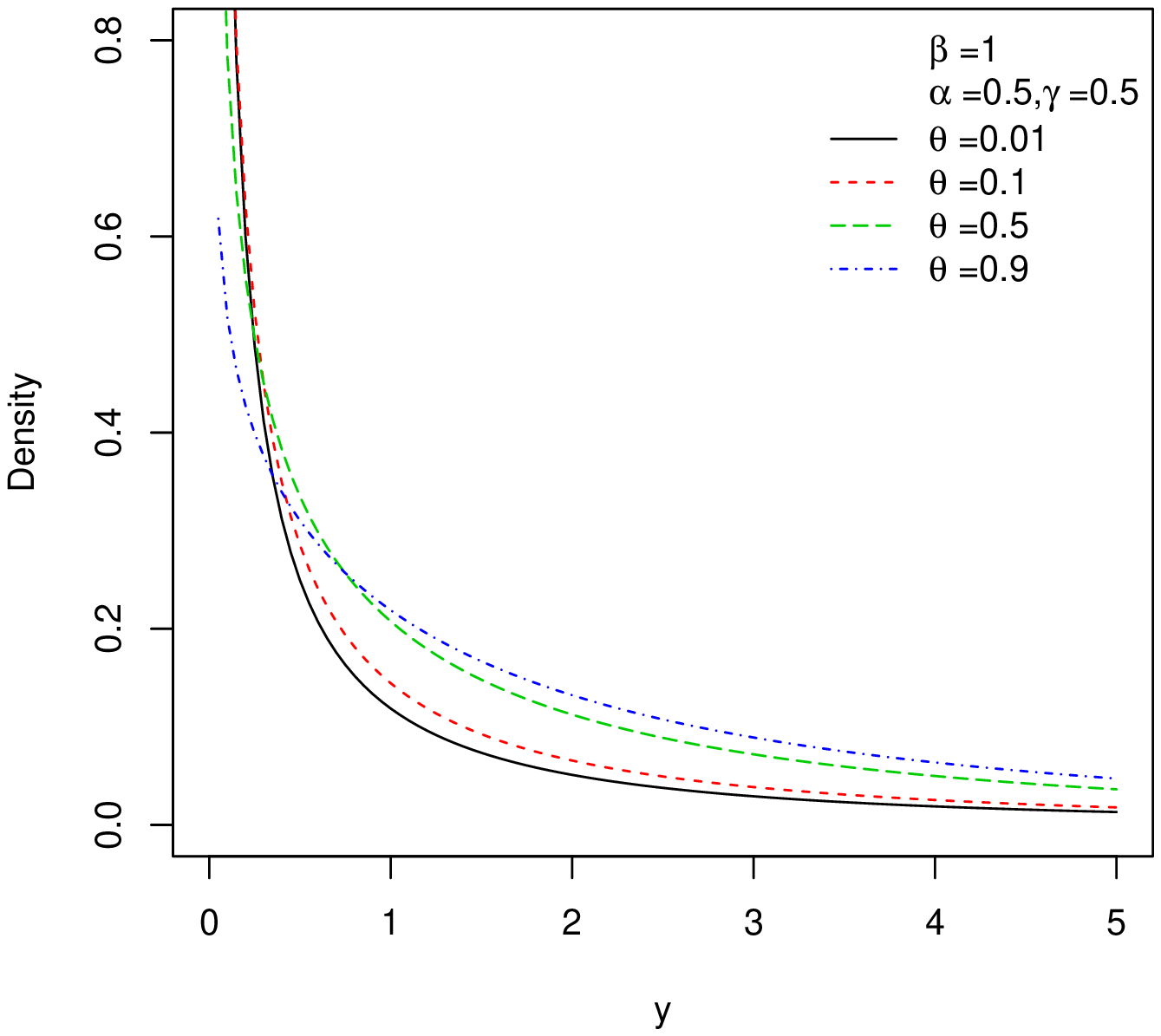}
\includegraphics[scale=0.25]{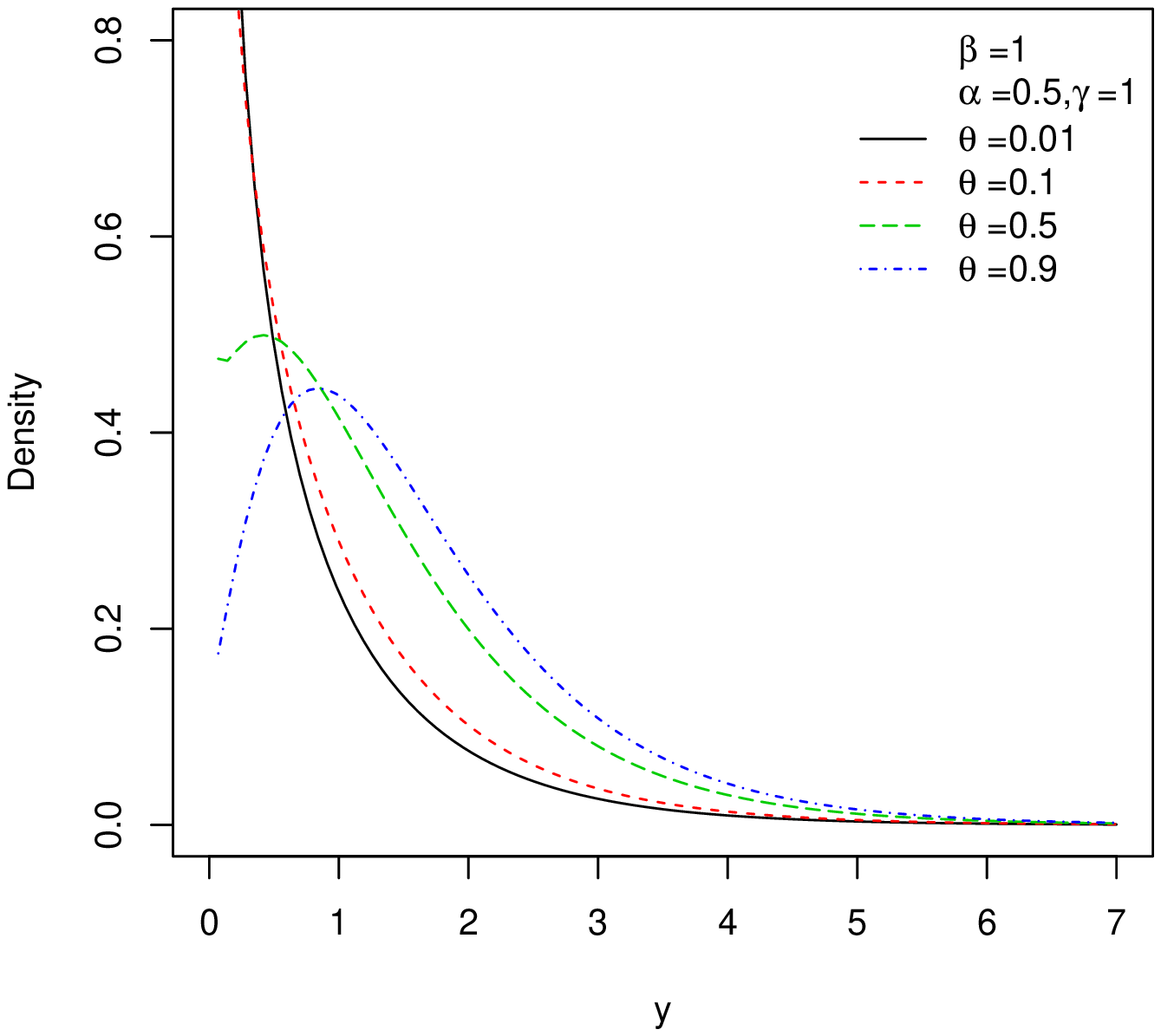}
\includegraphics[scale=0.25]{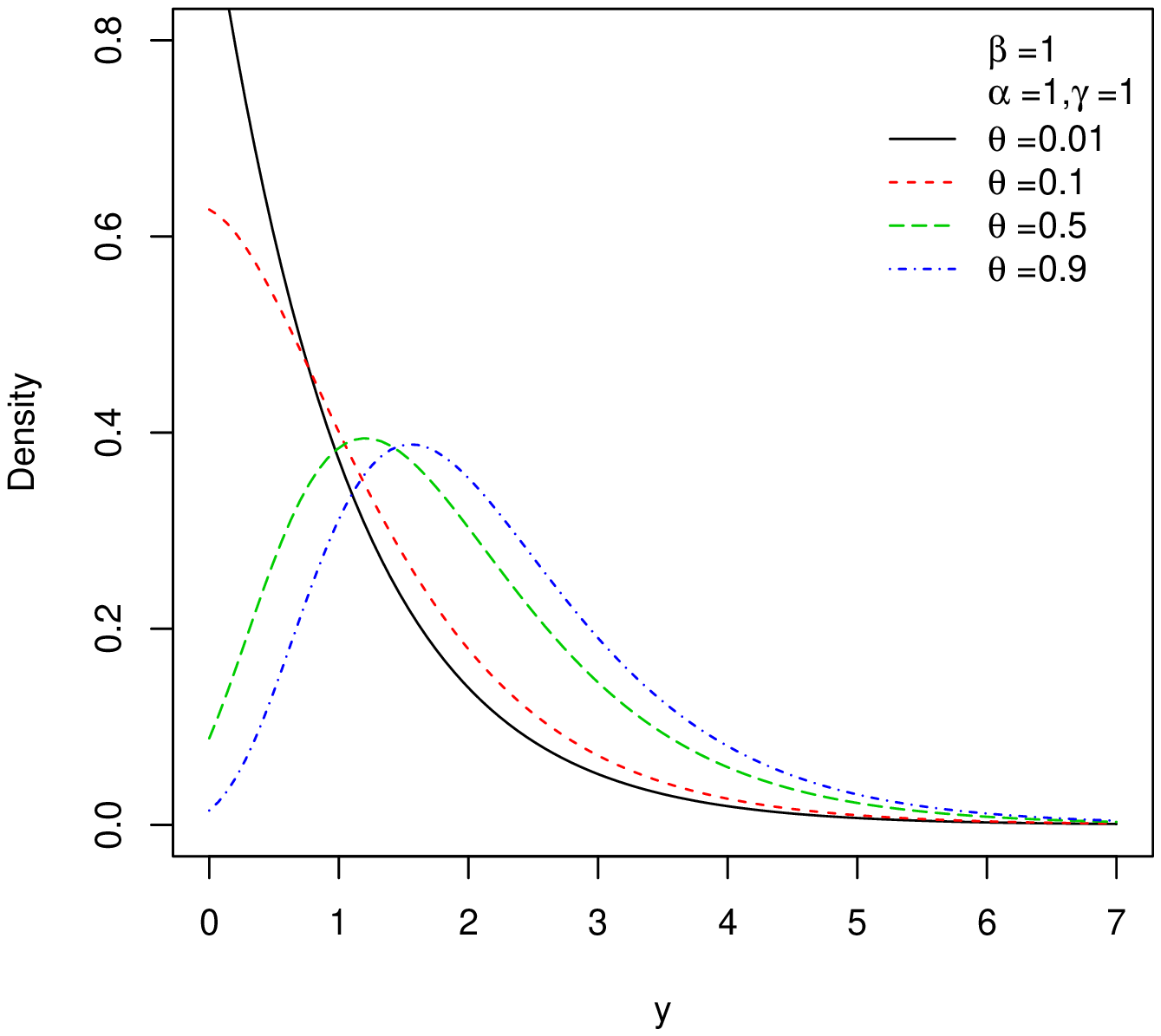}
\includegraphics[scale=0.25]{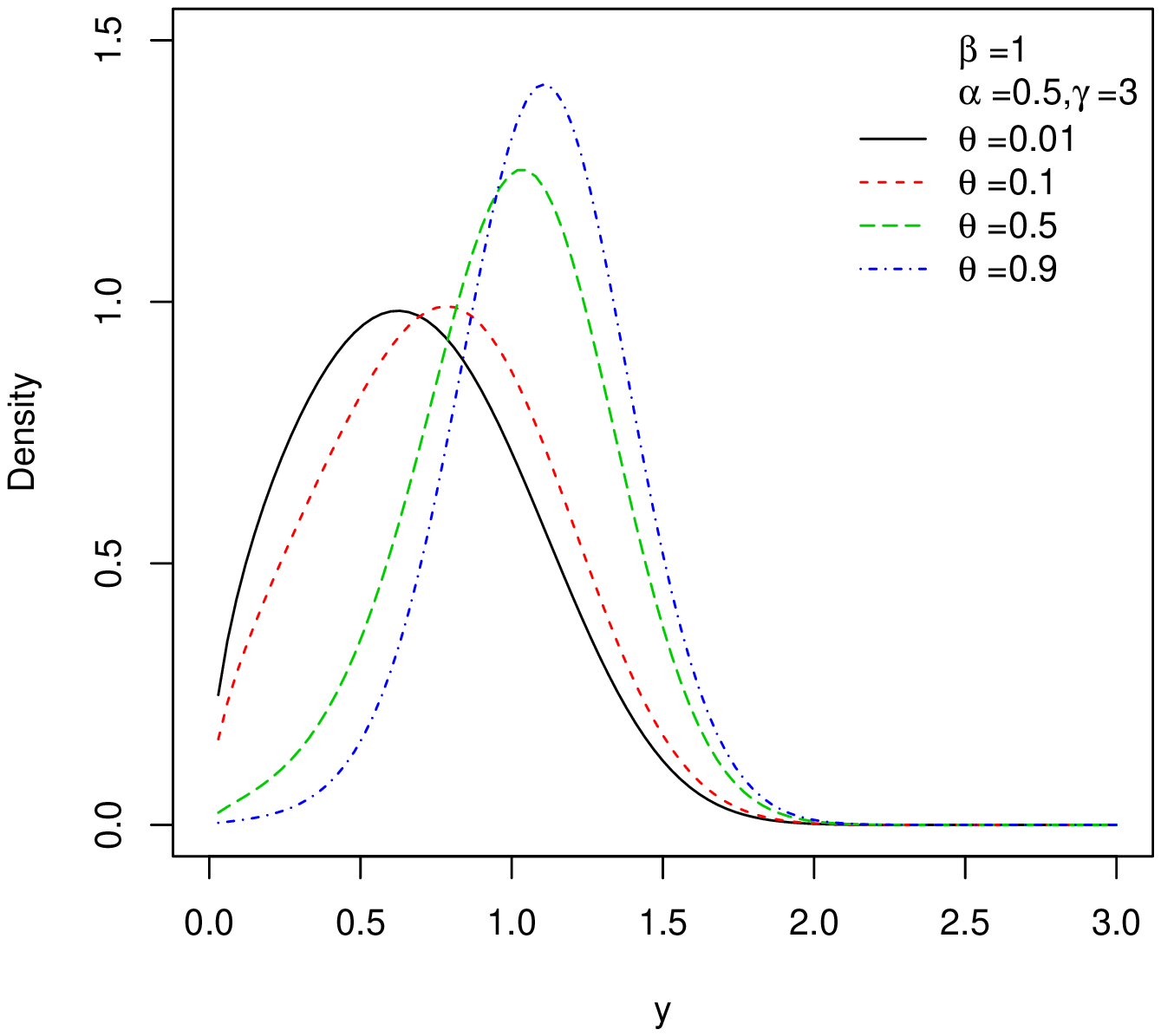}
\includegraphics[scale=0.25]{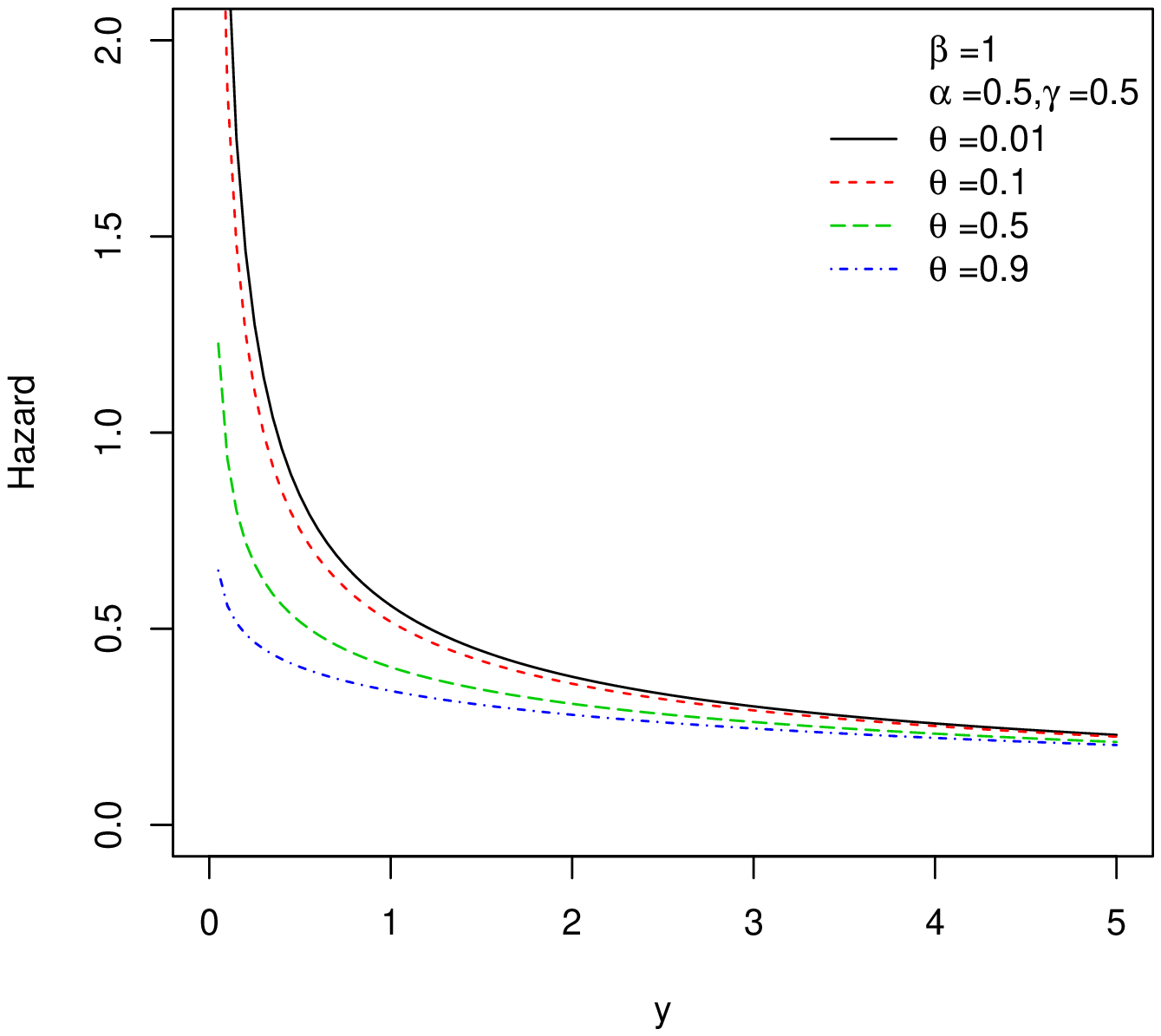}
\includegraphics[scale=0.25]{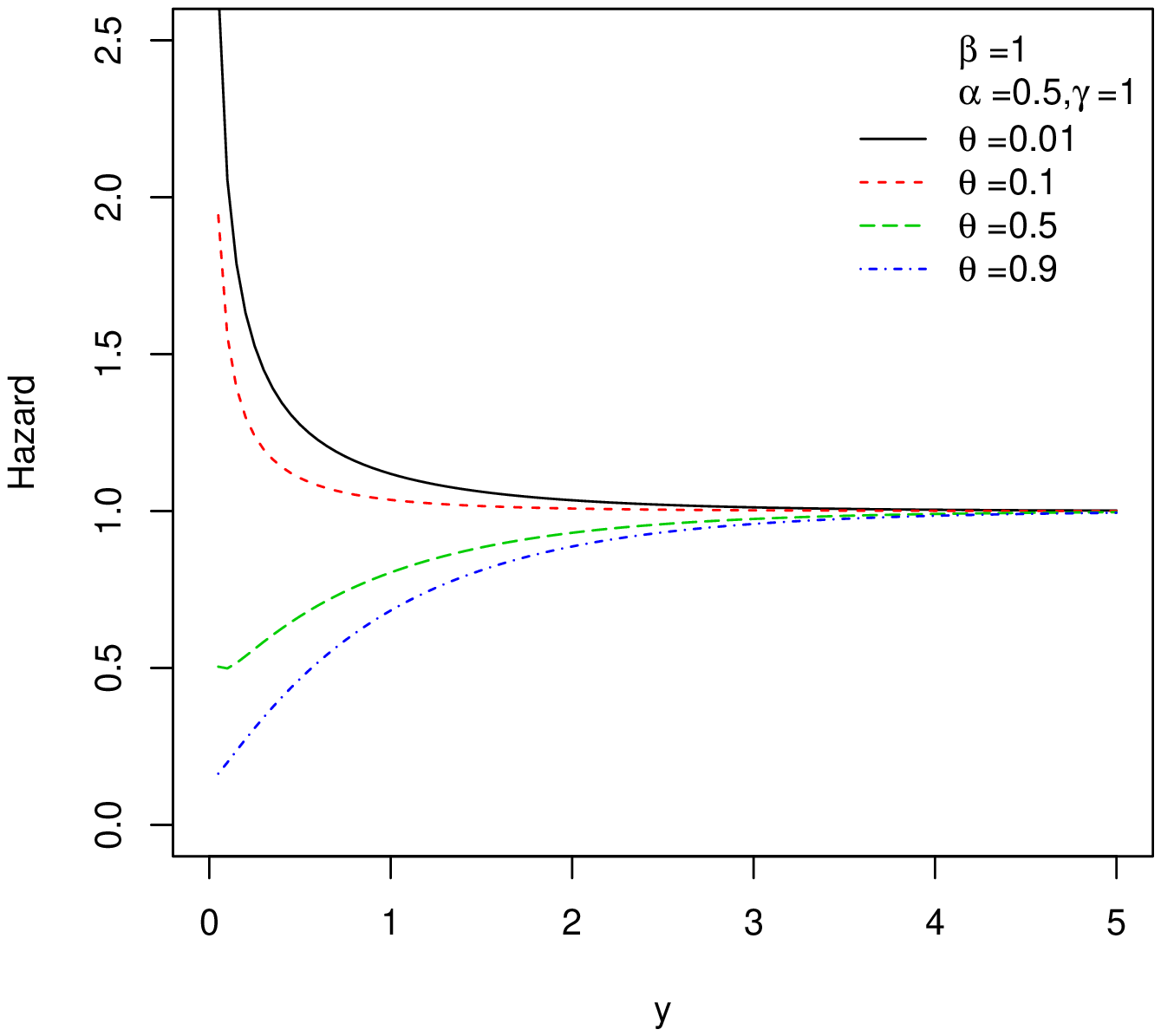}
\includegraphics[scale=0.25]{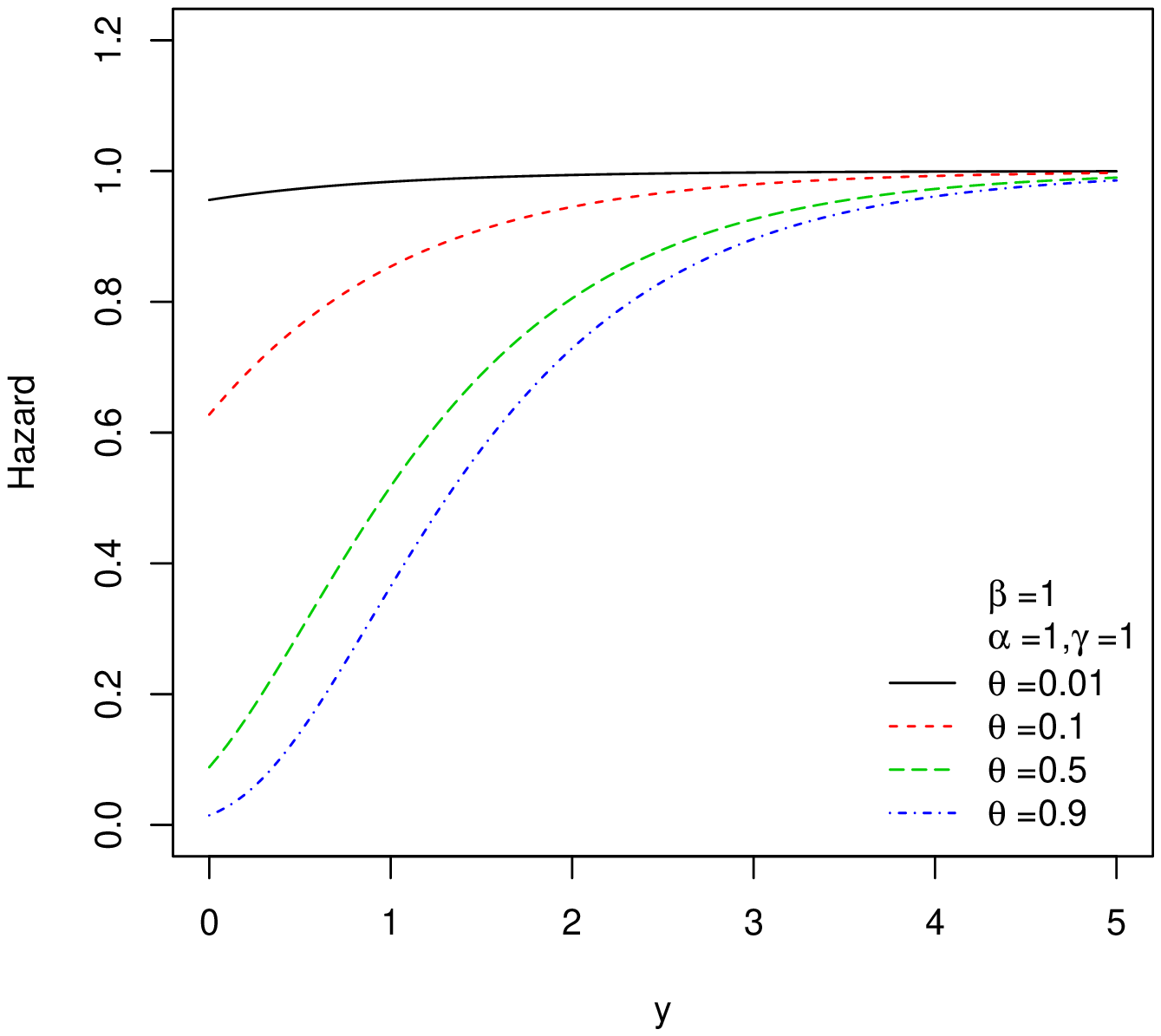}
\includegraphics[scale=0.25]{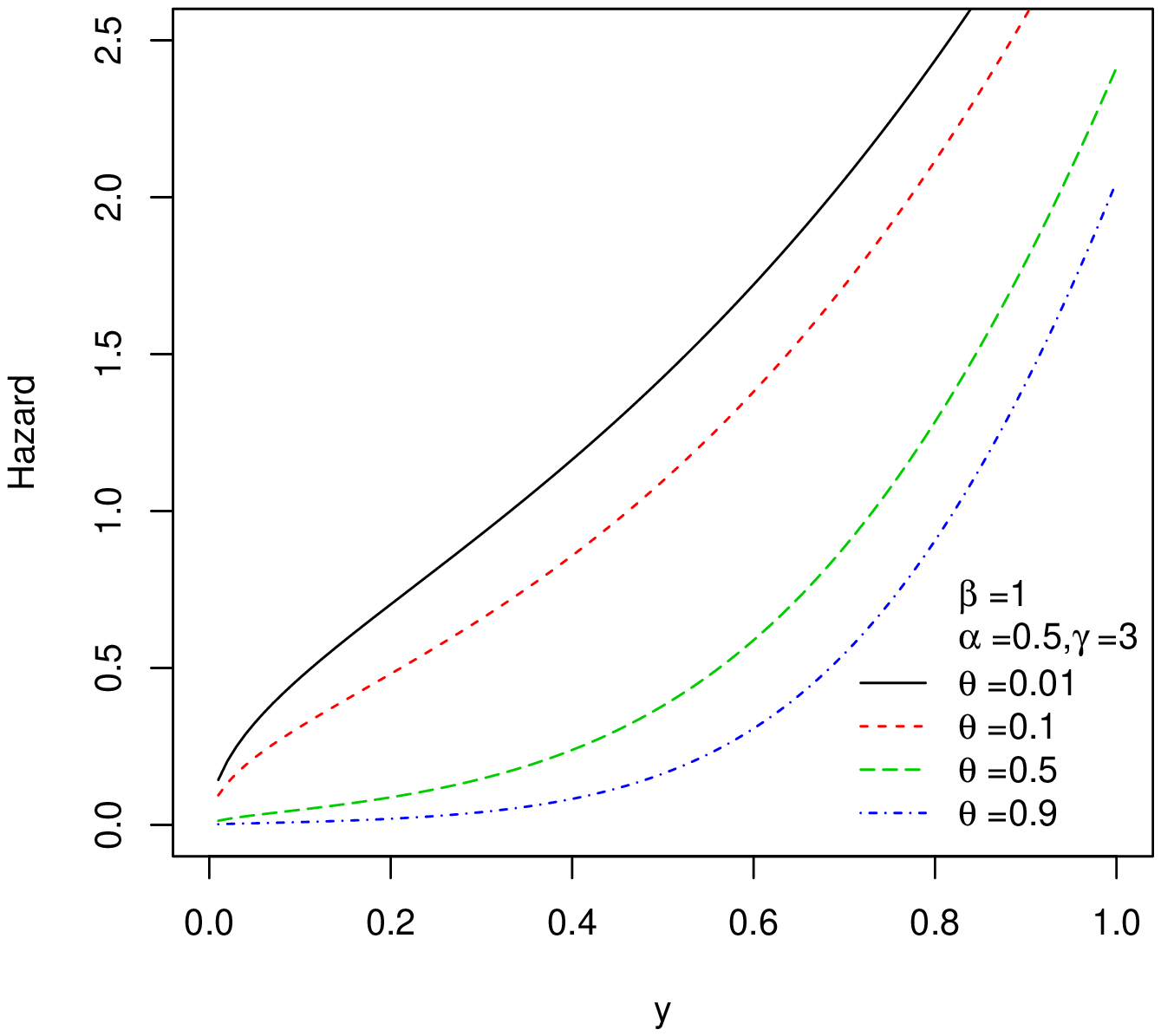}
\caption[]{Plots of pdf and hazard rate function of EWB for different values $‎\alpha, ‎\beta, ‎\gamma‎‎‎$ , $‎\theta‎$ and $m=10$.}
\end{figure}
From (\ref{mgf EWPS}), the moment generating function of EWB is
\begin{equation*}
 \begin{array}[b]{l}\label{mgf EWB}
M_{Y}(t)=‎\frac{‎\alpha‎ ‎\theta‎}{(‎\theta+1‎)^{m}-1}‎\sum^{\infty}_{i=0}\sum^{m}_{n=1}\sum^{\infty}_{j=0}‎\frac{(-1)^j}{i!}(‎\frac{t}{‎\beta‎}‎)‎^{i}‎\theta‎‎^{n-1}‎{m \choose n}{n\alpha-1 \choose j}‎\frac{n ‎\Gamma‎\left(1+\frac{i}{\gamma}\right)‎‎}{(j+1)^{1+\frac{i}{\gamma}}}.‎
\end{array}
\end{equation*}
\begin{equation*}
E(Y^k)=‎\frac{‎\alpha ‎\theta \Gamma‎\left(1+\frac{k}{\gamma}\right)‎‎}{‎\beta‎^{k} ‎\left[ (‎\theta+1‎)^{m}-1‎\right] ‎‎‎‎}‎\sum^{m}_{n=1}\sum^{\infty}_{j=0}(-1)^j‎\theta‎‎^{n-1}‎{m \choose n}{n\alpha-1 \choose j}‎\frac{n‎‎‎‎}{(j+1)^{1+\frac{k}{\gamma}}}.‎‎
\end{equation*}

\subsection{Exponentiated weibull poisson distribution}
The exponentiated weibull poisson distribution is a special case of power series distributions with
$a_{n}=n!^{-1}$ and $C(‎\theta‎)=‎‎e^{‎\theta‎}-1~(‎\theta‎>‎0‎‎)$.
Using the cdf in (\ref{cdf EWPS}), the cdf of exponentiated weibull poisson (EWP) distribution is given by

\begin{equation*}\label{cdf EWP}
F(y)=‎\frac{e^{‎\theta (1 -e^{-(‎\beta y‎)^{‎\gamma‎}}
)^{‎\alpha‎}}‎ -1}{e^{‎\theta‎}- 1}‎,
\end{equation*}

\begin{equation*}\label{pdf EWP}
f(y)=‎\frac{‎\alpha‎ \gamma‎ \theta‎‎‎ ‎\beta ^{‎\gamma‎} y^{‎\gamma -1‎}‎}{e^{‎\theta‎} - 1}
 e^{-(‎\beta  y‎)^{‎\gamma‎}}(1 -e^{-(‎\beta y‎)^{‎\gamma‎}})^
 {‎\alpha -1‎} e^{‎\theta (1 -e^{-(‎\beta y‎)^{‎\gamma‎}} )^{‎\alpha‎}‎} ‎,
\end{equation*}
and
\begin{equation*}\label{hazard EWP}
h(y)=\frac{‎\alpha‎ \gamma‎ \theta‎‎‎ ‎\beta
^{‎\gamma‎} y^{‎\gamma -1‎} e^{-(‎\beta
y‎)^{‎\gamma‎}}(1 -e^{-(‎\beta
y)^{‎\gamma‎}})^{‎\alpha -1‎} e^{‎\theta (1 -e^{-(‎\beta
y)^{‎\gamma‎}} )^{‎\alpha‎}‎}}{e^{‎\theta‎} -
e^{‎\theta (1 -e^{-(‎\beta y‎)^{‎\gamma‎}}
)^{‎\alpha}}}‎‎.
\end{equation*}
The plots of density and hazard rate function of EWP distribution for some values of $‎\alpha, ‎\beta, ‎\gamma‎‎‎$ and $‎\theta‎$ are given in Fig. 3. \\

\begin{figure}[t]
\centering
\includegraphics[scale=0.25]{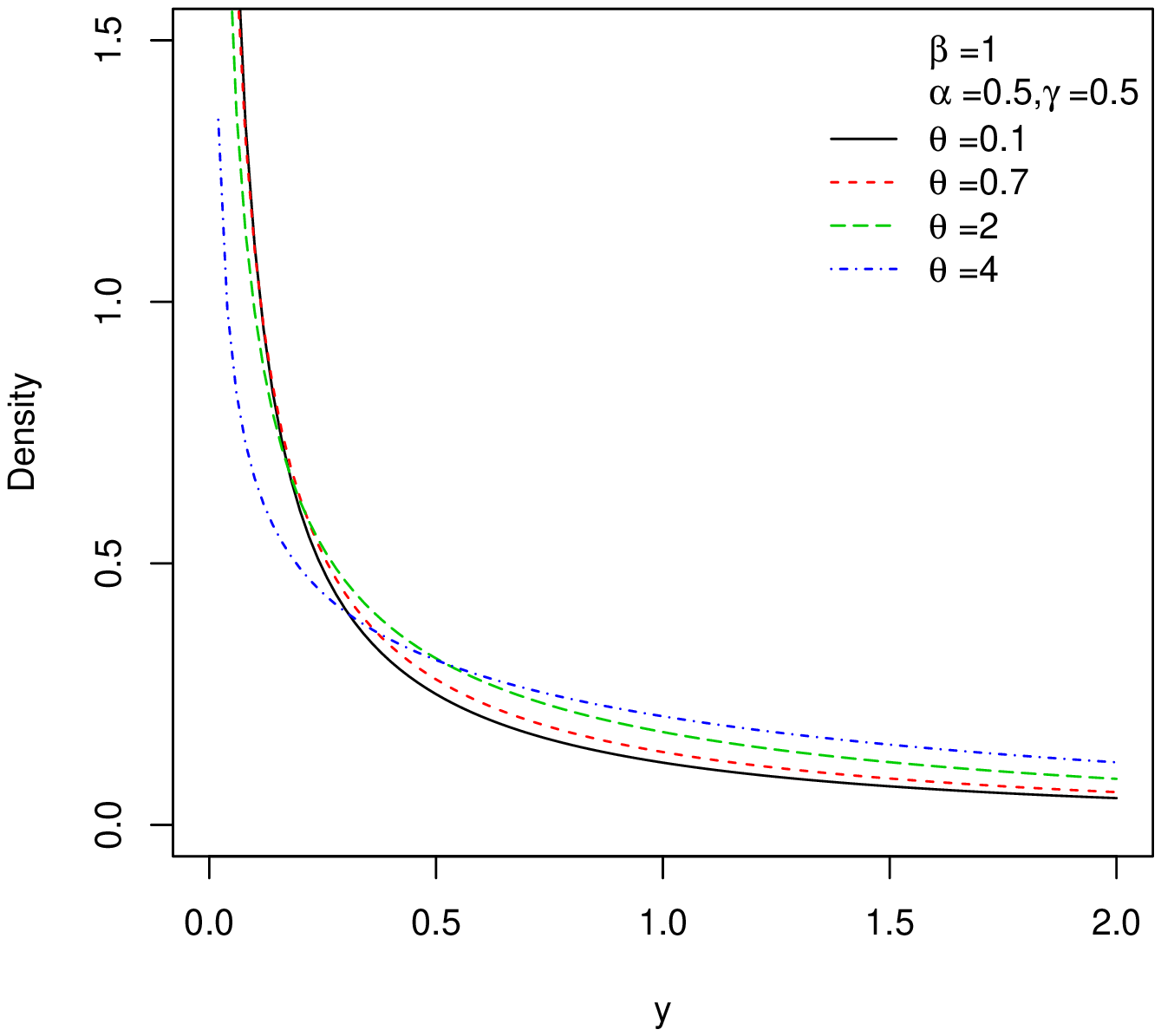}
\includegraphics[scale=0.25]{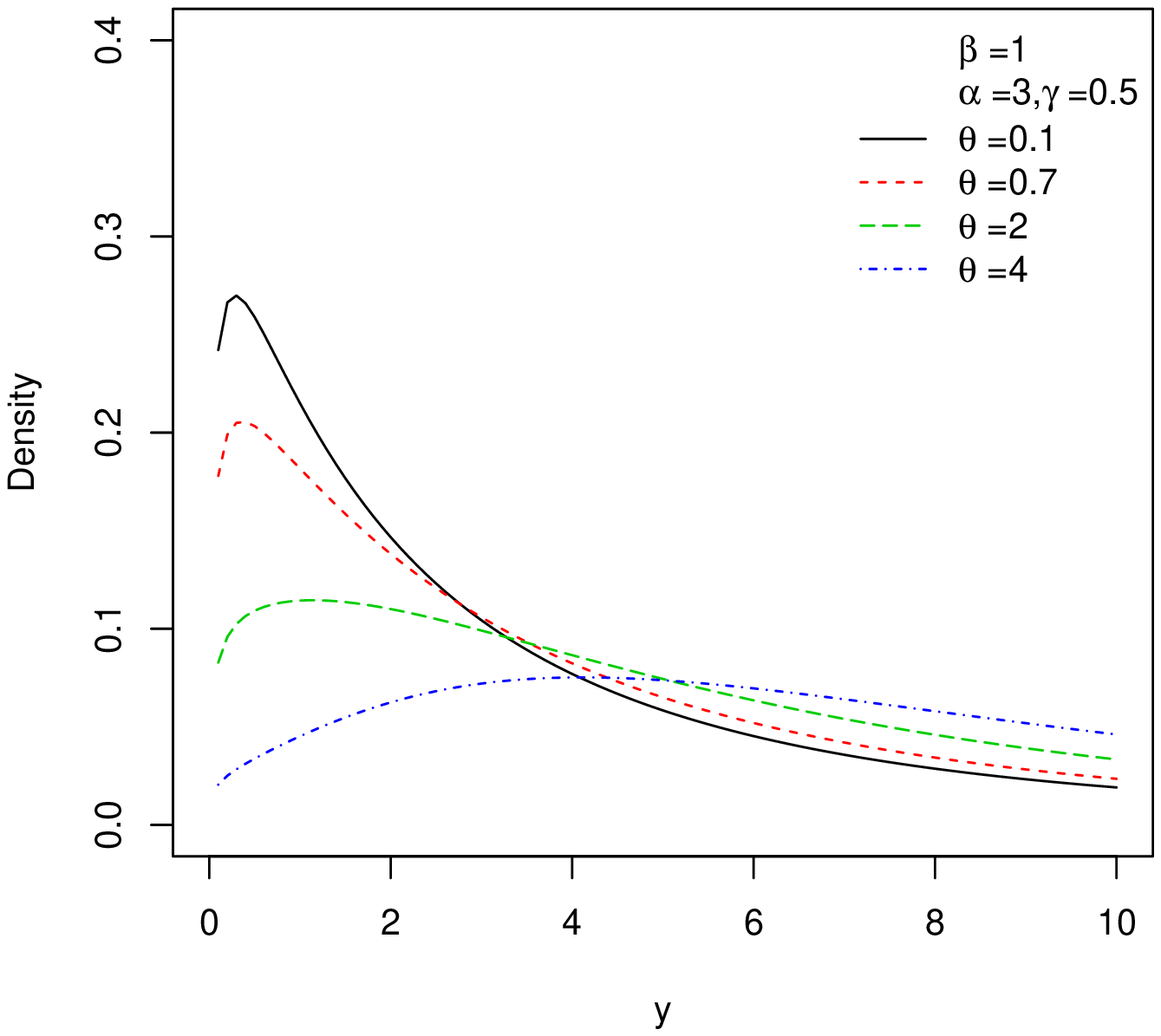}
\includegraphics[scale=0.25]{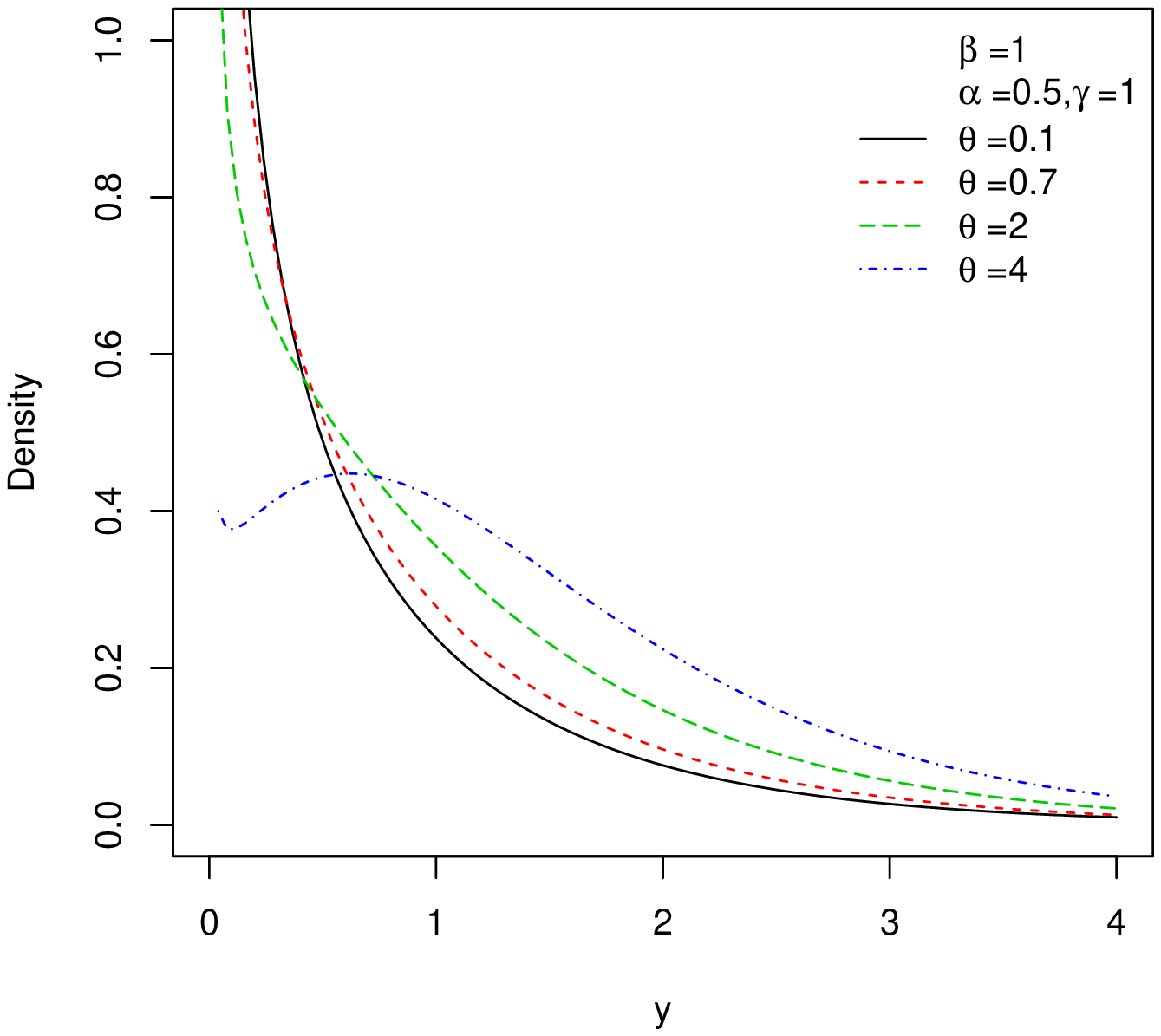}
\includegraphics[scale=0.25]{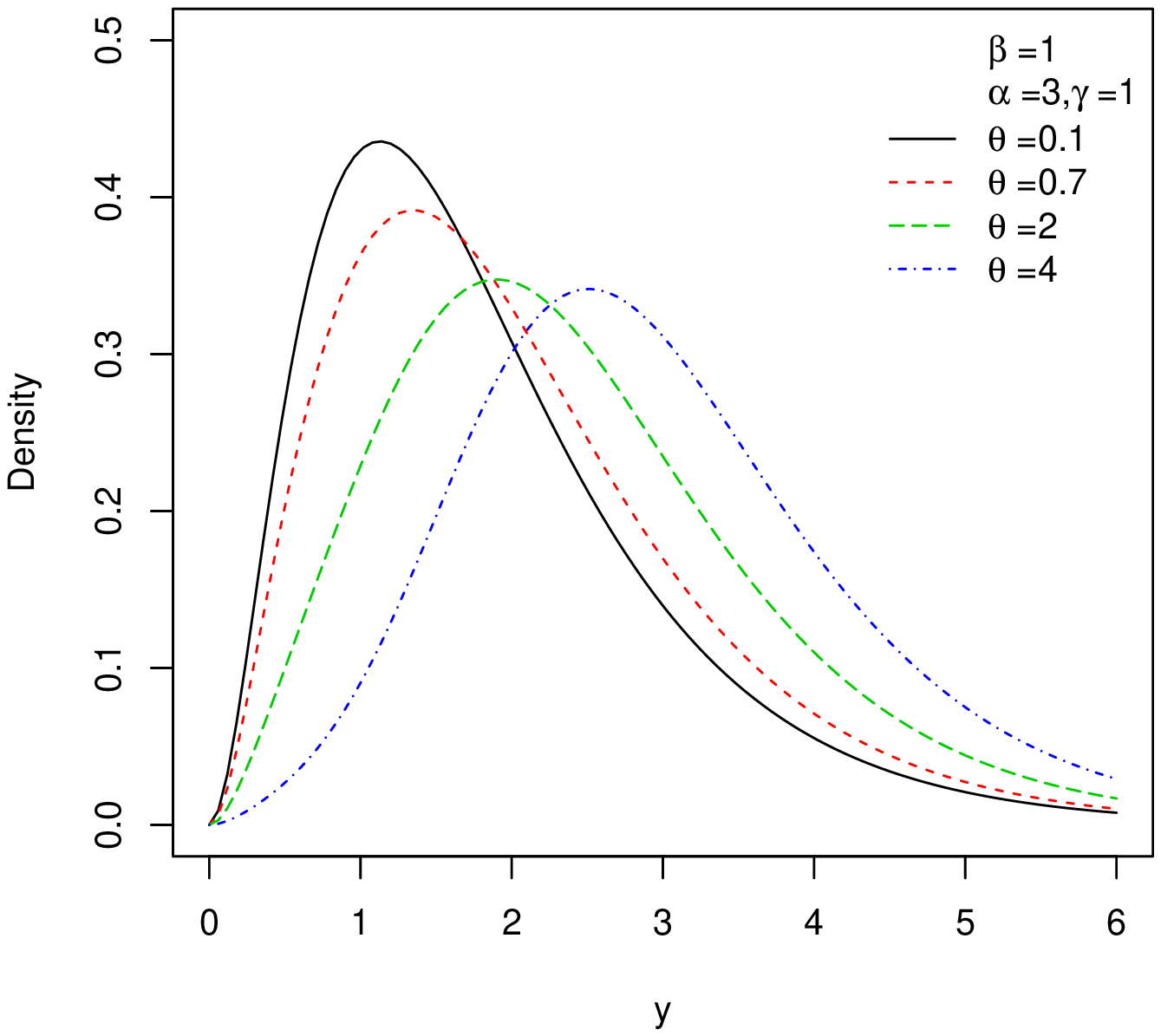}
\includegraphics[scale=0.25]{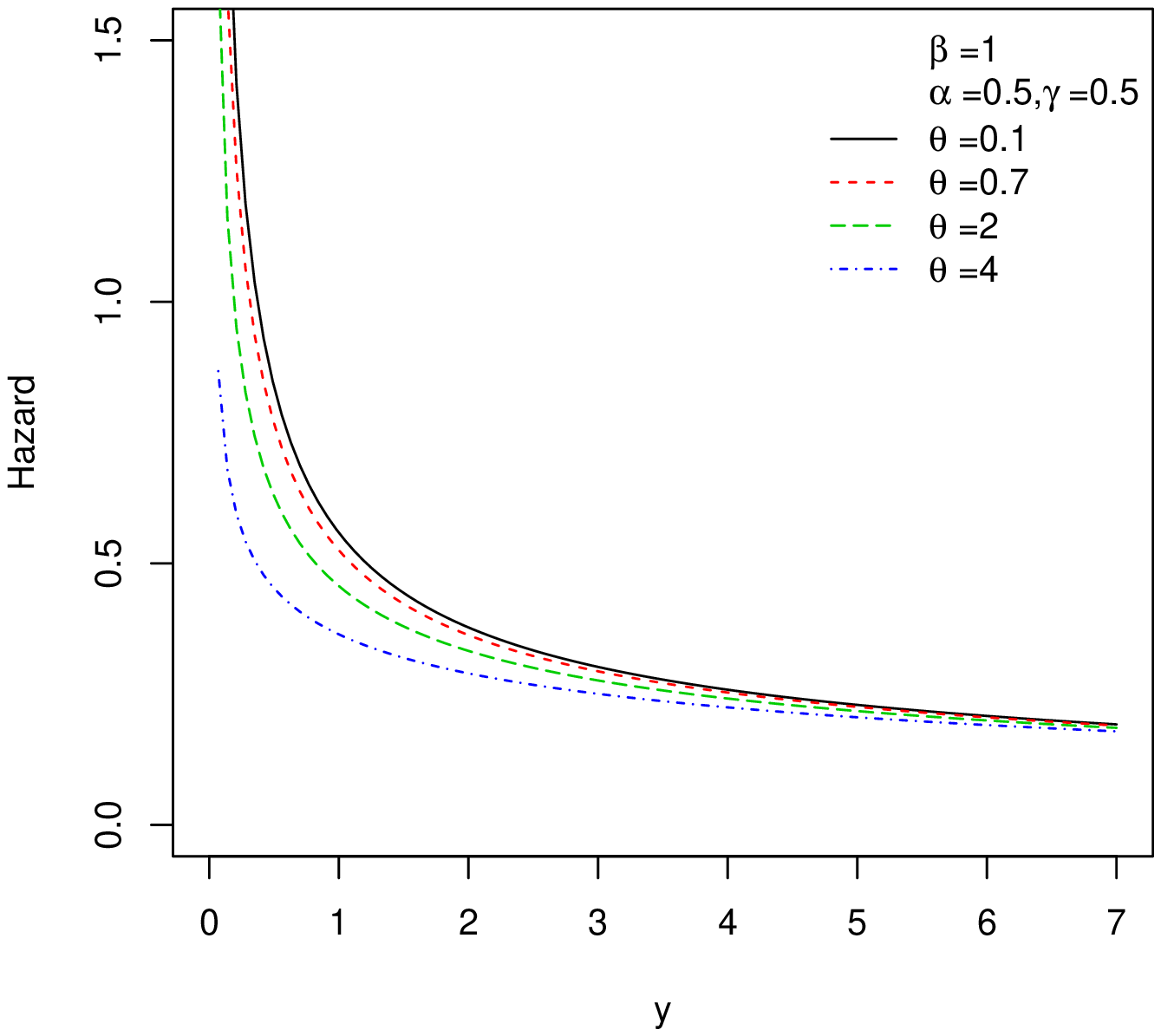}
\includegraphics[scale=0.25]{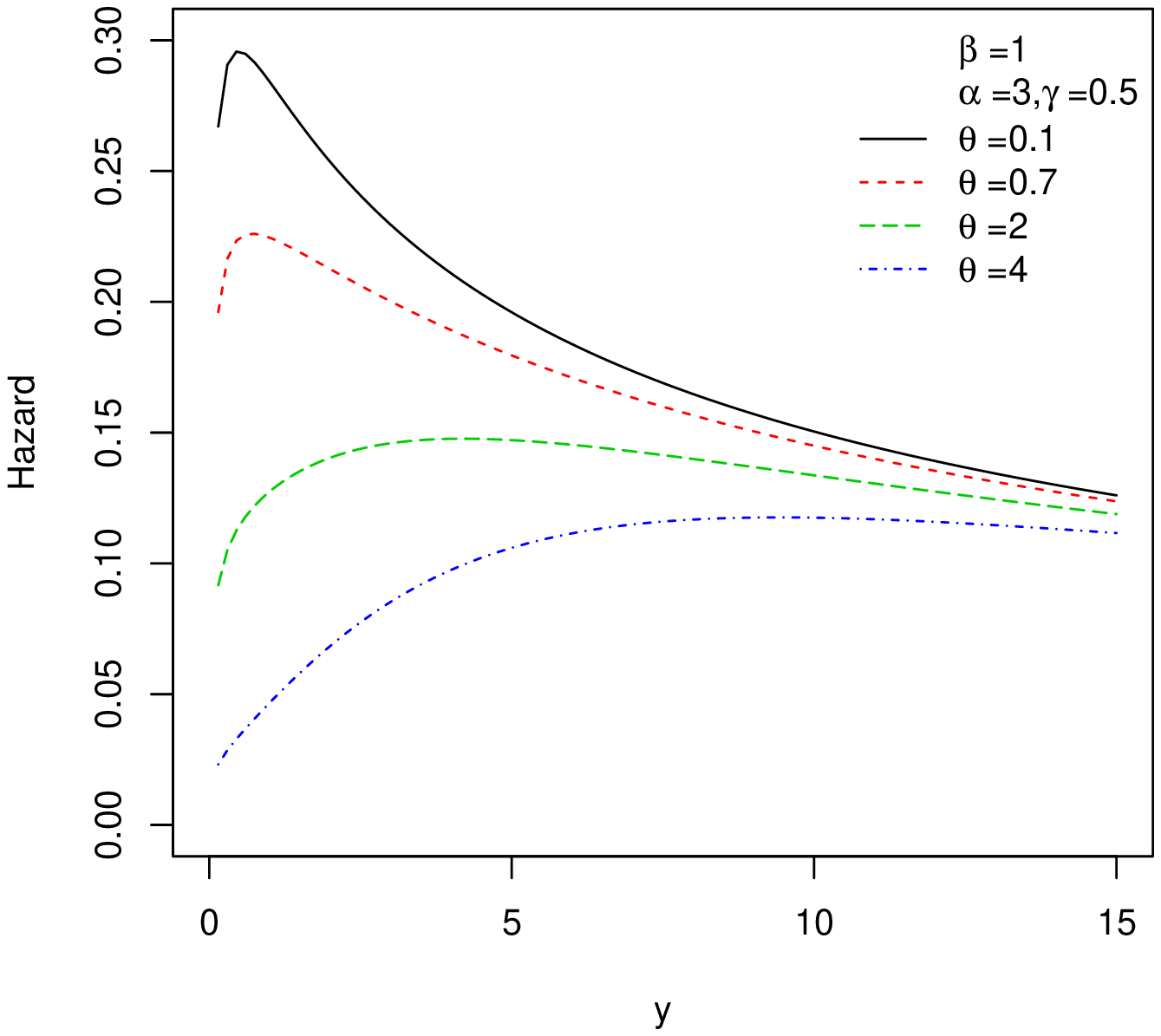}
\includegraphics[scale=0.25]{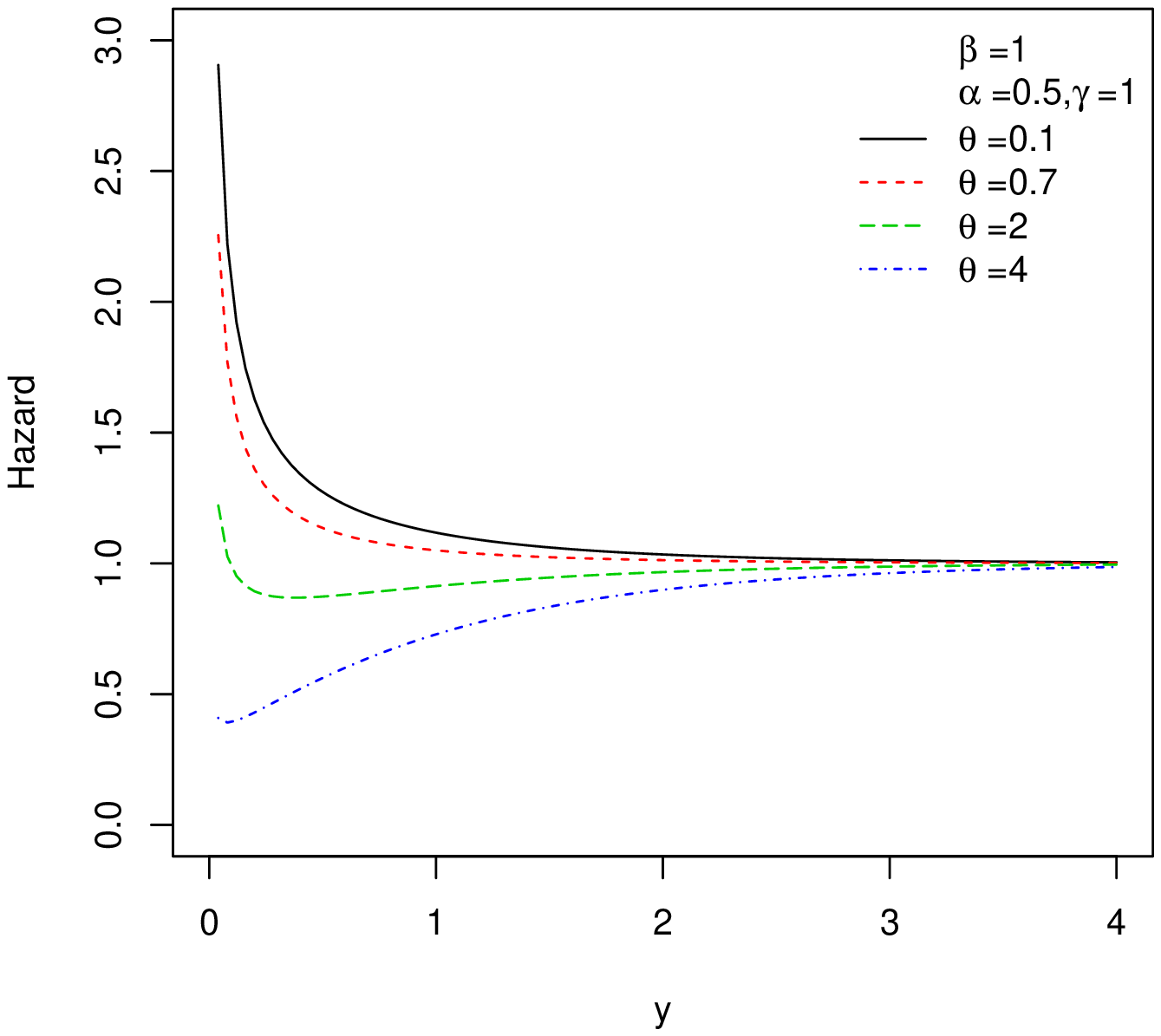}
\includegraphics[scale=0.25]{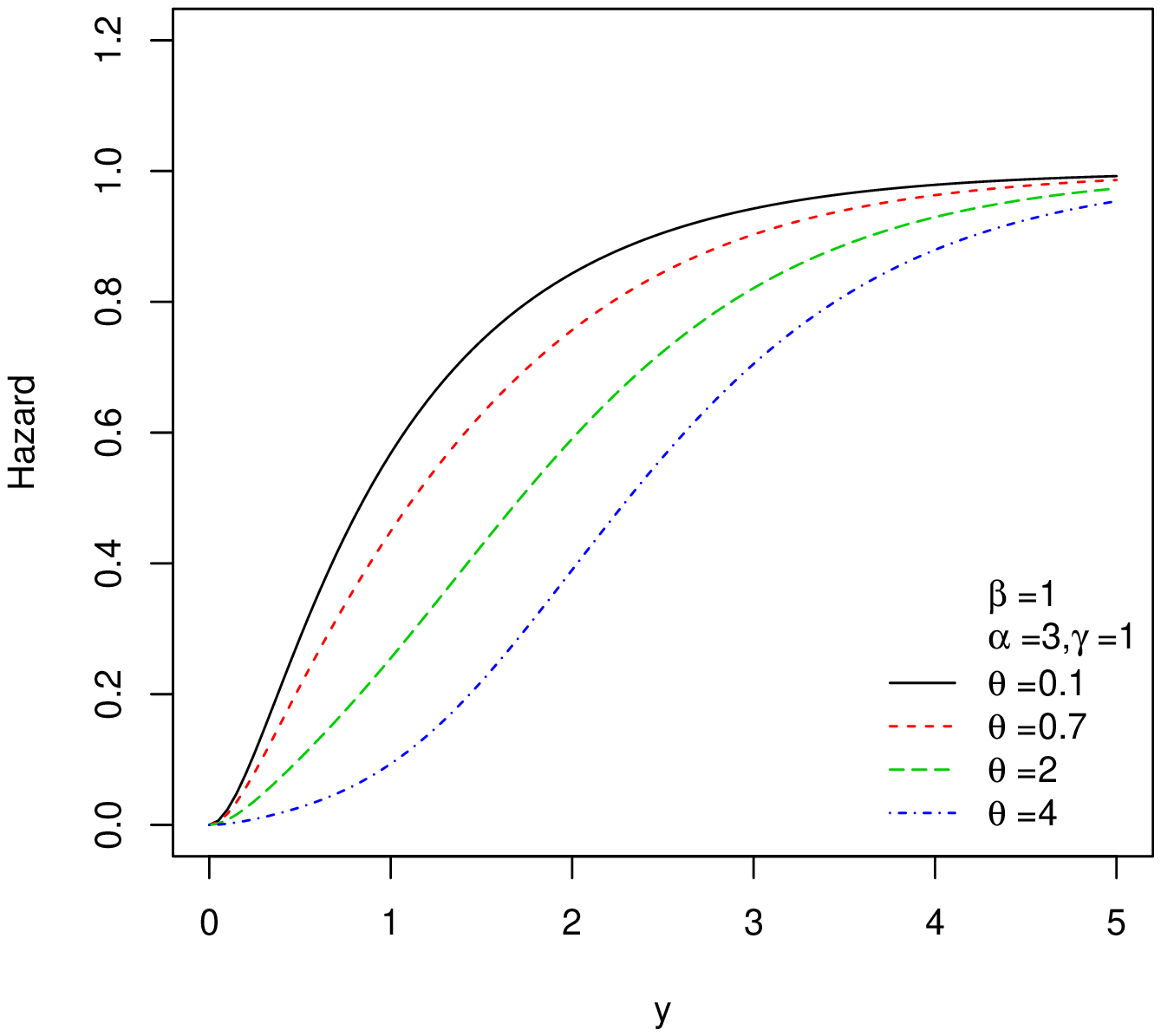}
\caption[]{Plots of pdf and hazard rate function of EWP for different values $‎\alpha, ‎\beta, ‎\gamma‎‎‎$ and $‎\theta‎$.}
\end{figure}
From (\ref{mgf EWPS}), the moment generating function of EWP is
\begin{equation*}
 \begin{array}[b]{l}\label{mgf EWP}
M_{Y}(t)=\frac{\alpha\theta‎‎}{(e^{\theta}-1)}\sum^{\infty}_{i=0}\sum^{\infty}_{n=1}\sum^{\infty}_{j=0}‎\frac{(-1)^{j}}{i!}‎\frac{\theta^{n-1}}{(n-1)!}(\frac{t}{‎\beta‎})‎^{i}‎
{n\alpha-1 \choose
j}‎\frac{\Gamma(1+\frac{i}{\gamma})}{‎(j+1)^{1+\frac{i}{\gamma}}}.
\end{array}
\end{equation*}
\begin{equation*}
E(Y^k)=‎\frac{\alpha\theta‎\Gamma(1+\frac{k}{\gamma})}{\beta^{k}(e^{\theta}-1)}\sum^{\infty}_{n=1}\sum^{\infty}_{j=0}(-1)^{j}{n\alpha-1 \choose j}\frac{\theta^{n-1}}{(n-1)!(j+1)^{1+\frac{k}{\gamma}}}.
\end{equation*}

\subsection{Exponentiated weibull geometric distribution}
The exponentiated weibull geometric distribution is a special case of power series distributions with
$a_{n}=1$ and $C(‎\theta‎)=\theta‎‎(1-‎\theta‎)^{-1}~(0<‎\theta‎<1)$.
Using the cdf in (\ref{cdf EWPS}), the cdf of exponentiated weibull poisson (EWG) distribution is given by

\begin{equation*}\label{cdf EWG}
F(y)=\frac{(1-\theta)\left(1-e^{-(\beta
y)^{\gamma}}\right)^{\alpha}}{1-\theta\left(1-e^{-(\beta
y)^{\gamma}}\right)^{\alpha}}‎,
\end{equation*}

\begin{equation*}\label{pdf EWG}
f(y)=‎\frac{(1-\theta)\alpha\gamma\beta^{\gamma}y^{\gamma-1}e^{-(\beta
y)^{\gamma}}\left(1-e^{-(\beta
y)^{\gamma}}\right)^{\alpha-1}}{\left[1-\theta\left(1-e^{-(\beta
y)^{\gamma}}\right)^{\alpha}\right]^2},
\end{equation*}
and
\begin{equation*}\label{hazard EWG}
h(y)=\frac{(1-\theta)\alpha\gamma\beta^{\gamma}y^{\gamma-1}e^{-(\beta y)^{\gamma}}
\left(1-e^{-(\beta y)^{\gamma}}\right)^{\alpha-1}}{\left[1-\theta\left(1-e^{-(\beta y)^
{\gamma}}\right)^{\alpha}\right]\left[1-\left(1-e^{-(\beta y)^{\gamma}}\right)^{\alpha}\right]}.
\end{equation*}
The plots of density and hazard rate function of EWG distribution for some values of $‎\alpha, ‎\beta, ‎\gamma‎‎‎$ and $‎\theta‎$ are given in Fig. 4. \\

\begin{figure}[t]
\centering
\includegraphics[scale=0.25]{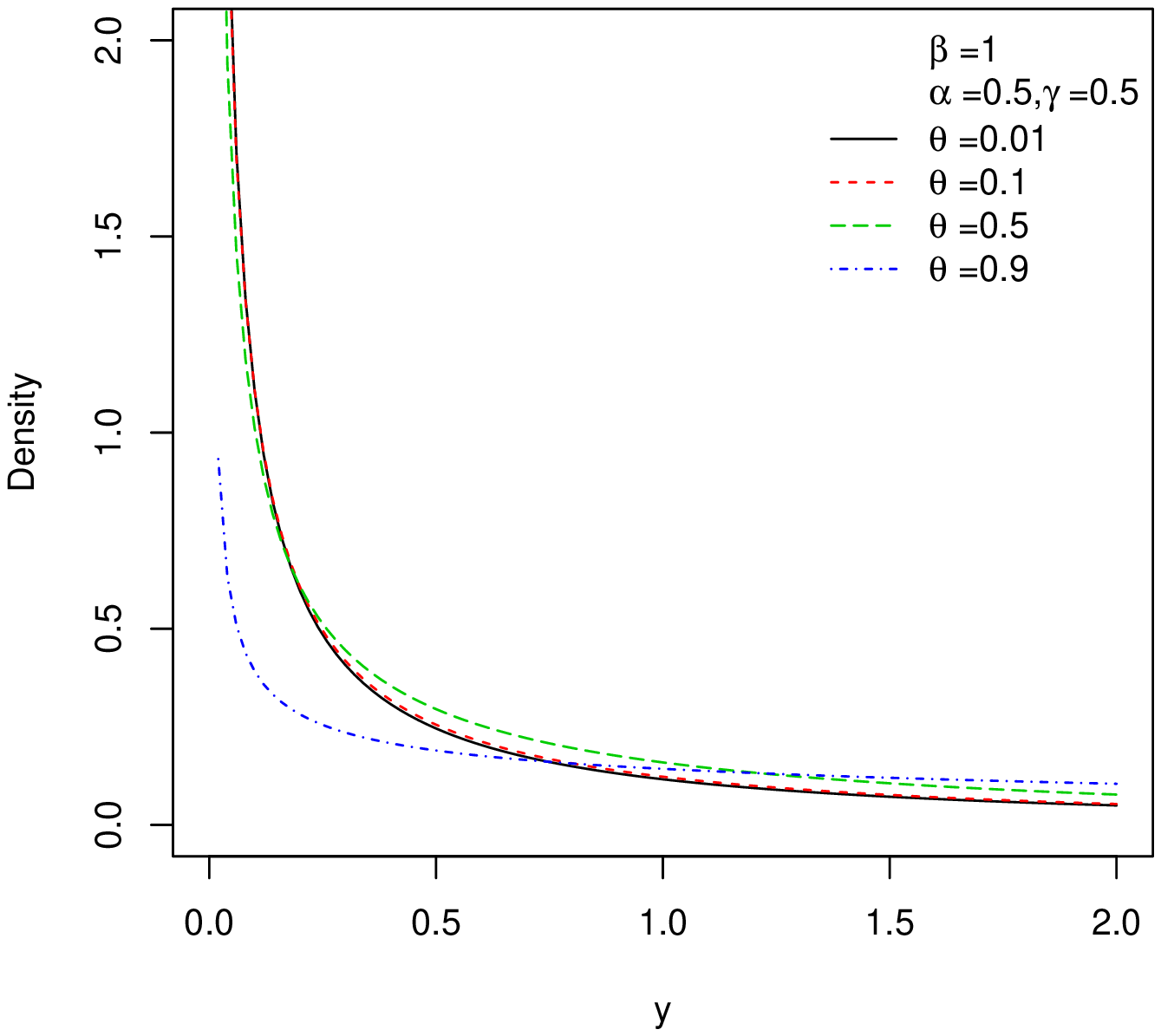}
\includegraphics[scale=0.25]{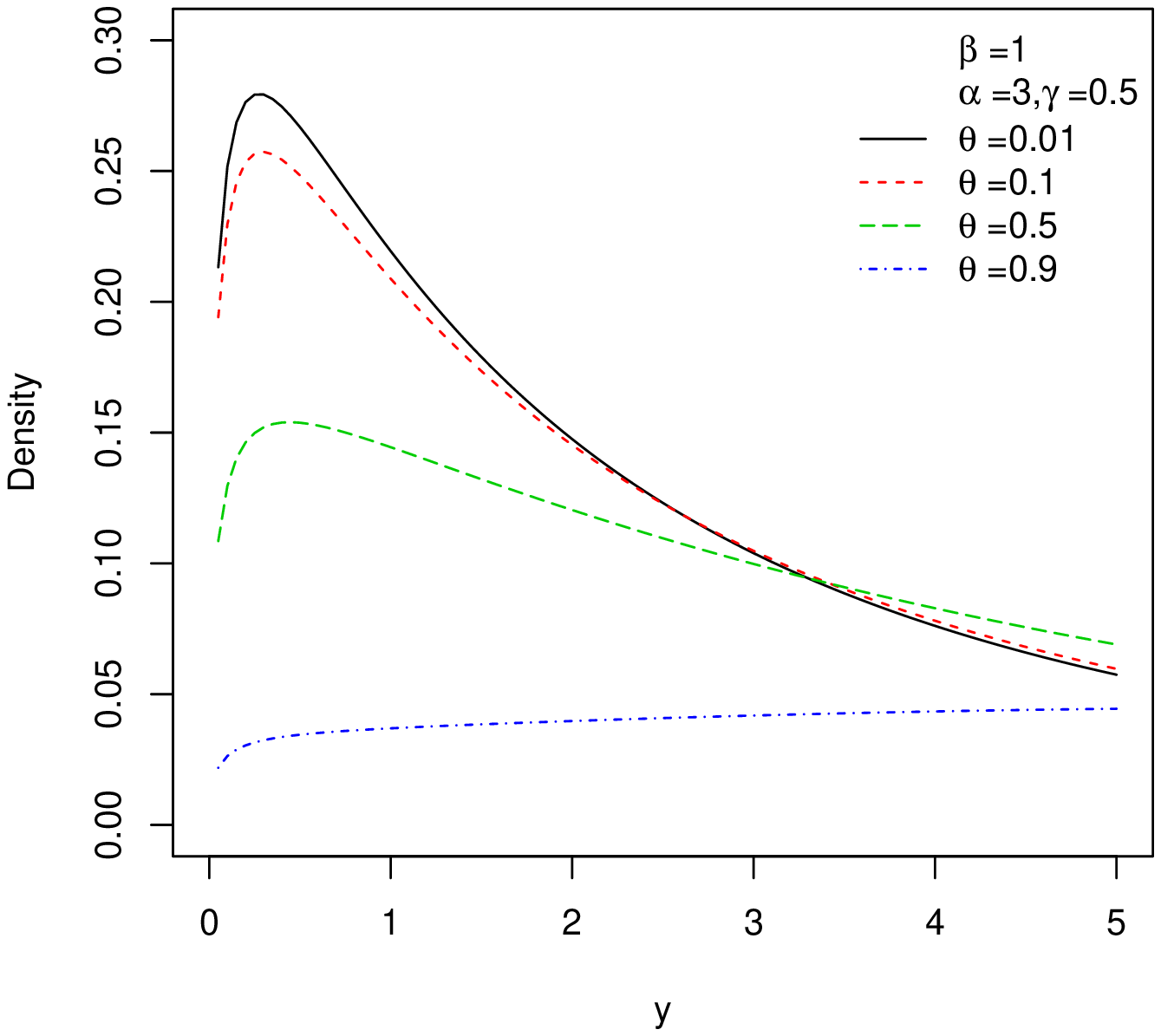}
\includegraphics[scale=0.25]{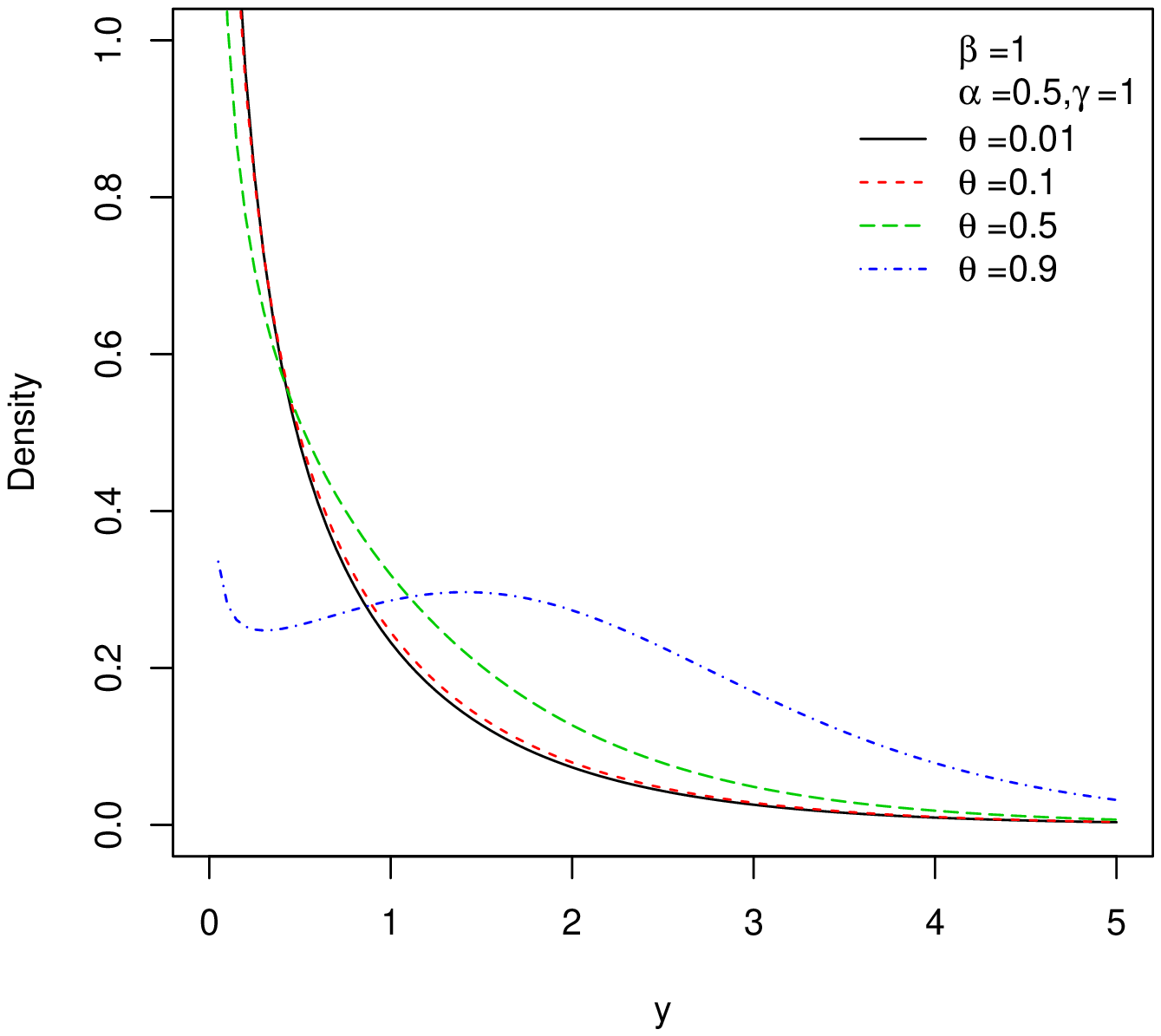}
\includegraphics[scale=0.25]{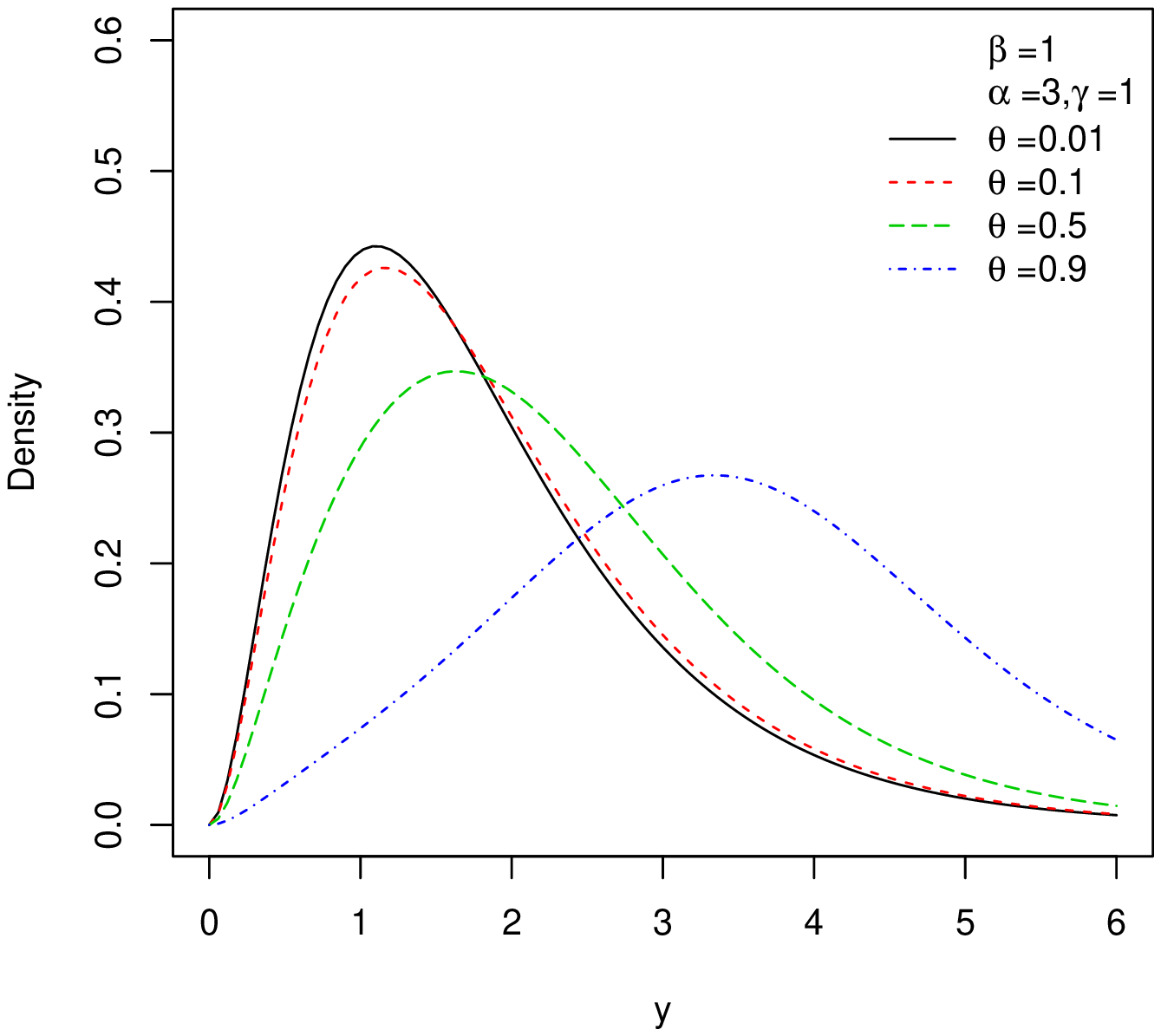}
\includegraphics[scale=0.25]{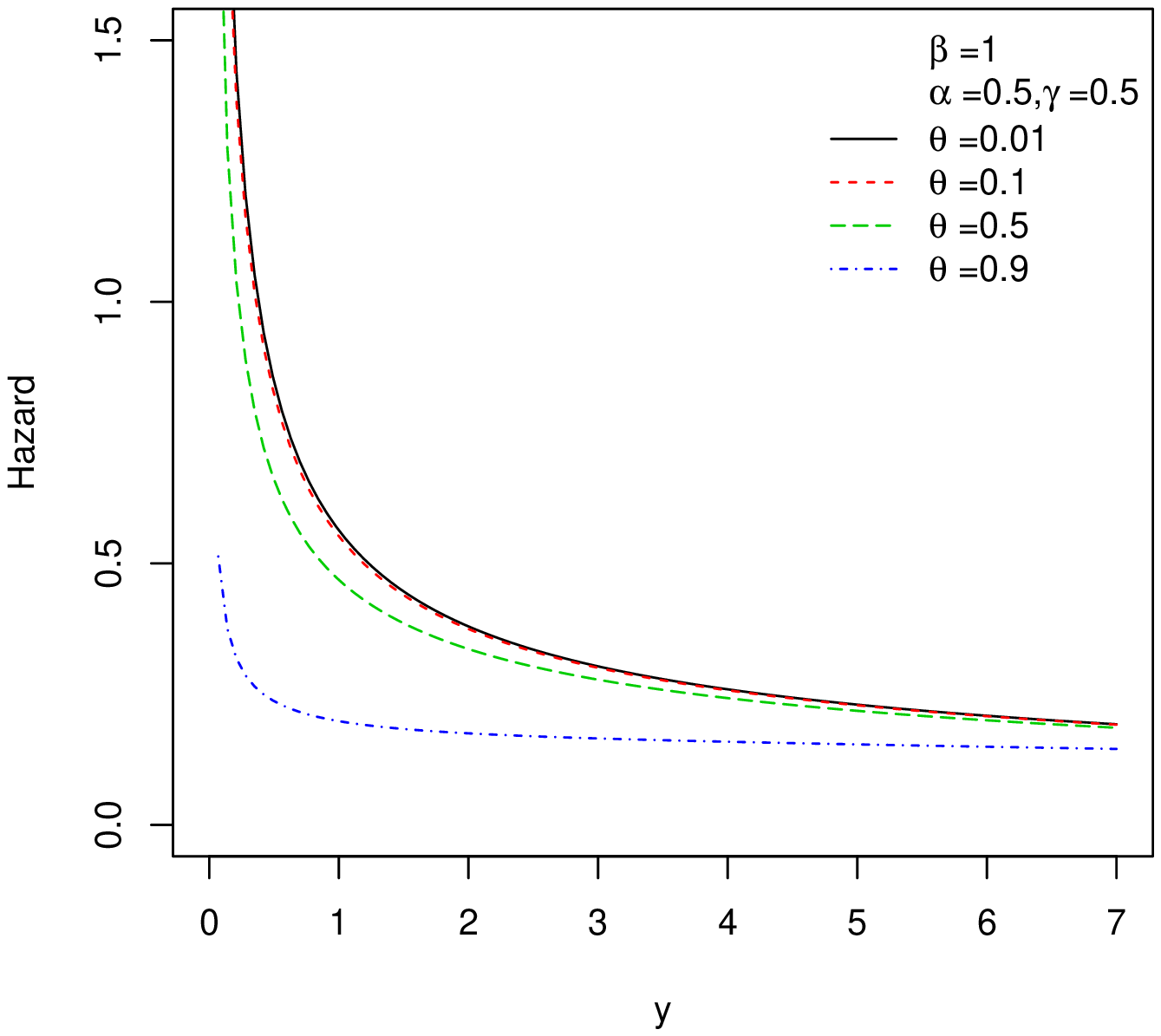}
\includegraphics[scale=0.25]{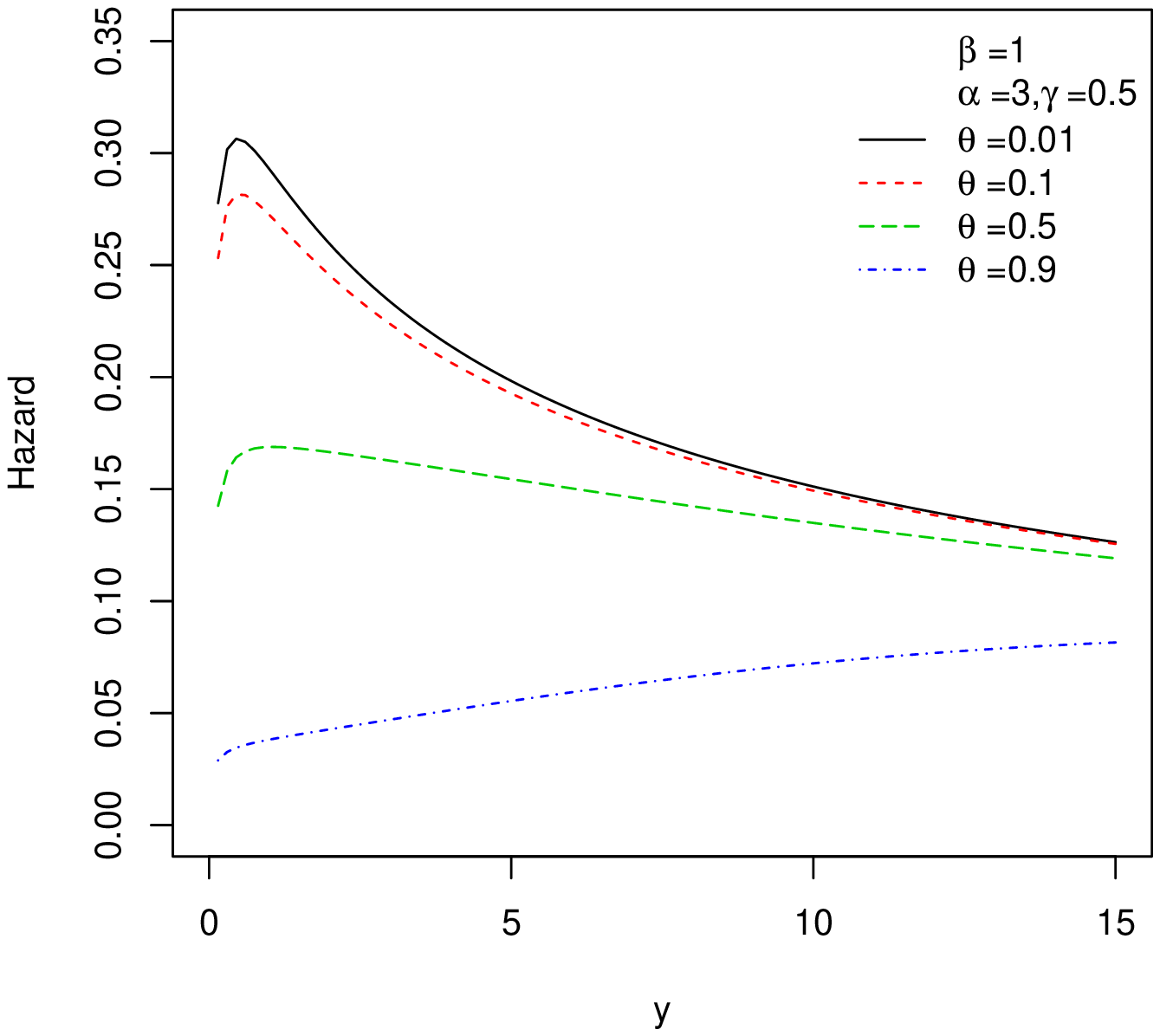}
\includegraphics[scale=0.25]{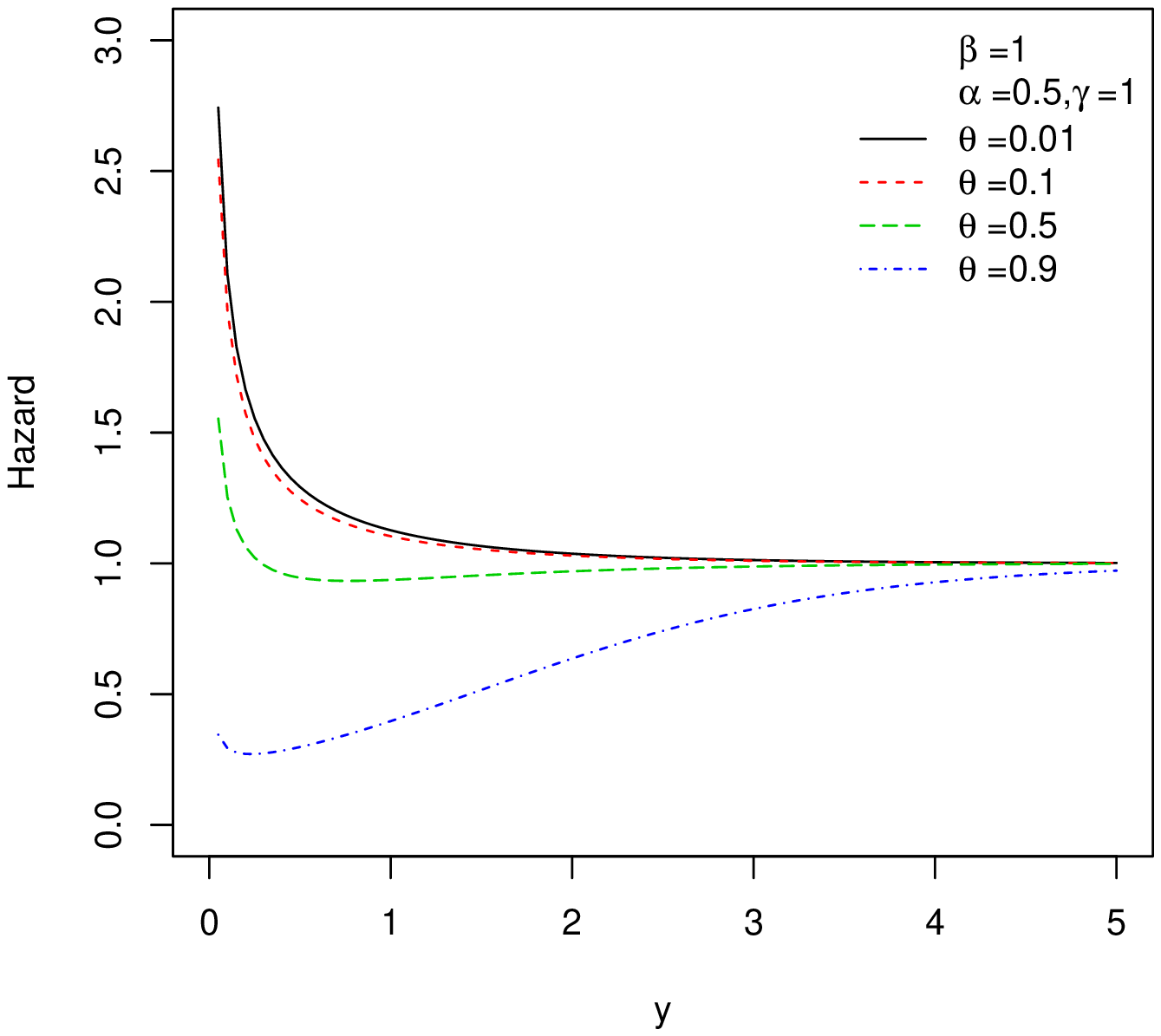}
\includegraphics[scale=0.25]{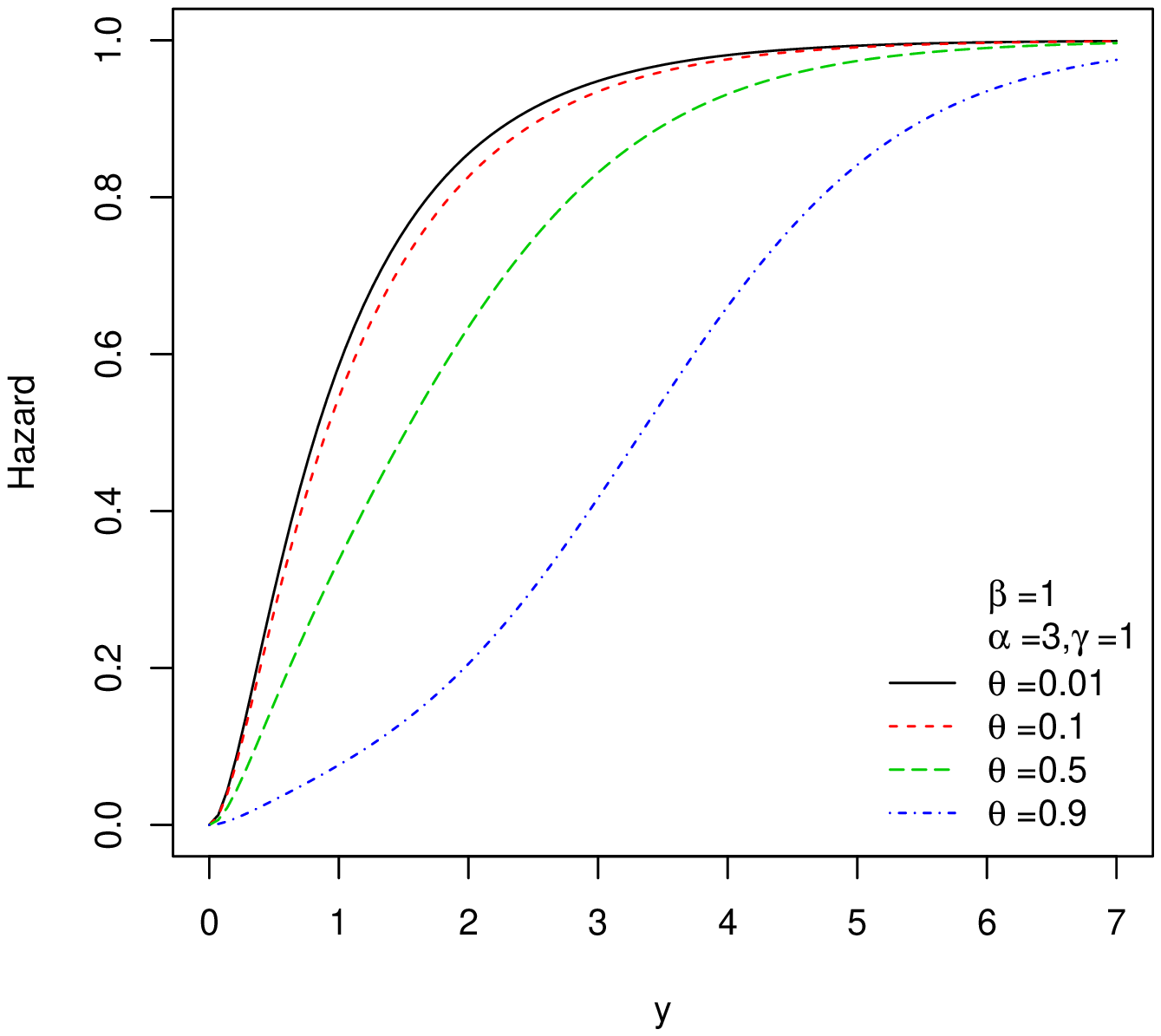}
\caption[]{Plots of pdf and hazard rate function of EWG for different values $‎\alpha, ‎\beta, ‎\gamma‎‎‎$ and $‎\theta‎$.}
\end{figure}
From (\ref{mgf EWPS}), the moment generating function of EWP is
\begin{equation*}
 \begin{array}[b]{l}\label{mgf EWG}
M_{Y}(t)=\alpha(1-\theta)\sum^{\infty}_{i=0}\sum^{\infty}_{n=1}\sum^{\infty}_{j=0}(-1)^j
{n\alpha-1 \choose
j}\Gamma\left(\frac{i}{\gamma}+1\right)\frac{(t/\beta)^{i}}{i!}n
\theta^{n-1}(j+1)^{-(\frac{i}{\gamma}+1)}.
\end{array}
\end{equation*}
\begin{equation*}
E(Y^k)=(1-\theta)\alpha
\beta^{-k}\Gamma\left(\frac{k}{\gamma}+1\right)\sum^{\infty}_{n=1}\sum^{\infty}_{j=0}(-1)^j
{n\alpha-1 \choose j} n \theta^{n-1}(j+1)^{-(\frac{k}{\gamma}+1)}.
\end{equation*}

\subsection{Exponentiated weibull logarithmic distribution}
The exponentiated weibull logarithmic distribution is a special case of power series distributions with
$a_{n}=n^{-1}$ and $C(‎\theta‎)=-‎‎\log‎(1-‎\theta‎)~(0<‎\theta‎<1)$.
Using the cdf in (\ref{cdf EWPS}), the cdf of exponentiated weibull poisson (EWL) distribution is given by

\begin{equation*}\label{cdf EWL}
F(y)=‎‎\frac{\log (1-‎\theta(1-e^{-(\beta
y)^{\gamma}})‎^{‎\alpha‎})}{\log (1-‎\theta‎)}‎,
\end{equation*}

\begin{equation*}\label{pdf EWL}
f(y)=‎\frac{‎\theta‎\alpha\gamma\beta^{\gamma}y^{\gamma-1}e^{-(\beta
y)^{\gamma}}\left(1-e^{-(\beta
y)^{\gamma}}\right)^{\alpha-1}}{\left[\theta\left(1-e^{-(\beta
y)^{\gamma}}\right)^{\alpha}-1\right]\log (1-‎\theta‎)},
\end{equation*}
and
\begin{equation*}\label{hazard EWL}
h(y)=\frac{‎\theta‎\alpha\gamma\beta^{\gamma}y^{\gamma-1}e^{-(\beta
y)^{\gamma}}\left(1-e^{-(\beta
y)^{\gamma}}\right)^{\alpha-1}}{\left[\theta\left(1-e^{-(\beta
y)^{\gamma}}\right)^{\alpha}-1\right]\log (‎\frac{1-‎\theta‎}{1-‎\theta(1-e^{-(\beta
y)^{\gamma}})‎^{‎\alpha‎}}‎)}.
\end{equation*}
The plots of density and hazard rate function of EWL distribution for some values of $‎\alpha, ‎\beta, ‎\gamma‎‎‎$ and $‎\theta‎$ are given in Fig. 5 \\

\begin{figure}[t]
\centering
\includegraphics[scale=0.25]{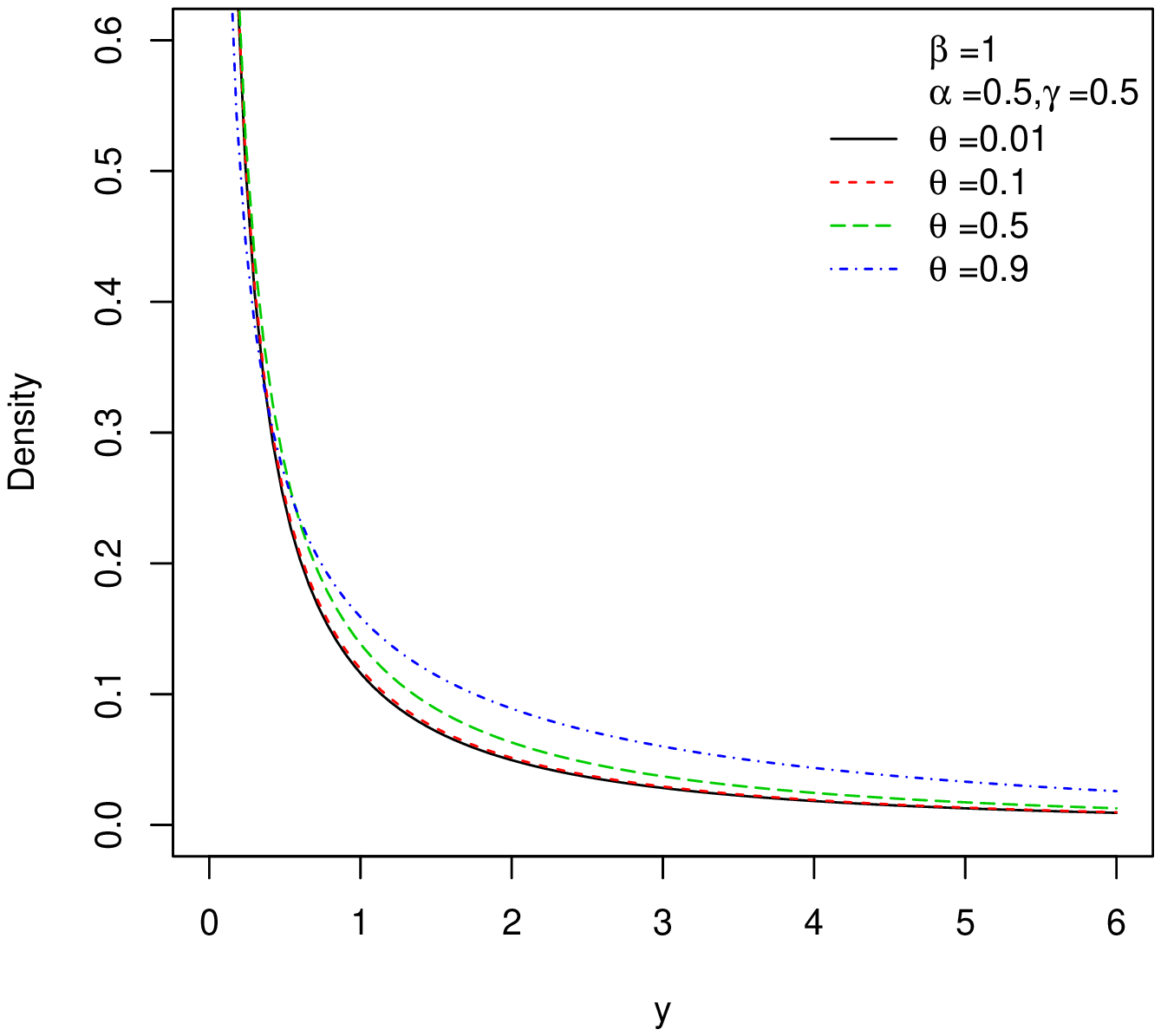}
\includegraphics[scale=0.25]{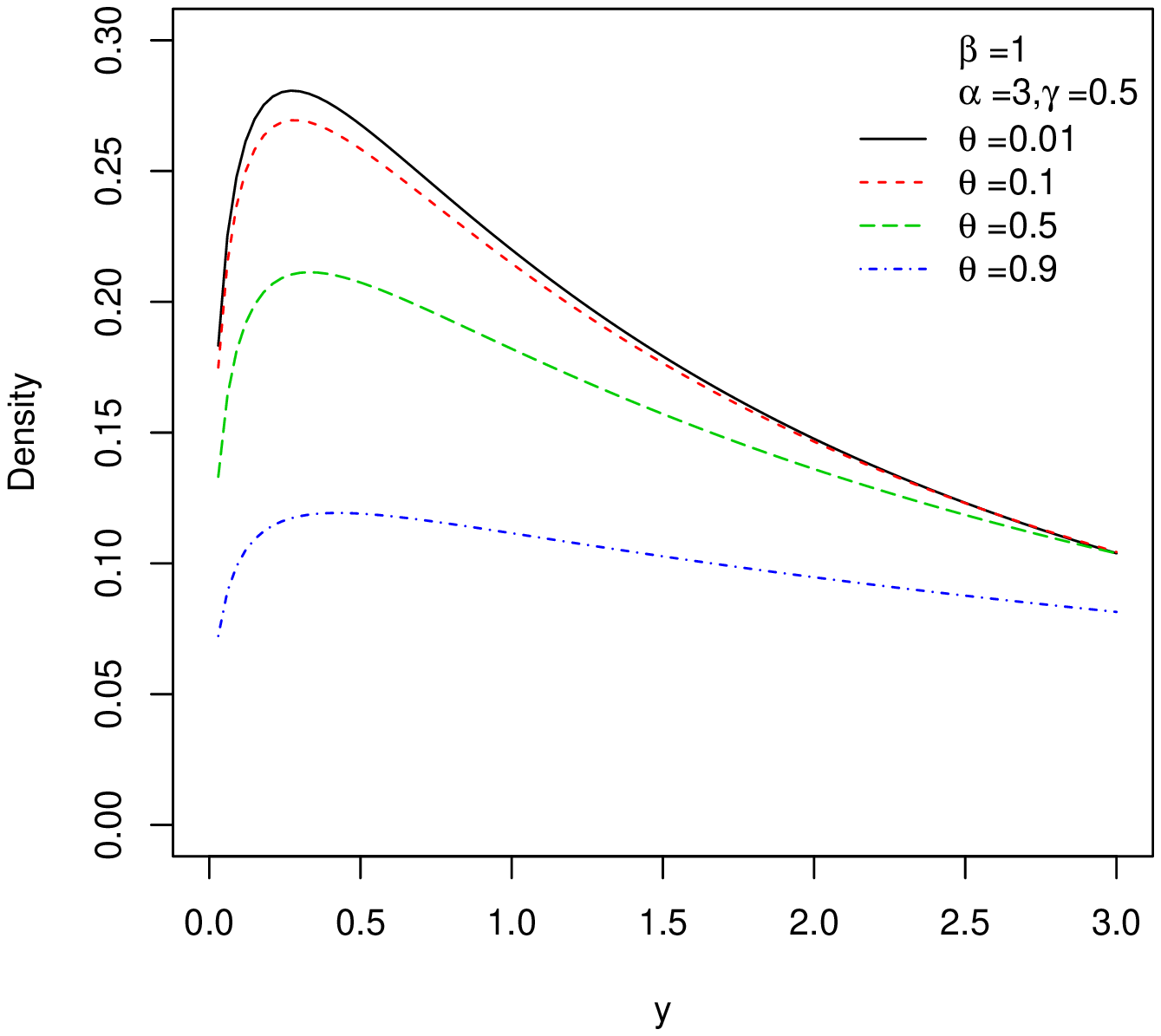}
\includegraphics[scale=0.25]{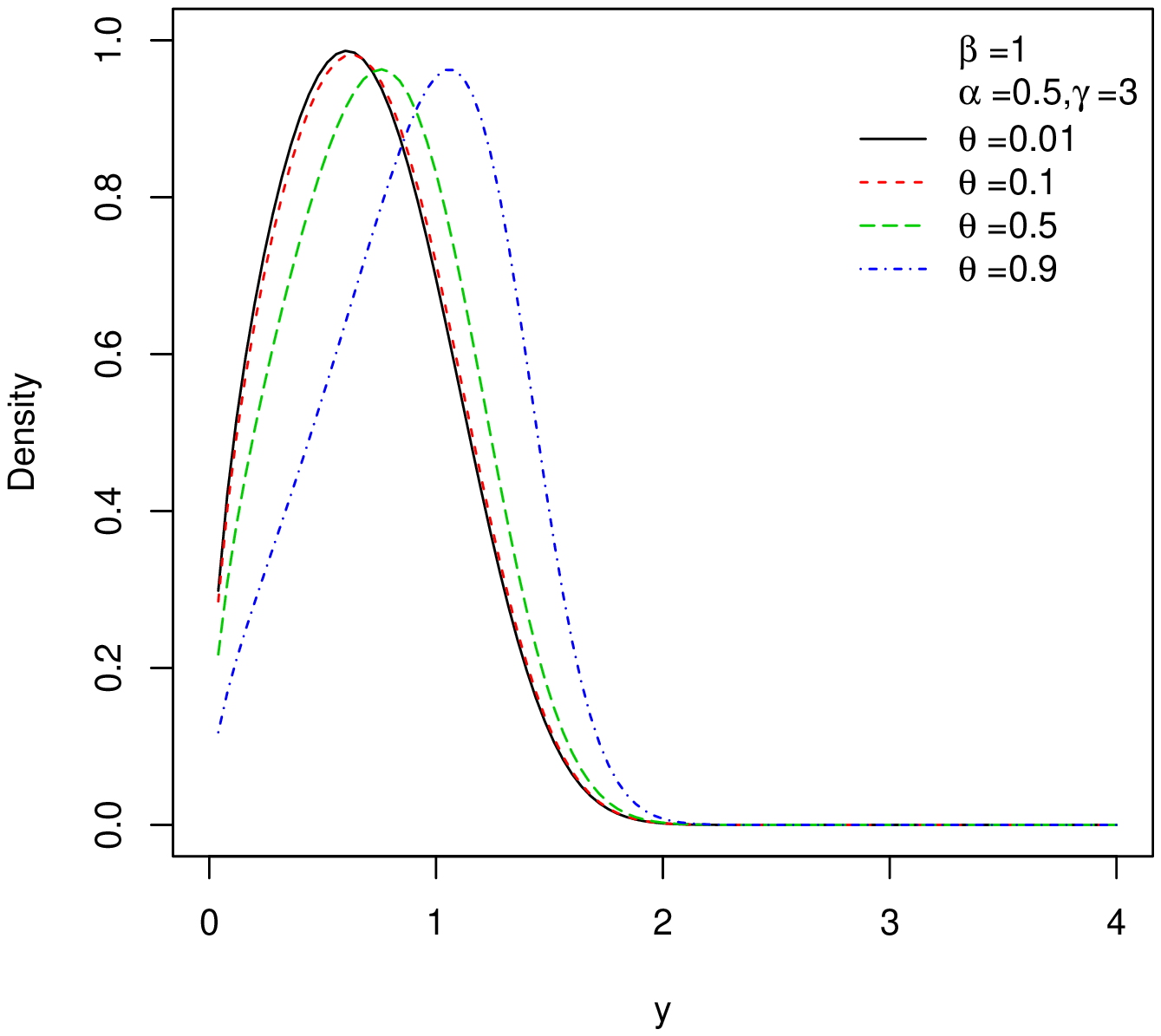}
\includegraphics[scale=0.25]{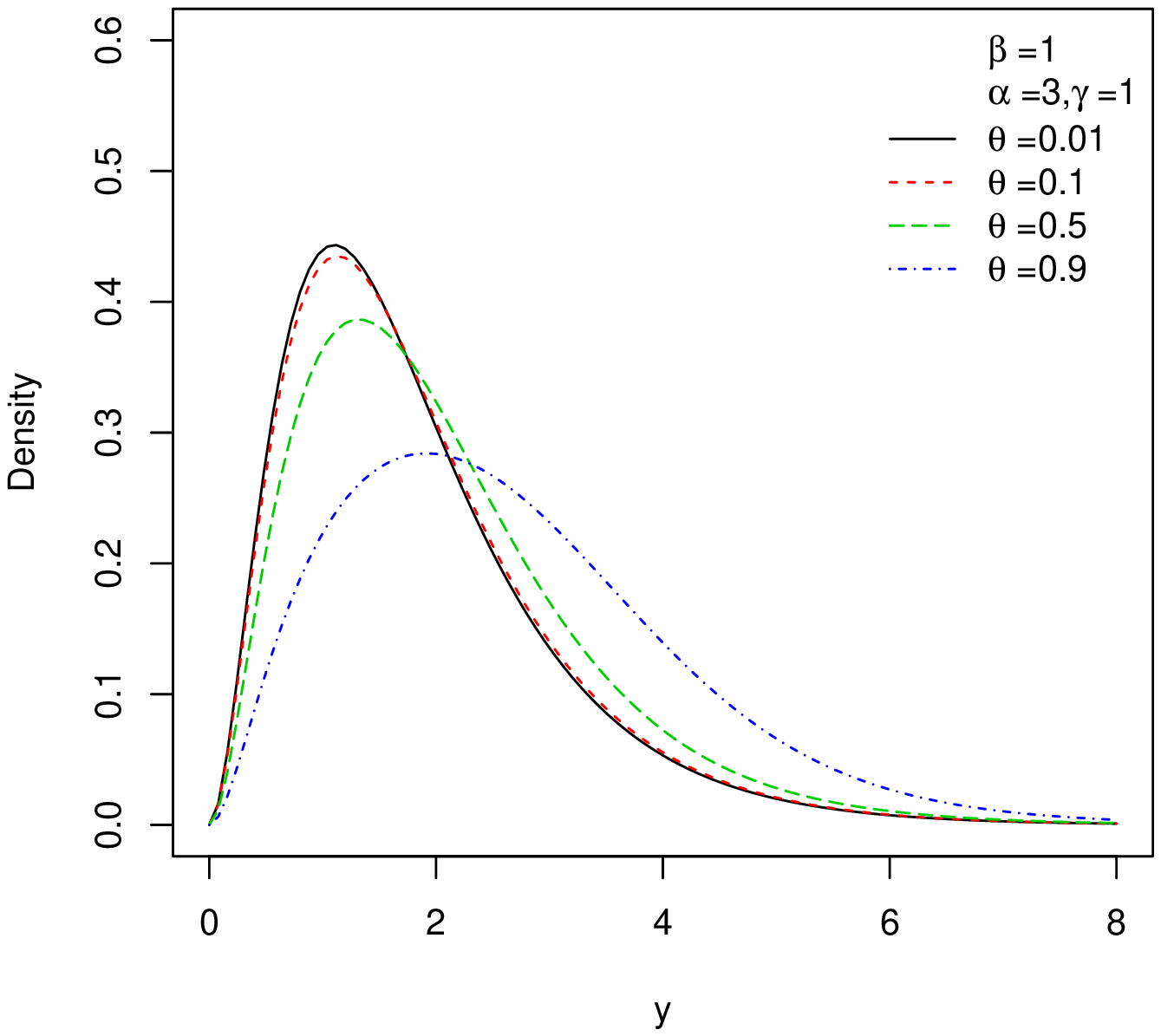}
\includegraphics[scale=0.25]{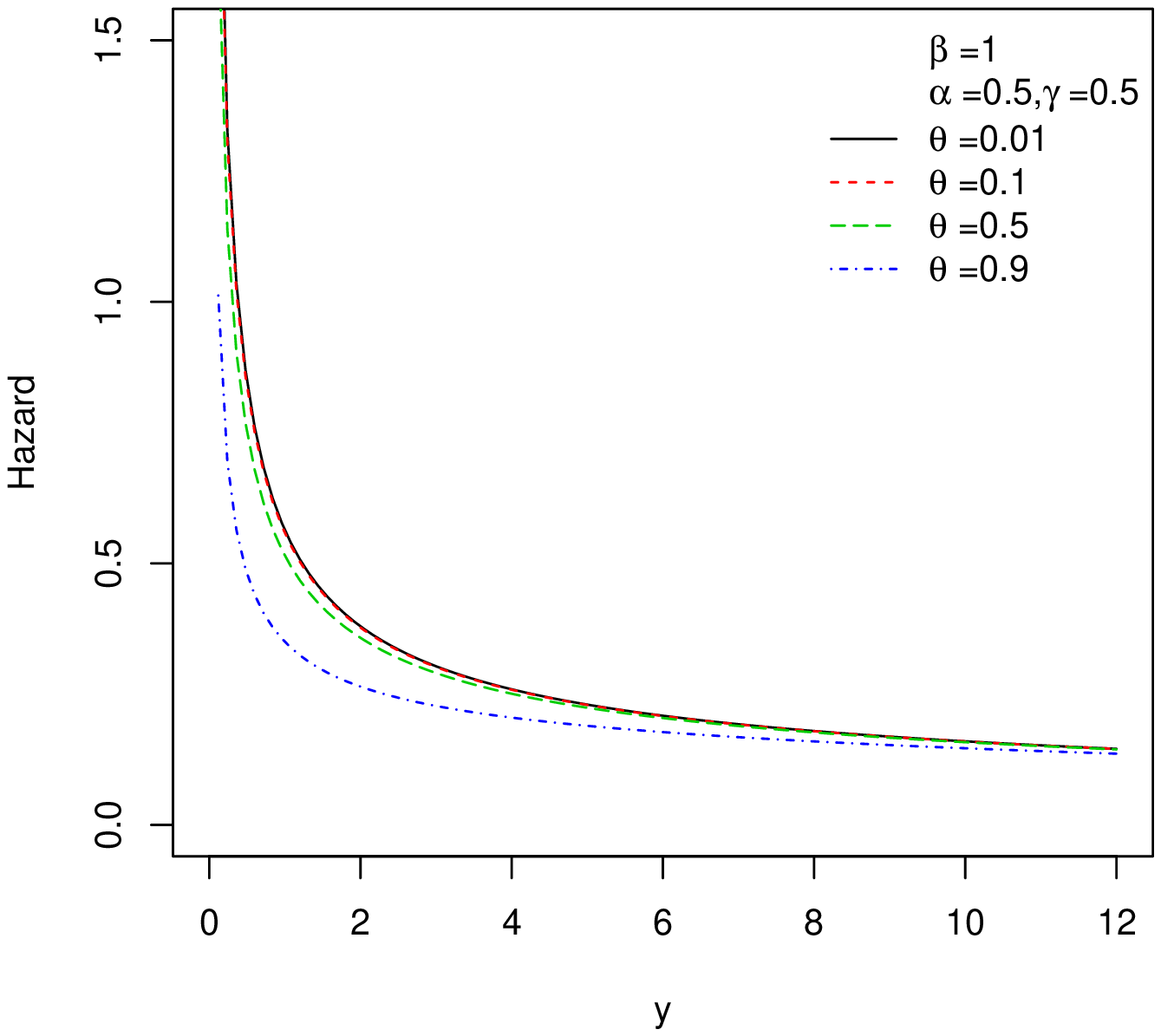}
\includegraphics[scale=0.25]{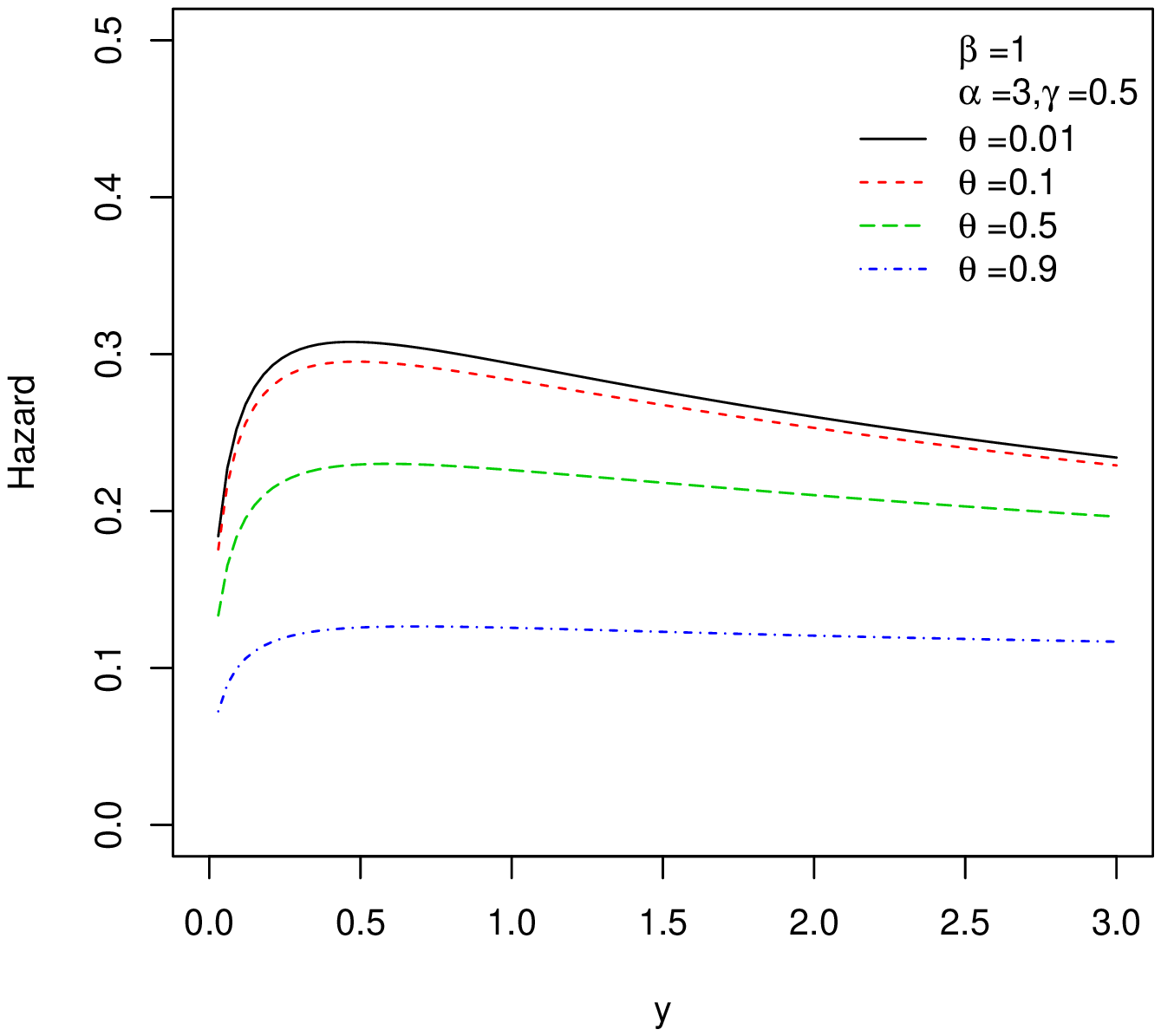}
\includegraphics[scale=0.25]{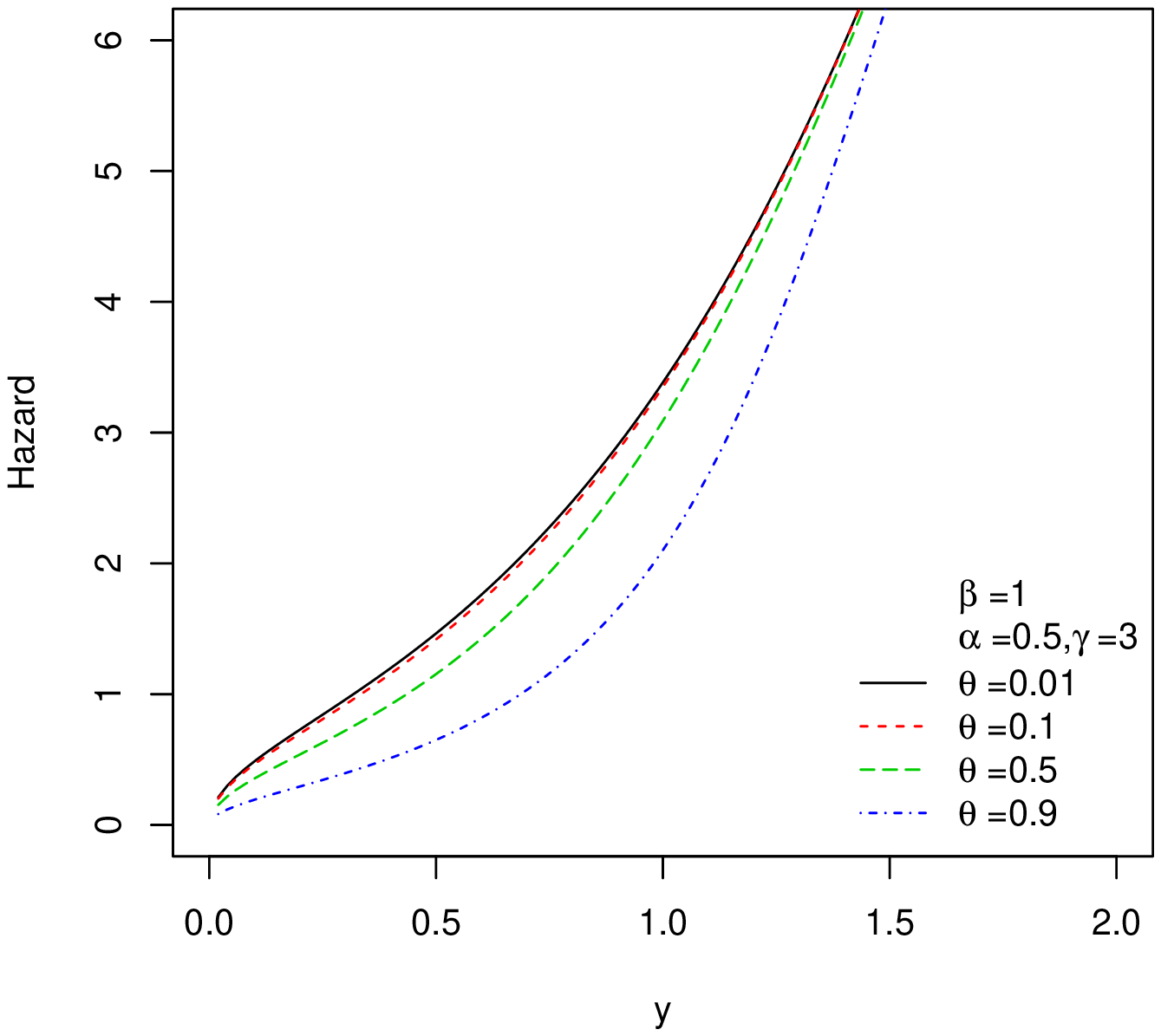}
\includegraphics[scale=0.25]{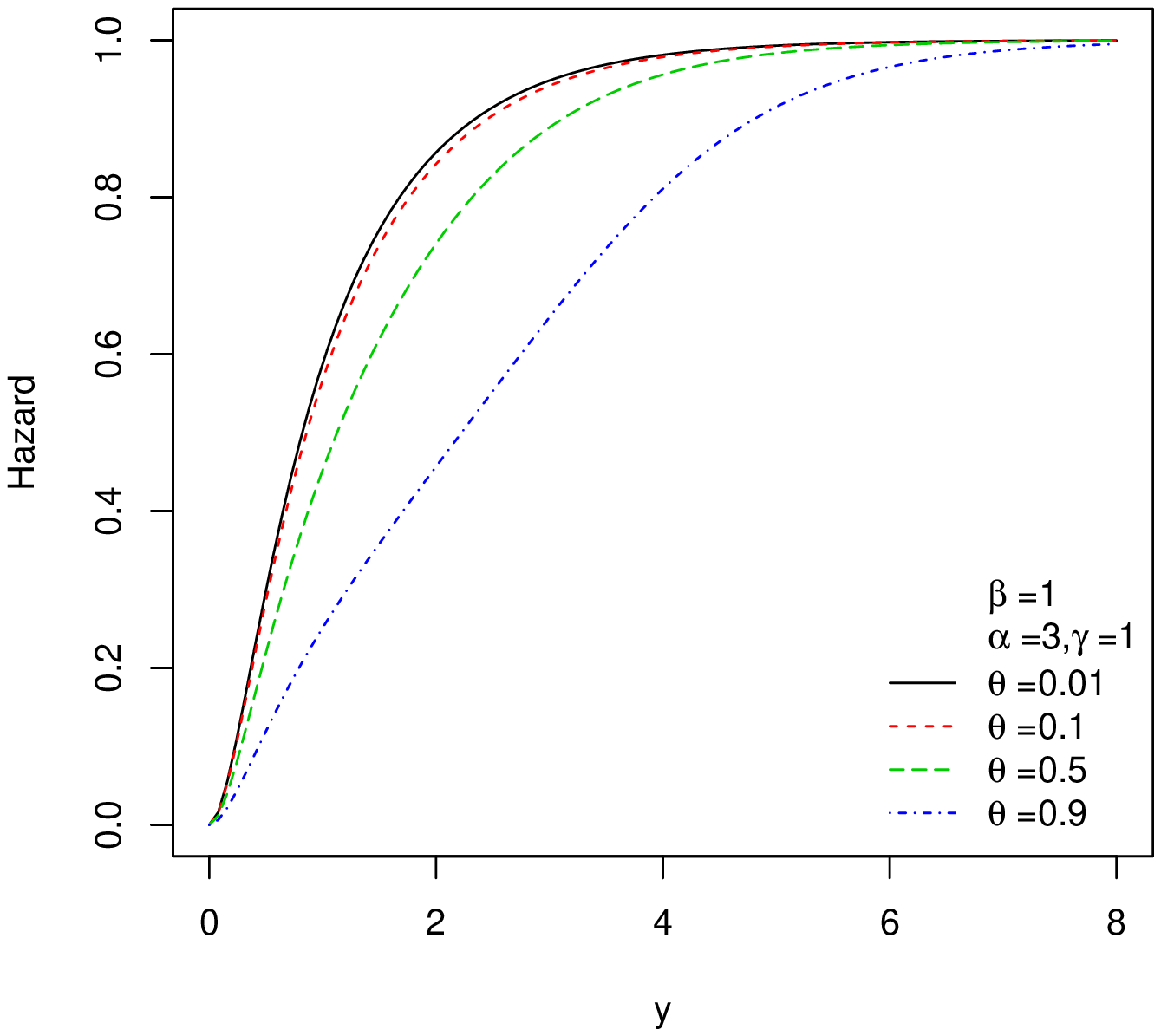}
\caption[]{Plots of pdf and hazard rate function of EWL for different values $‎\alpha, ‎\beta, ‎\gamma‎‎‎$ and $‎\theta‎$.}
\end{figure}
From (\ref{mgf EWPS}), the moment generating function of EWL is
\begin{equation*}
 \begin{array}[b]{l}\label{mgf EWL}
M_{Y}(t)=‎\frac{‎\alpha ‎\theta‎‎}{\log(1-‎\theta‎)}‎\sum^{\infty}_{i=0}\sum^{m}_{n=1}\sum^{\infty}_{j=0}(-1)^{j+1}‎‎\frac{‎\theta‎‎^{n-1}(‎\frac{t}{‎\beta‎}‎)‎^{i}‎}{i!}‎{n\alpha-1 \choose j}‎\frac{‎ ‎\Gamma‎\left(1+\frac{i}{\gamma}\right)‎‎}{(j+1)^{(1+\frac{i}{\gamma})}}.‎
\end{array}
\end{equation*}
\begin{equation*}
E(Y^k)=\frac{‎\theta ‎\Gamma‎\left(1+\frac{k}{\gamma}\right)‎‎‎}{‎\beta‎^{k}‎‎\log(1-‎\theta‎)}‎\sum^{m}_{n=1}\sum^{\infty}_{j=0}(-1)‎^{j+1}‎ {n\alpha-1 \choose j}‎\frac{\theta‎‎^{n-1} }{(j+1)^{(1+\frac{k}{\gamma})}}.‎‎
\end{equation*}

\section{Applications of the EWPS distribution}
In this section we present an application of the EWPS to three real data sets. The fit of EWG, EWP, and EWL on real data sets is examined by graghical methods using MLEs. They are also compared with the EW and Weibull models with respective densities.\\
The first data set is given by Barreto-Souza(2009), Morais and Cordeiro on the fatigue life (rounded to the nearest thousand cycles) for 67 specimens of Alloy T7987 that failed before having accumulated 300 thousand cycles of testing.\\
Now, we estimate the parameters of distributions and compare the p-values of Kolmogorov-Smirnov test and AIC (Akaike Information Criterion), AD (Anderson-Darling statistic) and CM (Cram�r-von Mises statistic) for these  distributions.\\
The empirical scaled TTT transform (Aarset, \cite{Aarset}) and Kaplan-Meier Curve can be used to identify the shape of the
hazard function.\\
The TTT plot and Kaplan-Meier curve for the first data in Fig. 6
shows an increasing hazard rate function.

\begin{figure}
\centering
\includegraphics[scale=0.45]{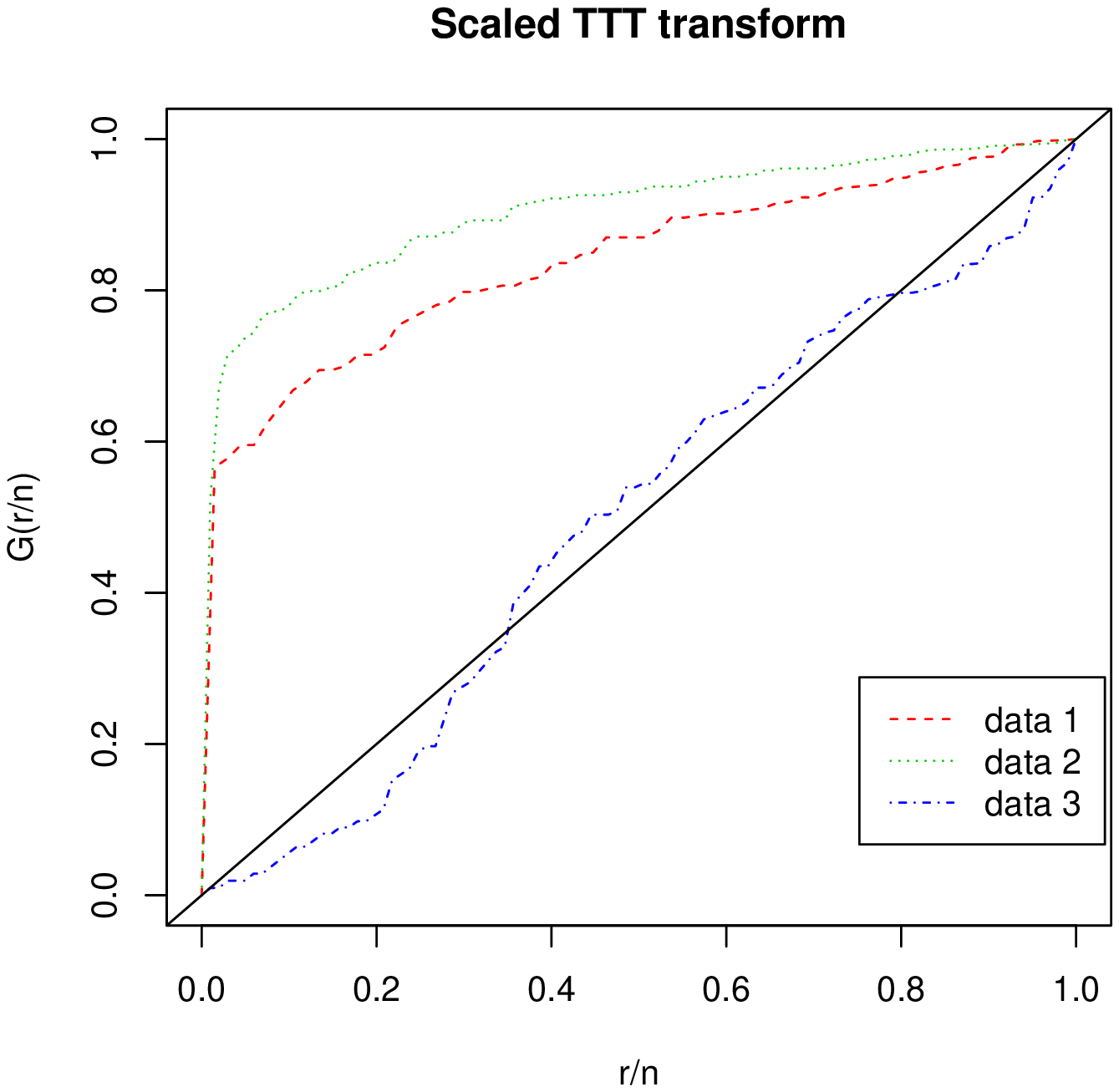}
\includegraphics[scale=0.45]{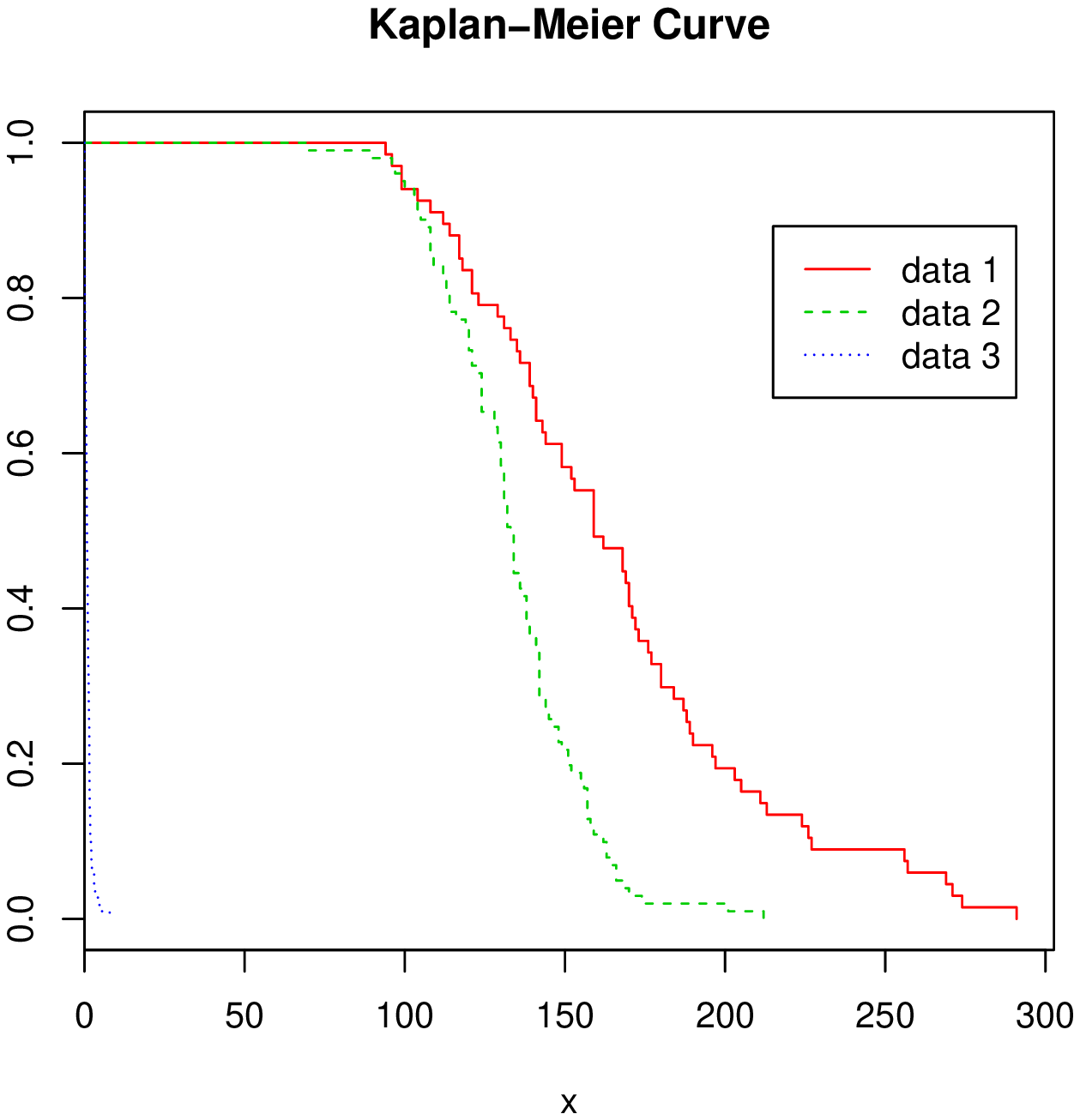}
\caption[]{TTT plots and Kaplan-Meier curves of data 1, data 2 and data 3 .}
\end{figure}
Table 2 lists the MLEs of the parameters, the
values of K-S (Kolmogorov-Smirnov) statistic with its respective
\textit{p}-value, -2log(L), AIC (Akaike Information Criterion), AD
(Anderson-Darling statistic) and CM (Cram�r-von Mises statistic) for
the first data. These values show that the EWG, EWL and EW distributions provide
a better fit than the EWP and Weibull for fitting the first data.\\
 We apply the Arderson-Darling (AD) and Cram�r-von Mises (CM) statistics, in order to verify which
 distribution fits better to this data. The AD and CM test statistics are described in details in
Chen and Balakrishnan \cite{Chen}. In general, the smaller the
values of AD and CM, the better the fit to the data.
\begin{table}[htp!]
\centering \caption{MLEs(stds.), K-S statistics, \textit{p}-values,
$-2\log(L)$ , AIC, AD and CM for data 1.}
\begin{small}
\begin{tabular}{|l|lcccccc|}
\hline
Dist.& MLEs & K-S  & \textit{p}-value &$-2\log(L)$& AIC& AD& CM  \\
\hline EWG & $\begin{array}{l}
\hat{\alpha}= 15.3396~,
\hat{\beta}=0.0154 \\
\hat{\gamma}=1.3155~,\hat{\theta}=0.1860
\end{array}$
&0.0486 & 0.9974 &695.9917&703.9917&0.1968&0.1029 \\
EWP & $\begin{array}{l}
\hat{\alpha}= 20.48  ~,
\hat{\beta}=0.0732 \\
\hat{\gamma}=0.7316~,\hat{\theta}=13.74
\end{array}$

& 0.0717&  0.8811&696.2272& 704.2272&0.2205&0.1128 \\
EWL & $\begin{array}{l}
\hat{\alpha}=14.0601~,
\hat{\beta}=0.0158 \\
\hat{\gamma}=1.3671~,\hat{\theta}=0.7721
\end{array}$
&0.0524 & 0.993 &696.8654& 704.8654&0.2956& 0.1165\\
EW & $\begin{array}{l}
\hat{\alpha}=12.1645~,
\hat{\beta}=0.0134 \\
\hat{‎\gamma‎}=1.4034
\end{array}$
  &0.0522 &‎0.9931&696.0166 &702.0166&0.19097& 0.1023 \\
Weibull &~ $\hat{\beta}$=0.0054~~~, $\hat{\gamma}$=3.7349~~~&0.1027 &0.4793& 706.598&710.598&1.1684&0.2541\\
\hline
\end{tabular}
\end{small}
\end{table}
As a second application, we consider the data show the fatigue life of 6061-T6 aluminum coupons cut parallel to the direction of rolling and oscillated at 18 cyclers per second. The pooled data, yielding a total of 101 observations, were first analyzed by Birnbaum and Saunders (1969).The TTT plot and Kaplan-Meier curve for this data in Fig. 6 shows  an increasing hazard rate function.\\
The MLEs of the parameters, the values of K-S statistic,
\textit{p}-value, -2log(L), AIC, AD and CM are listed in Table 3.
From these values, we note that the EWG model is better than the EWP,
EWL, EW and Weibull distributions in terms of fitting to this data.
\begin{table}[htp!]
\centering \caption{MLEs(stds.), K-S statistics, \textit{p}-values,
$-2\log(L)$ , AIC, AD and CM for data 2.}
\begin{small}
\begin{tabular}{|l|lcccccc|}
\hline
Dist.& MLEs & K-S  & \textit{p}-value &$-2\log(L)$& AIC& AD& CM  \\
\hline EWG & $\begin{array}{l}
\hat{\alpha}= 8.0516~,
\hat{\beta}=0.0129 \\
\hat{\gamma}=2.3695~,\hat{\theta}=0.7745
\end{array}$
&0.0618 &0.8352 &913.1816&921.1816&0.3426&0.1299 \\
EWP & $\begin{array}{l}
\hat{\alpha}= 14.022~,
\hat{\beta}=0.0135 \\
\hat{\gamma}=2.1176~,\hat{\theta}=1.059
\end{array}$

&0.0791& 0.552&913.4216& 921.4216&0.4363&0.1557 \\
EWL & $\begin{array}{l}
\hat{\alpha}=8.9561~,
\hat{\beta}=0.01143\\
\hat{\gamma}=2.4247~,\hat{\theta}=0.2769
\end{array}$
&0.0832 &0.4867 &913.7988&921.7988&0.5413& 0.1729\\
EW & $\begin{array}{l}
\hat{\alpha}=8.072~,
\hat{\beta}=0.0108 \\
\hat{‎\gamma‎}=2.5872
\end{array}$
  &0.082 &0.5049&913.498 &919.498&0.4597& 0.1616\\
Weibull &~ $\hat{\beta}$=0.0069~~~, $\hat{\gamma}$=6.0347~~~&0.1234&0.0923& 926.9108&930.9108&1.755&0.3657\\
\hline
\end{tabular}
\end{small}
\end{table}
The last data set consists 101 observations show the
stress-rupture life of kevlar 49/epoxy strands which were subjected
to constant sustained pressure at the 90 stress level until all had
failed. The failure times in hours are shown in Andrews and Herzberg
\cite{Andrews } and Barlow et al. \cite{Barlow}. The TTT plot and
Kaplan-Meier curve for this data in Fig. 6 shows bathtub-shaped
hazard rate function. The MLEs of the parameters, the values of K-S statistic,
\textit{p}-value, -2log(L), AIC, AD and CM are listed in Table 4.
From these values, we note that the EWG and EWP models are better than the EW
and Weibull distributions in terms of fitting to this data.

\begin{table}[htp!]
\centering \caption{MLEs(stds.), K-S statistics, \textit{p}-values,
$-2\log(L)$ , AIC, AD and CM for data 3.}
\begin{small}
\begin{tabular}{|l|lcccccc|}
\hline
Dist.& MLEs & K-S  & \textit{p}-value &$-2\log(L)$& AIC& AD& CM  \\
\hline EWG & $\begin{array}{l}
\hat{\alpha}= 1.0921~,
\hat{\beta}=3.1202\\
\hat{\gamma}=0.661~,\hat{\theta}=0.7559
\end{array}$
&0.0724 &0.6657 &203.66&211.66&0.7842&0.2019 \\
EWP & $\begin{array}{l}
\hat{\alpha}= 0.8589~,
\hat{\beta}=1.3032\\
\hat{\gamma}=0.8717~,\hat{\theta}=1.2661
\end{array}$

&0.0725&0.6638&204.6174& 212.6174&0.8409&0.2182 \\
EWL & $\begin{array}{l}
\hat{\alpha}=2.4513~,
\hat{\beta}=17.0129\\
\hat{\gamma}=0.4978~,\hat{\theta}=0.9918
\end{array}$
&0.0898 &0.3893 &202.4622&210.4622&0.8643&0.2455\\
EW & $\begin{array}{l}
\hat{\alpha}=0.7929~,
\hat{\beta}=0.8210 \\
\hat{‎\gamma‎}=1.0604
\end{array}$
  &0.0844 &0.468&205.5743 &211.5743&0.9554& 0.2473\\
Weibull &~ $\hat{\beta}$=1.0101~~~, $\hat{\gamma}$=0.9259~~~&0.0906&0.3778& 205.9536&209.9536&1.1221&0.2789\\
\hline
\end{tabular}
\end{small}
\end{table}

Plots of the densities and cumulative distribution functions of the EWG, EWP, EWL, EW
and Weibull models fitted to the data sets corresponding to Tables 2, 3
and 4, respectively, are given in Fig. 7, 8 and 9.

\begin{figure}
\centering
\includegraphics[scale=0.45]{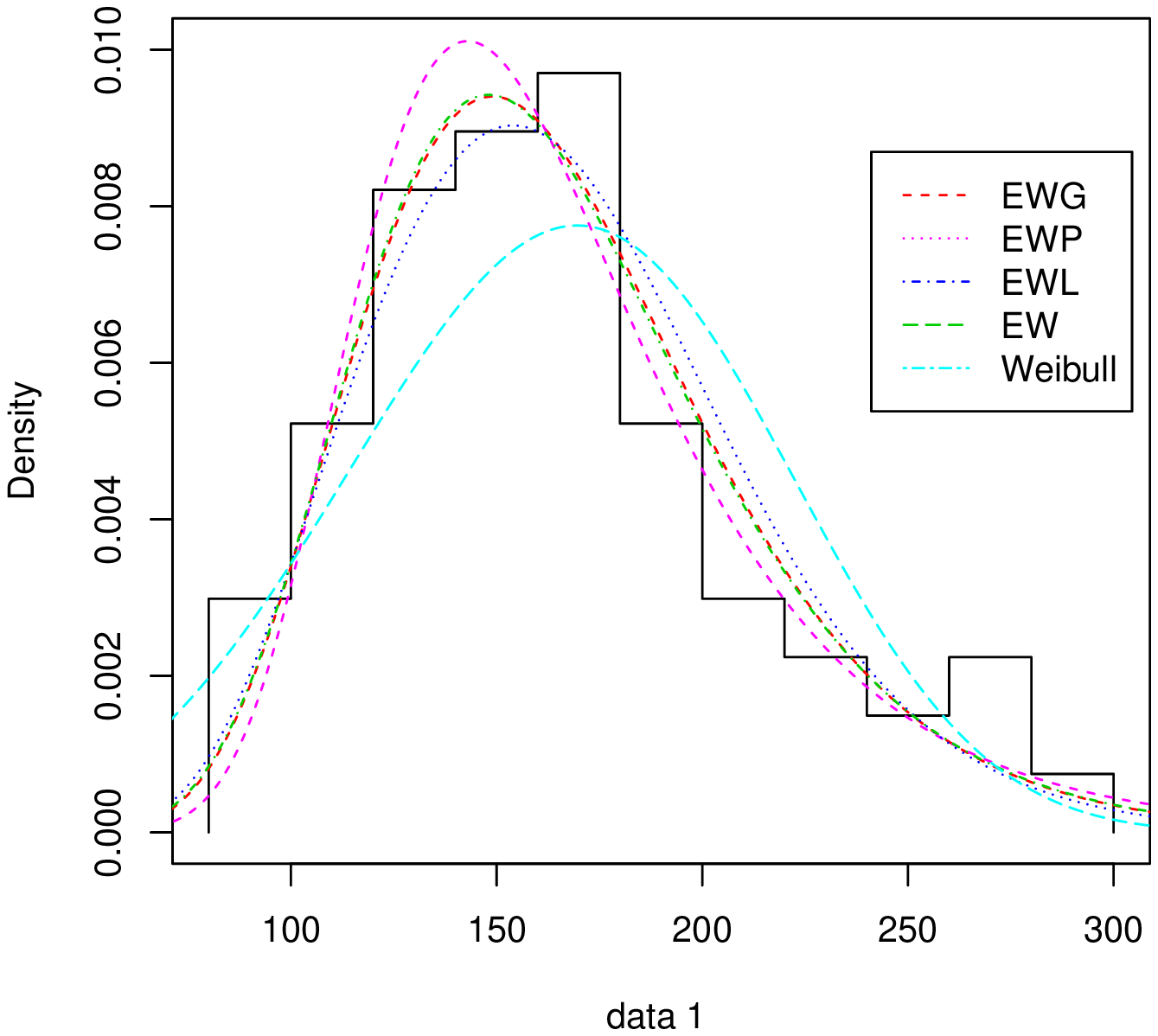}
\includegraphics[scale=0.45]{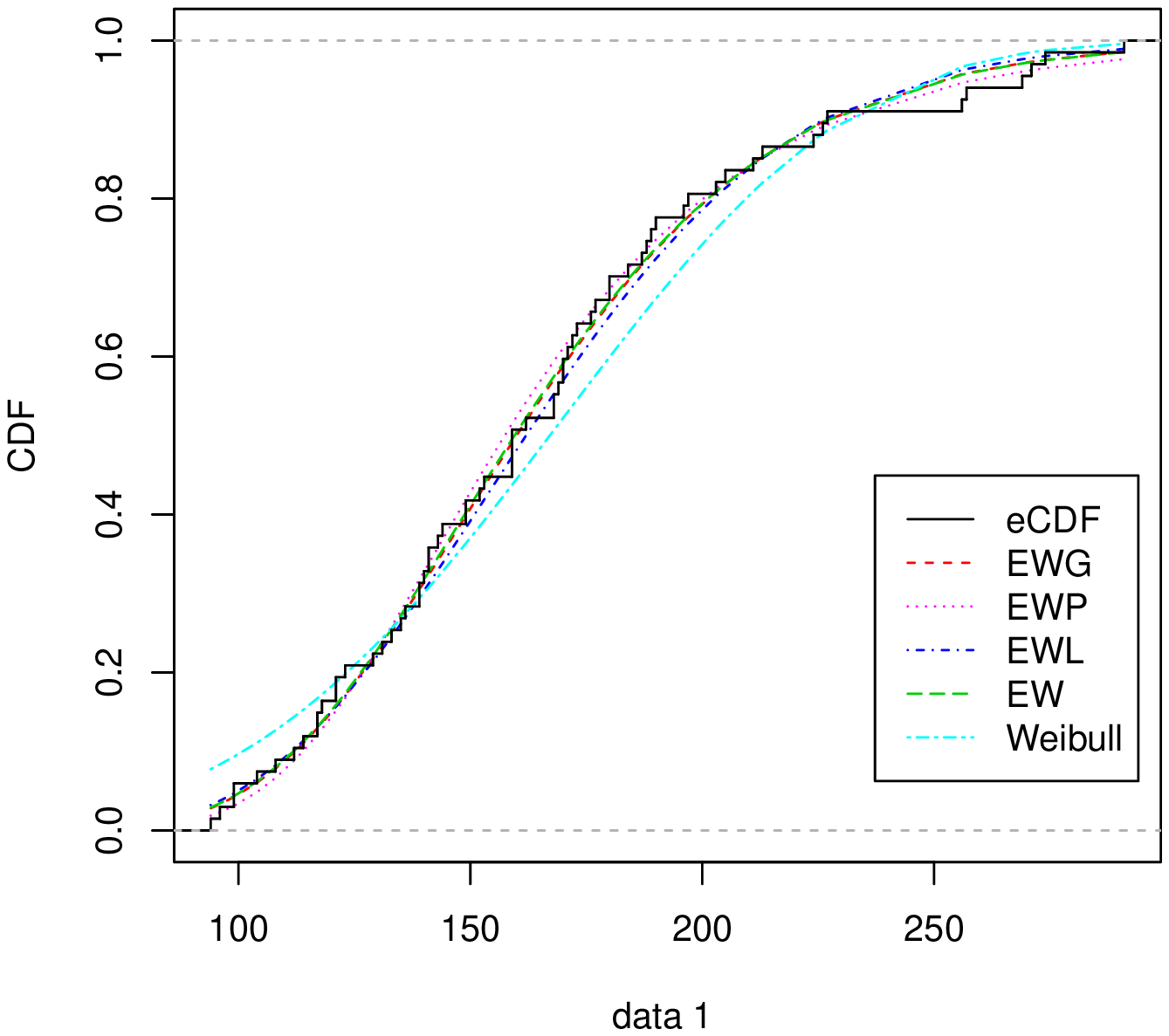}
\caption[]{Fitted cdf and survival function of the EWG, EWP, EWL, EW and
Weibull distributions for the data sets corresponding to Table 2.}
\end{figure}
\begin{figure}
\centering
\includegraphics[scale=0.45]{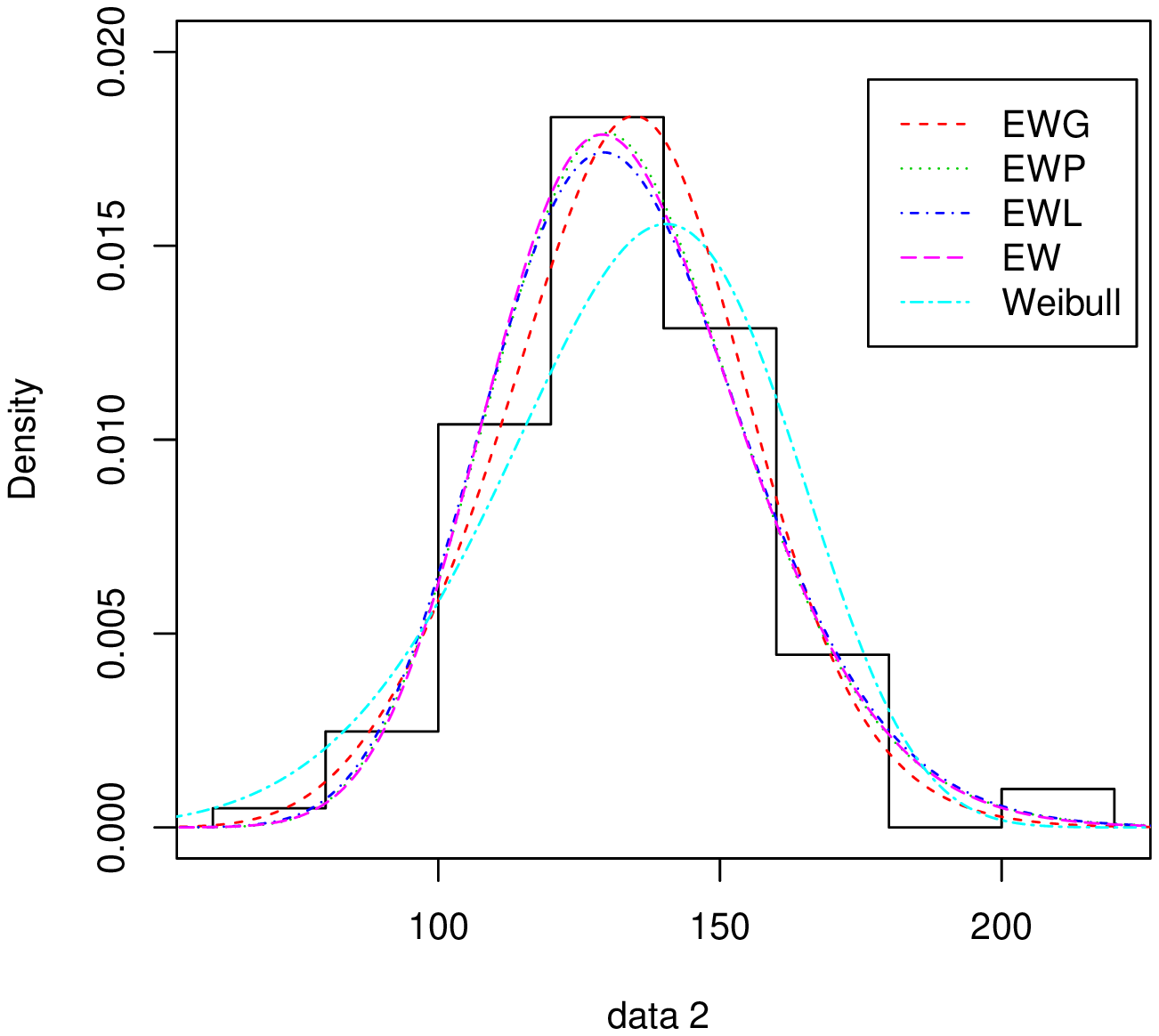}
\includegraphics[scale=0.45]{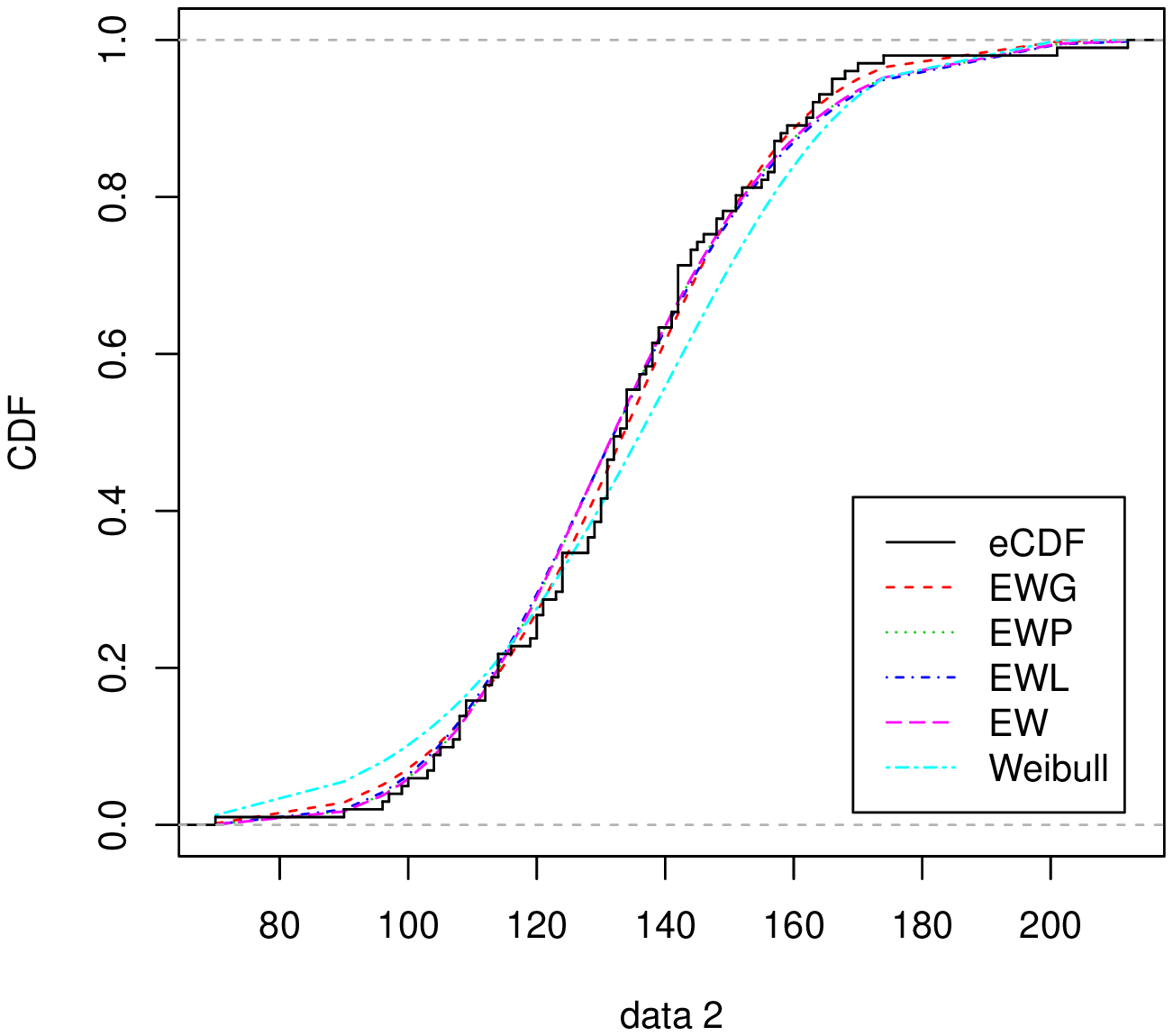}
\caption[]{Fitted cdf and survival function of the EWG, EWP, EWL, EW and
Weibull distributions for the data sets corresponding to  Table 3.}
\end{figure}
\begin{figure}
\centering
\includegraphics[scale=0.45]{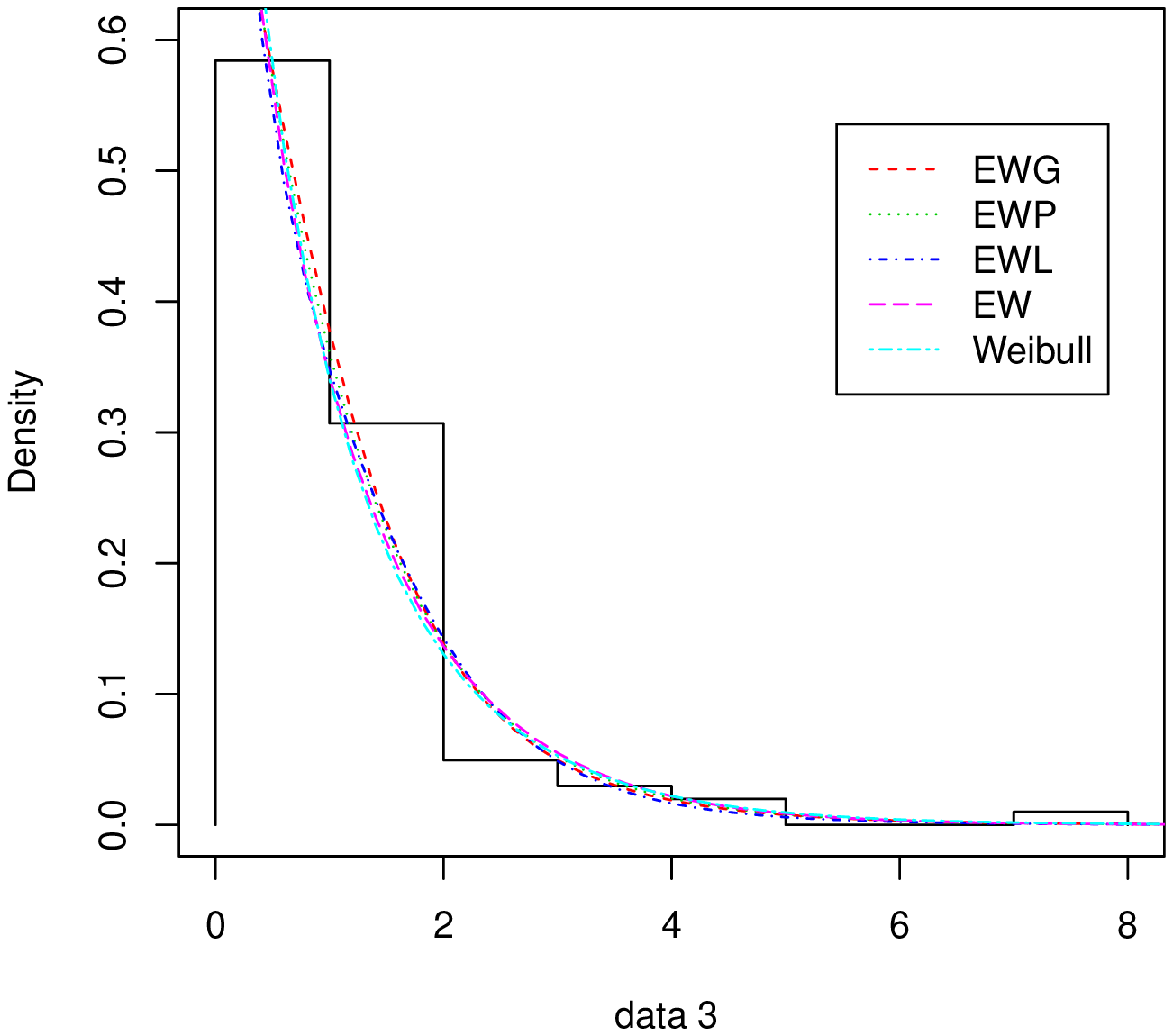}
\includegraphics[scale=0.45]{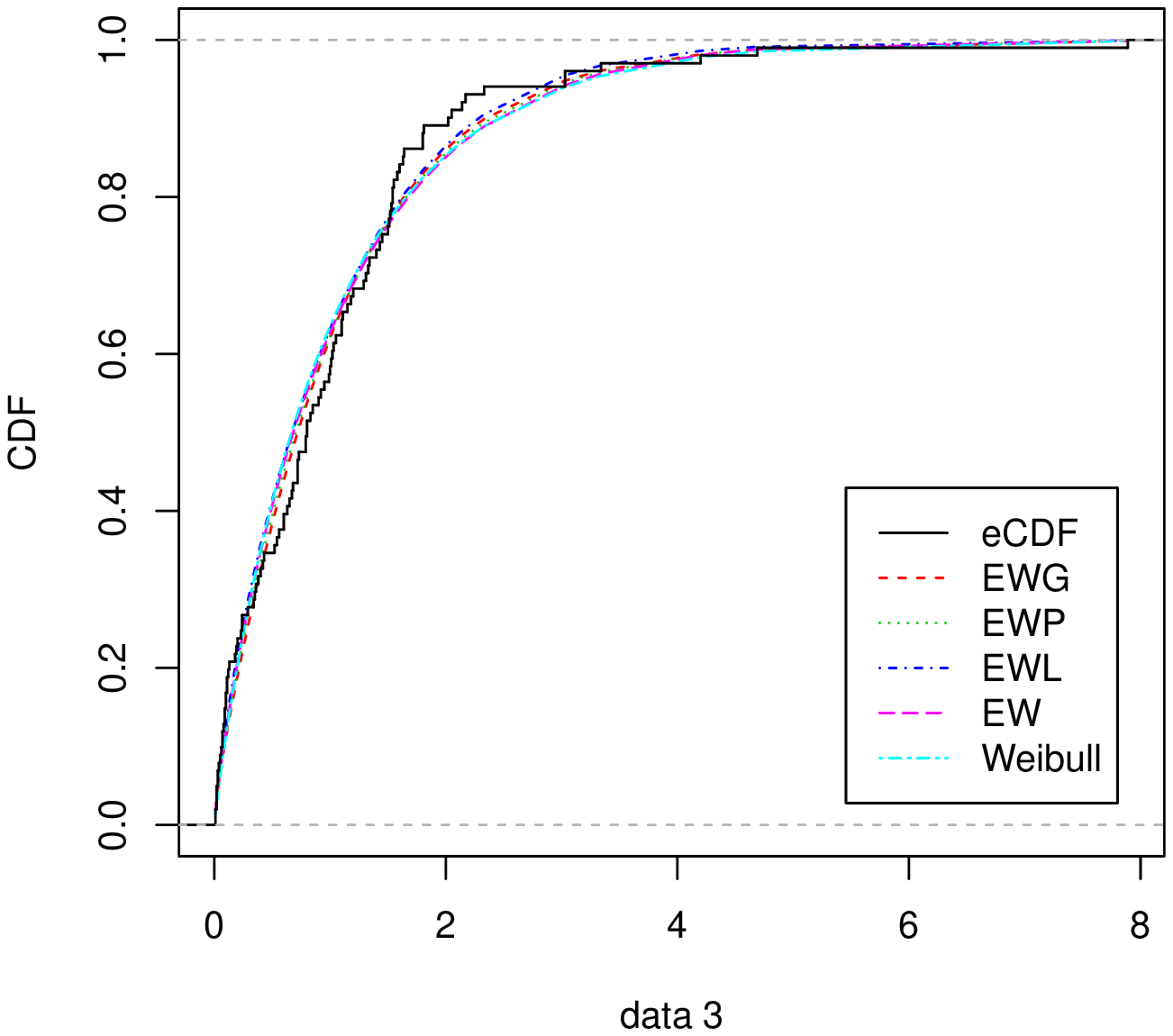}
\caption[]{Fitted cdf and survival function of the EWG, EWP, EWL, EW and
Weibull distributions for the data sets corresponding to  Table 4.}
\end{figure}

\newpage
\section{ Conclusion}
\section*{Appendix}

The elements of the $4\times 4$ observed information matrix
$I_n\left(\Theta \right)$ are given by

\begin{equation*}
\begin{array}{ll}
I_{\alpha\alpha}&=‎-\frac{n}{‎\alpha‎^{2}‎‎}+‎\sum^{n}_{i=1}‎\frac{‎\theta‎(\log (1-e^{-(‎\beta y‎_{i}‎‎)‎^{‎\gamma‎}‎}))‎^{2}(1-e^{-(‎\beta y‎_{i}‎‎)‎^{‎\gamma‎}‎})‎^{‎\alpha‎}C‎^{\prime‎ ‎\prime‎}‎ (‎\theta (1-e^{-(‎\beta y‎_{i}‎‎)‎^{‎\gamma‎}‎})‎^{‎\alpha‎} ‎)‎‎}{C ‎^{‎\prime} (‎\theta (1-e^{-(‎\beta y‎_{i}‎‎)‎^{‎\gamma‎}}‎)‎^{‎\alpha‎} ‎)‎‎}\medskip\\ &+\sum^{n}_{i=1}‎‎\frac{‎\theta‎^2‎\left[ ‎(1-e^{-(‎\beta y‎_{i}‎‎)‎^{‎\gamma‎}‎})‎^{‎\alpha‎}‎\log (1-e^{-(‎\beta y‎_{i}‎‎)‎^{‎\gamma‎}‎})\right]‎^2 C‎^{\prime \prime \prime}(‎\theta (1-e^{-(‎\beta y‎_{i}‎‎)‎^{‎\gamma‎}‎})‎^{‎\alpha‎})‎‎‎}{C ‎^{‎\prime} (‎\theta (1-e^{-(‎\beta y‎_{i}‎‎)‎^{‎\gamma‎}}‎)‎^{‎\alpha‎} ‎)}\medskip\\&-\sum^{n}_{i=1}‎\frac{\theta‎^2((1-e^{-(‎\beta y‎_{i}‎‎)‎^{‎\gamma‎}‎})‎^{‎\alpha‎})‎^{2}‎(\log (1-e^{-(‎\beta y‎_{i}‎‎)‎^{‎\gamma‎}‎}))‎^{2}(C‎^{\prime‎ ‎\prime‎}‎ (‎\theta (1-e^{-(‎\beta y‎_{i}‎‎)‎^{‎\gamma‎}‎})‎^{‎\alpha‎} ‎)‎‎)^2‎}{(C ‎^{‎\prime} (‎\theta (1-e^{-(‎\beta y‎_{i}‎‎)‎^{‎\gamma‎}}‎)‎^{‎\alpha‎} ‎))^2}‎‎‎,\bigskip\\\\

I_{\alpha\beta}&=\sum^{n}_{i=1}‎‎\frac{‎\gamma ‎\beta‎^{‎\gamma‎ -1}‎‎‎y‎_{i}‎‎^{‎\gamma‎}e^{-(‎\beta y‎_{i}‎‎)‎^{‎\gamma‎}‎}‎}{1-e^{-(‎\beta y‎_{i}‎‎)‎^{‎\gamma‎}‎}}‎\medskip\\&+‎\sum^{n}_{i=1}\frac{‎\theta \gamma ‎\beta‎^{‎\gamma‎ -1}‎‎‎y‎_{i}‎‎^{‎\gamma‎}e^{-(‎\beta y‎_{i}‎‎)‎^{‎\gamma‎}‎}‎(1-e^{-(‎\beta y‎_{i}‎‎)‎^{‎\gamma‎}‎})‎^{‎\alpha -1‎}C‎^{\prime‎ ‎\prime‎}‎ (‎\theta (1-e^{-(‎\beta y‎_{i}‎‎)‎^{‎\gamma‎}‎})‎^{‎\alpha‎} ‎)‎‎‎}{C ‎^{‎\prime} (‎\theta (1-e^{-(‎\beta y‎_{i}‎‎)‎^{‎\gamma‎}}‎)‎^{‎\alpha‎} ‎)‎‎}‎\medskip\\&+\sum^{n}_{i=1}‎\frac{‎\theta ‎\alpha‎ \gamma ‎\beta‎^{‎\gamma‎ -1}‎‎‎y‎_{i}‎‎^{‎\gamma‎}e^{-(‎\beta y‎_{i}‎‎)‎^{‎\gamma‎}‎}‎‎‎\log (1-e^{-(‎\beta y‎_{i}‎‎)‎^{‎\gamma‎}‎})‎(1-e^{-(‎\beta y‎_{i}‎‎)‎^{‎\gamma‎}‎})‎^{‎\alpha -1‎}C‎^{\prime‎ ‎\prime‎}‎ (‎\theta (1-e^{-(‎\beta y‎_{i}‎‎)‎^{‎\gamma‎}‎})‎^{‎\alpha‎} ‎)}{C ‎^{‎\prime} (‎\theta (1-e^{-(‎\beta y‎_{i}‎‎)‎^{‎\gamma‎}}‎)‎^{‎\alpha‎} ‎)‎‎}‎‎\medskip\\&+\sum^{n}_{i=1}‎\frac{\theta ‎^{2}‎‎\alpha‎ \gamma ‎\beta‎^{‎\gamma‎ -1}‎‎‎y‎_{i}‎‎^{‎\gamma‎}e^{-(‎\beta y‎_{i}‎‎)‎^{‎\gamma‎}‎}‎‎‎\log (1-e^{-(‎\beta y‎_{i}‎‎)‎^{‎\gamma‎}‎})‎(1-e^{-(‎\beta y‎_{i}‎‎)‎^{‎\gamma‎}‎})‎^{‎2\alpha -1‎}C‎^{\prime‎ ‎\prime  ‎\prime‎}‎ (‎\theta (1-e^{-(‎\beta y‎_{i}‎‎)‎^{‎\gamma‎}‎})‎^{‎\alpha‎} ‎)}{C ‎^{‎\prime} (‎\theta (1-e^{-(‎\beta y‎_{i}‎‎)‎^{‎\gamma‎}}‎)‎^{‎\alpha‎} ‎)‎‎}‎‎\medskip\\&-\sum^{n}_{i=1}‎\frac{\theta ‎^{2}‎‎\alpha‎ \gamma ‎\beta‎^{‎\gamma‎ -1}‎‎‎y‎_{i}‎‎^{‎\gamma‎}e^{-(‎\beta y‎_{i}‎‎)‎^{‎\gamma‎}‎}‎‎‎\log (1-e^{-(‎\beta y‎_{i}‎‎)‎^{‎\gamma‎}‎})‎(1-e^{-(‎\beta y‎_{i}‎‎)‎^{‎\gamma‎}‎})‎^{‎2\alpha -1‎}(C‎^{\prime‎ ‎\prime }‎ (‎\theta (1-e^{-(‎\beta y‎_{i}‎‎)‎^{‎\gamma‎}‎})‎^{‎\alpha‎} ‎))‎^{2}‎}{(C ‎^{‎\prime} (‎\theta (1-e^{-(‎\beta y‎_{i}‎‎)‎^{‎\gamma‎}}‎)‎^{‎\alpha‎} ‎))^2‎‎}‎‎‎,\bigskip\\\\

I_{\alpha‎\gamma‎}&=\sum^{n}_{i=1}‎\frac{‎\beta‎^{‎\gamma‎ }‎‎‎y‎_{i}‎‎^{‎\gamma‎}\log (‎\beta ‎‎‎y‎_{i}‎)e^{-(‎\beta y‎_{i}‎‎)‎^{‎\gamma‎}‎}‎‎‎}{1-e^{-(‎\beta y‎_{i}‎‎)‎^{‎\gamma‎}‎}}‎\medskip\\&+\sum^{n}_{i=1}‎\frac{‎\theta ‎\beta‎^{‎\gamma‎ }‎‎‎y‎_{i}‎‎^{‎\gamma‎}\log (‎\beta ‎‎‎y‎_{i}‎)e^{-(‎\beta y‎_{i}‎‎)‎^{‎\gamma‎}‎}‎(1-e^{-(‎\beta y‎_{i}‎‎)‎^{‎\gamma‎}‎})‎^{‎\alpha -1‎}C‎^{\prime‎ ‎\prime‎}‎ (‎\theta (1-e^{-(‎\beta y‎_{i}‎‎)‎^{‎\gamma‎}‎})‎^{‎\alpha‎} ‎)‎‎‎}{C ‎^{‎\prime} (‎\theta (1-e^{-(‎\beta y‎_{i}‎‎)‎^{‎\gamma‎}}‎)‎^{‎\alpha‎} ‎)‎‎}\medskip\\&+\sum^{n}_{i=1}‎‎\frac{‎\theta‎ ‎\alpha ‎‎\beta‎^{‎\gamma‎ }‎‎‎y‎_{i}‎‎^{‎\gamma‎}\log (‎\beta ‎‎‎y‎_{i}‎)e^{-(‎\beta y‎_{i}‎‎)‎^{‎\gamma‎}‎}\log (1-e^{-(‎\beta y‎_{i}‎‎)‎^{‎\gamma‎}‎})‎(1-e^{-(‎\beta y‎_{i}‎‎)‎^{‎\gamma‎}‎})‎^{‎\alpha -1‎}C‎^{\prime‎ ‎\prime‎}‎ (‎\theta (1-e^{-(‎\beta y‎_{i}‎‎)‎^{‎\gamma‎}‎})‎^{‎\alpha‎} ‎)}{C ‎^{‎\prime} (‎\theta (1-e^{-(‎\beta y‎_{i}‎‎)‎^{‎\gamma‎}}‎)‎^{‎\alpha‎} ‎)}‎\medskip\\&+\sum^{n}_{i=1}‎\frac{\theta‎ ^2 ‎\alpha ‎‎\beta‎^{‎\gamma‎ }‎‎‎y‎_{i}‎‎^{‎\gamma‎}\log (‎\beta ‎‎‎y‎_{i}‎)e^{-(‎\beta y‎_{i}‎‎)‎^{‎\gamma‎}‎}\log (1-e^{-(‎\beta y‎_{i}‎‎)‎^{‎\gamma‎}‎})‎(1-e^{-(‎\beta y‎_{i}‎‎)‎^{‎\gamma‎}‎})‎^{‎2 \alpha -1‎}C‎^{\prime‎ ‎\prime \prime‎‎}‎ (‎\theta (1-e^{-(‎\beta y‎_{i}‎‎)‎^{‎\gamma‎}‎})‎^{‎\alpha‎} ‎)}{C ‎^{‎\prime} (‎\theta (1-e^{-(‎\beta y‎_{i}‎‎)‎^{‎\gamma‎}}‎)‎^{‎\alpha‎} ‎)}‎‎\medskip\\&-\sum^{n}_{i=1}‎‎\frac{\theta‎ ^2 ‎\alpha ‎‎\beta‎^{‎\gamma‎ }‎‎‎y‎_{i}‎‎^{‎\gamma‎}\log (‎\beta ‎‎‎y‎_{i}‎)e^{-(‎\beta y‎_{i}‎‎)‎^{‎\gamma‎}‎}\log (1-e^{-(‎\beta y‎_{i}‎‎)‎^{‎\gamma‎}‎})‎(1-e^{-(‎\beta y‎_{i}‎‎)‎^{‎\gamma‎}‎})‎^{‎2 \alpha -1‎}(C‎^{\prime‎ ‎\prime ‎‎}‎ (‎\theta (1-e^{-(‎\beta y‎_{i}‎‎)‎^{‎\gamma‎}‎})‎^{‎\alpha‎} ‎))^2}{(C ‎^{‎\prime} (‎\theta (1-e^{-(‎\beta y‎_{i}‎‎)‎^{‎\gamma‎}}‎)‎^{‎\alpha‎} ‎))^2}‎‎‎‎,\bigskip\\\\

I_{\alpha‎‎\theta‎‎}&=\sum^{n}_{i=1}‎\frac{\log (1-e^{-(‎\beta y‎_{i}‎‎)‎^{‎\gamma‎}‎})(1-e^{-(‎\beta y‎_{i}‎‎)‎^{‎\gamma‎}}‎)‎^{‎\alpha‎}C‎^{\prime‎ ‎\prime‎}‎ (‎\theta (1-e^{-(‎\beta y‎_{i}‎‎)‎^{‎\gamma‎}‎})‎^{‎\alpha‎} ‎)}{C ‎^{‎\prime} (‎\theta (1-e^{-(‎\beta y‎_{i}‎‎)‎^{‎\gamma‎}}‎)‎^{‎\alpha‎} ‎)}‎\medskip\\&+\sum^{n}_{i=1}‎\frac{‎\theta \log (1-e^{-(‎\beta y‎_{i}‎‎)‎^{‎\gamma‎}‎})‎‎(1-e^{-(‎\beta y‎_{i}‎‎)‎^{‎\gamma‎}‎})‎^{‎2 \alpha ‎}C‎^{\prime‎ ‎\prime \prime‎}‎ (‎\theta (1-e^{-(‎\beta y‎_{i}‎‎)‎^{‎\gamma‎}‎})‎^{‎\alpha‎} ‎)}{C ‎^{‎\prime} (‎\theta (1-e^{-(‎\beta y‎_{i}‎‎)‎^{‎\gamma‎}}‎)‎^{‎\alpha‎} ‎)}\medskip\\&-\sum^{n}_{i=1}‎‎\frac{‎\theta \log (1-e^{-(‎\beta y‎_{i}‎‎)‎^{‎\gamma‎}‎})‎‎(1-e^{-(‎\beta y‎_{i}‎‎)‎^{‎\gamma‎}‎})‎^{‎2 \alpha ‎}(C‎^{\prime‎ ‎\prime ‎}‎ (‎\theta (1-e^{-(‎\beta y‎_{i}‎‎)‎^{‎\gamma‎}‎})‎^{‎\alpha‎} ‎))^2}{(C ‎^{‎\prime} (‎\theta (1-e^{-(‎\beta y‎_{i}‎‎)‎^{‎\gamma‎}}‎)‎^{‎\alpha‎} ‎))^2},\bigskip\\
‎
\end{array}
\end{equation*}
\begin{equation*}
\begin{array}{ll}
I_{‎\beta‎\beta‎‎‎‎}&=-‎\frac{n‎\gamma‎}{‎\beta ^2‎}‎-\sum^{n}_{i=1} ‎\gamma ‎(‎\gamma -1‎)\beta‎^{‎\gamma‎ -2} y‎_{i}‎‎^{‎\gamma‎}+\sum^{n}_{i=1}‎\frac{(‎\alpha -1‎)‎\gamma (‎\gamma -1‎)‎\beta‎^{‎\gamma‎ -2}e^{-(‎\beta y‎_{i}‎‎)^{‎\gamma‎}‎}}{1-e^{-(‎\beta y‎_{i}‎‎)‎^{‎\gamma‎}}‎}‎‎\medskip\\&-\sum^{n}_{i=1}‎\frac{‎(‎\alpha -1‎)‎ \gamma ^2 \beta‎^{2 ‎\gamma‎ -2} ‎y‎_{i}‎‎^{2 ‎\gamma‎}e^{-(‎\beta y‎_{i}‎‎)‎^{‎\gamma‎}}‎}{1-e^{-(‎\beta y‎_{i}‎‎)‎^{‎\gamma‎}}}-\sum^{n}_{i=1}‎\frac{\gamma ^2 \beta‎^{2 ‎\gamma‎ -2} ‎y‎_{i}‎‎^{2 ‎\gamma‎}(e^{-(‎\beta y‎_{i}‎‎)‎^{‎\gamma‎}})^2‎}{(1-e^{-(‎\beta y‎_{i}‎‎)‎^{‎\gamma‎}})^2}‎‎‎\medskip\\&+\sum^{n}_{i=1}‎\frac{‎\theta‎ ‎\alpha ‎\gamma (‎\gamma -1‎) ‎‎y‎_{i}‎‎^{‎\gamma‎}\beta‎^{ ‎\gamma‎ -2}e^{-(‎\beta y‎_{i}‎‎)‎^{‎\gamma‎}}(1-e^{-(‎\beta y‎_{i}‎‎)‎^{‎\gamma‎}}‎)‎^{‎\alpha -1‎}C‎^{\prime‎ ‎\prime ‎}‎ (‎\theta (1-e^{-(‎\beta y‎_{i}‎‎)‎^{‎\gamma‎}‎})‎^{‎\alpha‎} ‎) }{C ‎^{‎\prime} (‎\theta (1-e^{-(‎\beta y‎_{i}‎‎)‎^{‎\gamma‎}}‎)‎^{‎\alpha‎} ‎)}‎‎‎\medskip\\&-\sum^{n}_{i=1}‎\frac{‎\theta‎ ‎\alpha ‎\gamma ^2  ‎‎y‎_{i}‎‎^{‎ 2\gamma‎}\beta‎^{ 2 ‎\gamma‎ -2}e^{-(‎\beta y‎_{i}‎‎)‎^{‎\gamma‎}}(1-e^{-(‎\beta y‎_{i}‎‎)‎^{‎\gamma‎}}‎)‎^{‎\alpha -1‎}C‎^{\prime‎ ‎\prime ‎}‎ (‎\theta (1-e^{-(‎\beta y‎_{i}‎‎)‎^{‎\gamma‎}‎})‎^{‎\alpha‎})}{C ‎^{‎\prime} (‎\theta (1-e^{-(‎\beta y‎_{i}‎‎)‎^{‎\gamma‎}}‎)‎^{‎\alpha‎} ‎)}‎‎‎‎\medskip\\&+\sum^{n}_{i=1}‎\frac{ (‎‎\alpha ‎-1‎) \theta‎ ‎\alpha ‎\gamma ^2 ‎‎y‎_{i}‎‎^{‎2 \gamma‎}\beta‎^{ 2 ‎\gamma‎ -2}(e^{-(‎\beta y‎_{i}‎‎)‎^{‎\gamma‎}})^2(1-e^{-(‎\beta y‎_{i}‎‎)‎^{‎\gamma‎}}‎)‎^{‎\alpha -2‎}C‎^{\prime‎ ‎\prime ‎}‎ (‎\theta (1-e^{-(‎\beta y‎_{i}‎‎)‎^{‎\gamma‎}‎})‎^{‎\alpha‎} ‎) }{C ‎^{‎\prime} (‎\theta  (1-e^{-(‎\beta y‎_{i}‎‎)‎^{‎\gamma‎}}‎)‎^{‎\alpha‎} ‎)}‎‎‎‎\medskip\\&+\sum^{n}_{i=1}‎\frac{\theta ‎‎^{2}‎ ‎\alpha ‎^{2}‎ ‎\gamma ^2 ‎‎y‎_{i}‎‎^{‎2 \gamma‎}\beta‎^{ 2 ‎\gamma‎ -2}(e^{-(‎\beta y‎_{i}‎‎)‎^{‎\gamma‎}})^2(1-e^{-(‎\beta y‎_{i}‎‎)‎^{‎\gamma‎}}‎)‎^{‎2 \alpha -2‎}C‎^{\prime‎ ‎\prime \prime‎}‎ (‎\theta (1-e^{-(‎\beta y‎_{i}‎‎)‎^{‎\gamma‎}‎})‎^{‎\alpha‎} ‎) }{C ‎^{‎\prime} (‎\theta  (1-e^{-(‎\beta y‎_{i}‎‎)‎^{‎\gamma‎}}‎)‎^{‎\alpha‎} ‎)}‎‎‎‎\medskip\\&-\sum^{n}_{i=1}‎\frac{\theta ‎‎^{2}‎ ‎\alpha ‎^{2}‎ ‎\gamma ^2 ‎‎y‎_{i}‎‎^{‎2 \gamma‎}\beta‎^{ 2 ‎\gamma‎ -2}(e^{-(‎\beta y‎_{i}‎‎)‎^{‎\gamma‎}})^2(1-e^{-(‎\beta y‎_{i}‎‎)‎^{‎\gamma‎}}‎)‎^{‎2 \alpha -2‎}(C‎^{\prime‎ ‎\prime ‎}‎ (‎\theta (1-e^{-(‎\beta y‎_{i}‎‎)‎^{‎\gamma‎}‎})‎^{‎\alpha‎} ‎))‎^{2}‎ }{(C ‎^{‎\prime} (‎\theta  (1-e^{-(‎\beta y‎_{i}‎‎)‎^{‎\gamma‎}}‎)‎^{‎\alpha‎} ‎))‎^{2}‎},\bigskip\\\\

I_{‎\beta‎‎\gamma‎‎‎‎‎}&=‎\frac{1}{‎\beta‎}‎-\sum^{n}_{i=1}‎\gamma ‎\beta‎^{‎\gamma‎ -1} y‎_{i}‎‎^{‎\gamma‎}‎\log‎(‎\beta y‎_{i} ‎)-\sum^{n}_{i=1}‎ ‎\beta‎^{‎\gamma‎ -1} y‎_{i}‎‎^{‎\gamma‎}\medskip\\&+\sum^{n}_{i=1}‎\frac{‎\gamma (‎\alpha -1‎) ‎\beta‎^{‎\gamma‎ -1 }‎‎‎y‎_{i}‎‎^{‎\gamma‎}\log (‎\beta ‎‎‎y‎_{i}‎)e^{-(‎\beta y‎_{i}‎‎)‎^{‎\gamma‎}‎}‎‎‎}{1-e^{-(‎\beta y‎_{i}‎‎)‎^{‎\gamma‎}‎}}‎+\sum^{n}_{i=1}‎‎‎\frac{(‎\alpha -1‎) ‎\beta‎^{‎\gamma‎ -1 }‎‎‎y‎_{i}‎‎^{‎\gamma‎}e^{-(‎\beta y‎_{i}‎‎)‎^{‎\gamma‎}‎}‎‎‎}{1-e^{-(‎\beta y‎_{i}‎‎)‎^{‎\gamma‎}‎}}\medskip\\&-\sum^{n}_{i=1}‎‎\frac{(‎\alpha -1‎) ‎\gamma ‎‎\beta‎^{2 ‎\gamma‎ -1 }‎‎‎y‎_{i}‎‎^{‎2 \gamma‎}\log‎(‎\beta y‎_{i} ‎) e^{-(‎\beta y‎_{i}‎‎)‎^{‎\gamma‎}‎}‎‎‎}{1-e^{-(‎\beta y‎_{i}‎‎)‎^{‎\gamma‎}‎}}‎-\sum^{n}_{i=1}‎\frac{(‎\alpha -1‎) ‎\gamma ‎‎\beta‎^{2 ‎\gamma‎ -1 }‎‎‎y‎_{i}‎‎^{‎2 \gamma‎}\log‎(‎\beta y‎_{i} ‎) (e^{-(‎\beta y‎_{i}‎‎)‎^{‎\gamma‎}‎}‎‎‎)^2}{(1-e^{-(‎\beta y‎_{i}‎‎)‎^{‎\gamma‎}‎})^2}‎\medskip\\&+\sum^{n}_{i=1}‎‎\frac{‎\theta‎ ‎\alpha ‎\gamma ‎‎y‎_{i}‎‎^{‎\gamma‎}\beta‎^{ ‎\gamma‎ -1}\log‎(‎\beta y‎_{i} ‎)e^{-(‎\beta y‎_{i}‎‎)‎^{‎\gamma‎}}(1-e^{-(‎\beta y‎_{i}‎‎)‎^{‎\gamma‎}}‎)‎^{‎\alpha -1‎}C‎^{\prime‎ ‎\prime ‎}‎ (‎\theta (1-e^{-(‎\beta y‎_{i}‎‎)‎^{‎\gamma‎}‎})‎^{‎\alpha‎} ‎) }{C ‎^{‎\prime} (‎\theta (1-e^{-(‎\beta y‎_{i}‎‎)‎^{‎\gamma‎}}‎)‎^{‎\alpha‎} ‎)}‎‎‎\medskip\\&+\sum^{n}_{i=1}‎‎‎\frac{‎\theta‎ ‎\alpha ‎ ‎‎y‎_{i}‎‎^{‎\gamma‎}\beta‎^{ ‎\gamma‎ -1}e^{-(‎\beta y‎_{i}‎‎)‎^{‎\gamma‎}}(1-e^{-(‎\beta y‎_{i}‎‎)‎^{‎\gamma‎}}‎)‎^{‎\alpha -1‎}C‎^{\prime‎ ‎\prime ‎}‎ (‎\theta (1-e^{-(‎\beta y‎_{i}‎‎)‎^{‎\gamma‎}‎})‎^{‎\alpha‎} ‎) }{C ‎^{‎\prime} (‎\theta (1-e^{-(‎\beta y‎_{i}‎‎)‎^{‎\gamma‎}}‎)‎^{‎\alpha‎} ‎)}‎‎‎\medskip\\&-\sum^{n}_{i=1}‎‎‎\frac{‎\theta‎ ‎\alpha ‎\gamma ‎‎y‎_{i}‎‎^{‎2 \gamma‎}\beta‎^{ ‎2 \gamma‎ -1}\log‎(‎\beta y‎_{i} ‎)e^{-(‎\beta y‎_{i}‎‎)‎^{‎\gamma‎}}(1-e^{-(‎\beta y‎_{i}‎‎)‎^{‎\gamma‎}}‎)‎^{‎\alpha -1‎}C‎^{\prime‎ ‎\prime ‎}‎ (‎\theta (1-e^{-(‎\beta y‎_{i}‎‎)‎^{‎\gamma‎}‎})‎^{‎\alpha‎} ‎) }{C ‎^{‎\prime} (‎\theta (1-e^{-(‎\beta y‎_{i}‎‎)‎^{‎\gamma‎}}‎)‎^{‎\alpha‎} ‎)}‎‎‎\medskip\\&+\sum^{n}_{i=1}‎‎‎‎‎‎\frac{‎(‎\alpha -1‎)\theta‎ ‎\alpha ‎\gamma ‎‎y‎_{i}‎‎^{‎2 \gamma‎}\beta‎^{ ‎2 \gamma‎ -1}\log‎(‎\beta y‎_{i} ‎)(e^{-(‎\beta y‎_{i}‎‎)‎^{‎\gamma‎}})^2(1-e^{-(‎\beta y‎_{i}‎‎)‎^{‎\gamma‎}}‎)‎^{‎\alpha -2‎}C‎^{\prime‎ ‎\prime ‎}‎ (‎\theta (1-e^{-(‎\beta y‎_{i}‎‎)‎^{‎\gamma‎}‎})‎^{‎\alpha‎} ‎) }{C ‎^{‎\prime} (‎\theta (1-e^{-(‎\beta y‎_{i}‎‎)‎^{‎\gamma‎}}‎)‎^{‎\alpha‎} ‎)}‎‎‎\medskip\\&+\sum^{n}_{i=1}‎‎‎‎‎‎‎‎‎\frac{‎\theta‎ ^2 ‎\alpha ^2 ‎\gamma ‎‎y‎_{i}‎‎^{‎2 \gamma‎}\beta‎^{ ‎2 \gamma‎ -1}\log‎(‎\beta y‎_{i} ‎)(e^{-(‎\beta y‎_{i}‎‎)‎^{‎\gamma‎}})^2(1-e^{-(‎\beta y‎_{i}‎‎)‎^{‎\gamma‎}}‎)‎^{‎2 \alpha -2‎}C‎^{\prime‎ ‎\prime \prime‎}‎ (‎\theta (1-e^{-(‎\beta y‎_{i}‎‎)‎^{‎\gamma‎}‎})‎^{‎\alpha‎} ‎) }{C ‎^{‎\prime} (‎\theta (1-e^{-(‎\beta y‎_{i}‎‎)‎^{‎\gamma‎}}‎)‎^{‎\alpha‎} ‎)}‎‎‎\medskip\\&-\sum^{n}_{i=1}‎‎‎\frac{‎\theta‎ ^2 ‎\alpha ^2 ‎\gamma ‎‎y‎_{i}‎‎^{‎2 \gamma‎}\beta‎^{ ‎2 \gamma‎ -1}\log‎(‎\beta y‎_{i} ‎)(e^{-(‎\beta y‎_{i}‎‎)‎^{‎\gamma‎}})^2(1-e^{-(‎\beta y‎_{i}‎‎)‎^{‎\gamma‎}}‎)‎^{‎2 \alpha -2‎}(C‎^{\prime‎ ‎\prime ‎}‎ (‎\theta (1-e^{-(‎\beta y‎_{i}‎‎)‎^{‎\gamma‎}‎})‎^{‎\alpha‎} ‎))^2 }{(C ‎^{‎\prime} (‎\theta (1-e^{-(‎\beta y‎_{i}‎‎)‎^{‎\gamma‎}}‎)‎^{‎\alpha‎} ‎))^2},\bigskip\\\\

I_{‎\beta‎\theta‎‎‎‎‎‎}&=\sum^{n}_{i=1}‎‎‎‎‎‎‎‎‎‎‎\frac{‎ ‎\alpha ‎\gamma ‎‎y‎_{i}‎‎^{‎\gamma‎}\beta‎^{ ‎\gamma‎ -1}e^{-(‎\beta y‎_{i}‎‎)‎^{‎\gamma‎}}(1-e^{-(‎\beta y‎_{i}‎‎)‎^{‎\gamma‎}}‎)‎^{‎\alpha -1‎}C‎^{\prime‎ ‎\prime ‎}‎ (‎\theta (1-e^{-(‎\beta y‎_{i}‎‎)‎^{‎\gamma‎}‎})‎^{‎\alpha‎} ‎) }{C ‎^{‎\prime} (‎\theta (1-e^{-(‎\beta y‎_{i}‎‎)‎^{‎\gamma‎}}‎)‎^{‎\alpha‎} ‎)}‎‎‎\medskip\\&+\sum^{n}_{i=1}‎‎‎‎‎\frac{‎\theta‎ ‎\alpha ‎\gamma ‎‎y‎_{i}‎‎^{‎\gamma‎}\beta‎^{ ‎\gamma‎ -1}e^{-(‎\beta y‎_{i}‎‎)‎^{‎\gamma‎}}(1-e^{-(‎\beta y‎_{i}‎‎)‎^{‎\gamma‎}}‎)‎^{2 ‎\alpha -1‎}C‎^{\prime‎ ‎\prime \prime‎}‎ (‎\theta (1-e^{-(‎\beta y‎_{i}‎‎)‎^{‎\gamma‎}‎})‎^{‎\alpha‎} ‎) }{C ‎^{‎\prime} (‎\theta (1-e^{-(‎\beta y‎_{i}‎‎)‎^{‎\gamma‎}}‎)‎^{‎\alpha‎} ‎)}‎‎‎\medskip\\&-\sum^{n}_{i=1}\frac{‎\theta‎ ‎\alpha ‎\gamma ‎‎y‎_{i}‎‎^{‎\gamma‎}\beta‎^{ ‎\gamma‎ -1}e^{-(‎\beta y‎_{i}‎‎)‎^{‎\gamma‎}}(1-e^{-(‎\beta y‎_{i}‎‎)‎^{‎\gamma‎}}‎)‎^{2 ‎\alpha -1‎}(C‎^{\prime‎ ‎\prime }‎ (‎\theta (1-e^{-(‎\beta y‎_{i}‎‎)‎^{‎\gamma‎}‎})‎^{‎\alpha‎} ‎))^2 }{(C ‎^{‎\prime} (‎\theta (1-e^{-(‎\beta y‎_{i}‎‎)‎^{‎\gamma‎}}‎)‎^{‎\alpha‎} ‎))^2},\bigskip\\\\
\end{array}
\end{equation*}
\begin{equation*}
\begin{array}{ll}
I_{‎‎\gamma‎\gamma‎‎‎‎‎‎‎‎}&=-‎\frac{n}{‎\gamma ^2‎}‎-\sum^{n}_{i=1}‎\beta‎^{‎\gamma‎ } y‎_{i}‎‎^{‎\gamma‎}(‎\log‎(‎\beta y‎_{i} ‎))^2+\sum^{n}_{i=1}‎\frac{‎(‎\alpha -1‎)\beta‎^{‎\gamma‎ } y‎_{i}‎‎^{‎\gamma‎}(‎\log‎(‎\beta y‎_{i} ‎))^2 e^{-(‎\beta y‎_{i}‎‎)‎^{‎\gamma‎}}}{1-e^{-(‎\beta y‎_{i}‎‎)‎^{‎\gamma‎}‎}}‎‎‎‎\medskip\\&-\sum^{n}_{i=1}‎\frac{‎(‎\alpha -1‎)\beta‎^{2 ‎\gamma‎ } y‎_{i}‎‎^{‎2 \gamma‎}(‎\log‎(‎\beta y‎_{i} ‎))^2 e^{-(‎\beta y‎_{i}‎‎)‎^{‎\gamma‎}}}{1-e^{-(‎\beta y‎_{i}‎‎)‎^{‎\gamma‎}‎}}-\sum^{n}_{i=1}\frac{‎(‎\alpha -1‎)\beta‎^{2 ‎\gamma‎ } y‎_{i}‎‎^{‎2 \gamma‎}(‎\log‎(‎\beta y‎_{i} ‎))^2 e^{-(‎\beta y‎_{i}‎‎)‎^{‎2 \gamma‎}}}{(1-e^{-(‎\beta y‎_{i}‎‎)‎^{‎\gamma‎}‎})^2}‎‎‎‎\medskip\\&+\sum^{n}_{i=1}\frac{‎\theta‎ ‎\alpha ‎ ‎‎\beta‎^{ ‎\gamma‎ }y‎_{i}‎‎^{‎\gamma‎}(\log‎(‎\beta y‎_{i} ‎))^2e^{-(‎\beta y‎_{i}‎‎)‎^{‎\gamma‎}}(1-e^{-(‎\beta y‎_{i}‎‎)‎^{‎\gamma‎}}‎)‎^{‎\alpha -1‎}C‎^{\prime‎ ‎\prime ‎}‎ (‎\theta (1-e^{-(‎\beta y‎_{i}‎‎)‎^{‎\gamma‎}‎})‎^{‎\alpha‎} ‎) }{C ‎^{‎\prime} (‎\theta (1-e^{-(‎\beta y‎_{i}‎‎)‎^{‎\gamma‎}}‎)‎^{‎\alpha‎} ‎)}‎‎‎\medskip\\&-\sum^{n}_{i=1}\frac{‎\theta‎ ‎\alpha ‎ ‎‎\beta‎^{ 2‎\gamma‎ }y‎_{i}‎‎^{‎2\gamma‎}(\log‎(‎\beta y‎_{i} ‎))^2e^{-(‎\beta y‎_{i}‎‎)‎^{‎\gamma‎}}(1-e^{-(‎\beta y‎_{i}‎‎)‎^{‎\gamma‎}}‎)‎^{‎\alpha -1‎}C‎^{\prime‎ ‎\prime ‎}‎ (‎\theta (1-e^{-(‎\beta y‎_{i}‎‎)‎^{‎\gamma‎}‎})‎^{‎\alpha‎} ‎) }{C ‎^{‎\prime} (‎\theta (1-e^{-(‎\beta y‎_{i}‎‎)‎^{‎\gamma‎}}‎)‎^{‎\alpha‎} ‎)}‎‎‎\medskip\\&+\sum^{n}_{i=1}\frac{‎\theta‎ ‎\alpha (‎\alpha -1‎)‎ ‎‎\beta‎^{ 2‎\gamma‎ }y‎_{i}‎‎^{‎2\gamma‎}(\log‎(‎\beta y‎_{i} ‎))^2 (e^{-(‎\beta y‎_{i}‎‎)‎^{‎\gamma‎}})^2(1-e^{-(‎\beta y‎_{i}‎‎)‎^{‎\gamma‎}}‎)‎^{‎\alpha -2‎}C‎^{\prime‎ ‎\prime ‎}‎ (‎\theta (1-e^{-(‎\beta y‎_{i}‎‎)‎^{‎\gamma‎}‎})‎^{‎\alpha‎} ‎) }{C ‎^{‎\prime} (‎\theta (1-e^{-(‎\beta y‎_{i}‎‎)‎^{‎\gamma‎}}‎)‎^{‎\alpha‎} ‎)}‎‎‎\medskip\\&+\sum^{n}_{i=1}
\frac{‎\theta‎ ^2‎\alpha ^2 ‎ ‎‎\beta‎^{ 2‎\gamma‎ }y‎_{i}‎‎^{‎2\gamma‎}(\log‎(‎\beta y‎_{i} ‎))^2 (e^{-(‎\beta y‎_{i}‎‎)‎^{‎\gamma‎}})^2(1-e^{-(‎\beta y‎_{i}‎‎)‎^{‎\gamma‎}}‎)‎^{‎2\alpha -2‎}C‎^{\prime‎ ‎\prime \prime‎}‎ (‎\theta (1-e^{-(‎\beta y‎_{i}‎‎)‎^{‎\gamma‎}‎})‎^{‎\alpha‎} ‎) }{C ‎^{‎\prime} (‎\theta (1-e^{-(‎\beta y‎_{i}‎‎)‎^{‎\gamma‎}}‎)‎^{‎\alpha‎} ‎)}‎‎‎\medskip\\&-\sum^{n}_{i=1}\frac{‎\theta‎ ^2‎\alpha ^2 ‎ ‎‎\beta‎^{ 2‎\gamma‎ }y‎_{i}‎‎^{‎2\gamma‎}(\log‎(‎\beta y‎_{i} ‎))^2 (e^{-(‎\beta y‎_{i}‎‎)‎^{‎\gamma‎}})^2(1-e^{-(‎\beta y‎_{i}‎‎)‎^{‎\gamma‎}}‎)‎^{‎2\alpha -2‎}(C‎^{\prime‎ ‎\prime ‎}‎ (‎\theta (1-e^{-(‎\beta y‎_{i}‎‎)‎^{‎\gamma‎}‎})‎^{‎\alpha‎} ‎))^2 }{(C ‎^{‎\prime} (‎\theta (1-e^{-(‎\beta y‎_{i}‎‎)‎^{‎\gamma‎}}‎)‎^{‎\alpha‎} ‎))^2},\bigskip\\\\

I_{‎‎\gamma‎‎\theta‎}&=\sum^{n}_{i=1}‎\frac{‎\alpha ‎\beta‎^{‎\gamma‎ } y‎_{i}‎‎^{‎\gamma‎}‎\log‎(‎\beta y‎_{i} ‎)e^{-(‎\beta y‎_{i}‎‎)‎^{‎\gamma‎}}(1-e^{-(‎\beta y‎_{i}‎‎)‎^{‎\gamma‎}}‎)‎^{‎\alpha -1‎}C‎^{\prime‎ ‎\prime ‎}‎ (‎\theta (1-e^{-(‎\beta y‎_{i}‎‎)‎^{‎\gamma‎}‎})‎^{‎\alpha‎} ‎) }{C ‎^{‎\prime} (‎\theta (1-e^{-(‎\beta y‎_{i}‎‎)‎^{‎\gamma‎}}‎)‎^{‎\alpha‎} ‎)}‎‎‎‎\medskip\\&+\sum^{n}_{i=1}‎\frac{‎\theta‎ ‎\alpha \beta‎^{‎\gamma‎ } y‎_{i}‎‎^{‎\gamma‎}‎\log‎(‎\beta y‎_{i} ‎)‎e^{-(‎\beta y‎_{i}‎‎)‎^{‎\gamma‎}}(1-e^{-(‎\beta y‎_{i}‎‎)‎^{‎\gamma‎}}‎)‎^{2‎\alpha -1‎}C‎^{\prime‎ ‎\prime ‎\prime}‎ (‎\theta (1-e^{-(‎\beta y‎_{i}‎‎)‎^{‎\gamma‎}‎})‎^{‎\alpha‎} ‎) }{C ‎^{‎\prime} (‎\theta (1-e^{-(‎\beta y‎_{i}‎‎)‎^{‎\gamma‎}}‎)‎^{‎\alpha‎} ‎)}‎‎‎‎‎\medskip\\&-\sum^{n}_{i=1}‎\frac{‎\theta‎ ‎\alpha ‎ ‎‎\beta‎^{ ‎\gamma‎ }y‎_{i}‎‎^{‎\gamma‎}\log‎(‎\beta y‎_{i} ‎)e^{-(‎\beta y‎_{i}‎‎)‎^{‎\gamma‎}}(1-e^{-(‎\beta y‎_{i}‎‎)‎^{‎\gamma‎}}‎)‎^{2‎\alpha -1‎}(C‎^{\prime‎ ‎\prime ‎}‎ (‎\theta (1-e^{-(‎\beta y‎_{i}‎‎)‎^{‎\gamma‎}‎})‎^{‎\alpha‎}) ‎)^2}{(C ‎^{‎\prime} (‎\theta (1-e^{-(‎\beta y‎_{i}‎‎)‎^{‎\gamma‎}}‎)‎^{‎\alpha‎} ‎))^2}‎,\bigskip\\\\

I_{‎\theta‎‎‎\theta‎}&=-‎\frac{n}{‎\theta‎^{2}‎‎}‎+\sum^{n}_{i=1}‎\frac{(1-e^{-(‎\beta y‎_{i}‎‎)‎^{‎\gamma‎}}‎)‎^{‎2\alpha ‎}C‎^{\prime‎ ‎\prime \prime‎}‎ (‎\theta (1-e^{-(‎\beta y‎_{i}‎‎)‎^{‎\gamma‎}‎})‎^{‎\alpha‎} ‎)}{C ‎^{‎\prime} (‎\theta (1-e^{-(‎\beta y‎_{i}‎‎)‎^{‎\gamma‎}}‎)‎^{‎\alpha‎} ‎)}‎‎‎‎‎\medskip\\&-\sum^{n}_{i=1}‎\frac{(1-e^{-(‎\beta y‎_{i}‎‎)‎^{‎\gamma‎}}‎)‎^{‎2\alpha ‎}(C‎^{\prime‎ ‎\prime ‎}‎ (‎\theta (1-e^{-(‎\beta y‎_{i}‎‎)‎^{‎\gamma‎}‎})‎^{‎\alpha‎} ‎))^2}{(C ‎^{‎\prime} (‎\theta (1-e^{-(‎\beta y‎_{i}‎‎)‎^{‎\gamma‎}}‎)‎^{‎\alpha‎} ‎))^2}-‎\frac{nC‎^{\prime‎ ‎\prime }‎(‎\theta‎)}{C (‎\theta)}‎+‎\frac{n(C ‎^{‎\prime}(‎\theta‎))^2}{(C (‎\theta))^2}‎.
\end{array}
\end{equation*}

\end{document}